\documentclass[12pt]{iopart}
\usepackage{epsfig}
\usepackage{epstopdf}
\usepackage{caption}
\usepackage{color}
\usepackage{enumerate}
\usepackage{dsfont}
\usepackage{cite}
\usepackage{mathrsfs}
\newcommand{\ket}[1]{{\left | {#1} \right\rangle}}
\newcommand{\bra}[1]{{\left\langle {#1} \right |}}

\begin{document}

\title{\bf 
Quantum Information Processing with Superconducting Circuits: a Review}
\author{G. Wendin}
\address{Department of Microtechnology and Nanoscience - MC2, \\
Chalmers University of Technology, \\SE-41296 Gothenburg, Sweden}

\date{\today}

\begin{abstract}

During the last ten years, superconducting circuits have passed from being interesting physical devices to becoming contenders for near-future useful and scalable quantum information processing (QIP). Advanced quantum simulation experiments have been shown with up to nine qubits, while a demonstration of Quantum Supremacy with fifty qubits is anticipated in just a few years. Quantum Supremacy means that the quantum system can no longer be simulated by the most powerful classical supercomputers. Integrated classical-quantum computing systems are already emerging that can be used for software development and experimentation, even via web interfaces.

Therefore, the time is  ripe for describing some of the recent development of superconducting devices, systems and applications. As such, the discussion of superconducting qubits and circuits is limited to devices that are proven useful for current or near future applications. Consequently, the centre of interest is the practical applications of QIP, such as computation and simulation in Physics and Chemistry. \\

\end{abstract}

\maketitle

\tableofcontents

\newpage

\section{Introduction}

Quantum Computing is the art of controlling the time evolution of highly complex, entangled quantum states in physical hardware registers for computation and simulation. Quantum Supremacy is a recent term for an old ambition - to prove and demonstrate that quantum computers can outperform conventional classical computers \cite {Preskill2012}.

Since the 1980's, quantum computer science has been way ahead of experiment, driving the development of quantum information processing (QIP) at abstract and formal levels. This situation may now be changing, due to recent experimental advances to scale up and operate highly coherent and operational qubit platforms. In particular, one can expect superconducting quantum hardware systems with 50 qubits or more during the next few years. 

The near-term goal is to operate a physical quantum device that a classical computer cannot simulate \cite {Boixo2016b}, therefore demonstrating Quantum Supremacy.
For QIP to be of interest one often requests "killer applications", outperforming classical supercomputers on real-world applications like factorisation and code breaking \cite{Shor1994}. 
However, this is not a realistic way to look upon the power of QIP. During the last seventy years, classical information processing has progressed via continuous development and improvement of more or less efficient algorithms to solve specific tasks, in tune with the development of increasingly powerful hardware. The same will certainly apply also to QIP, the really useful applications arriving along the way. 

There is, in fact, already a clear short-term QIP perspective, recognising the importance of quantum technologies (QT) and quantum engineering for driving the present and near-future development \cite{Qmanifesto2016}. This is necessary for developing large scale devices to achieve the long-term goals of useful quantum computing. To this end, investigations of quantum device physics have been essential for the present efforts to scale up of multi-qubit platforms.
The work on improving coherence has demonstrated that qubits are extremely sensitive to a variety of noise sources. The requirement of making measurements without destroying the coherence has led to the development of a range of quantum limited superconducting amplifiers. As a result, advanced quantum sensors and quantum measurement may give rise to a new quantum technology:  "a qubit at the tip of a scanning probe", greatly enhancing the sensitivity of magnetic measurements \cite{Neumann2013,Bienfait2015,Bienfait2016}.

The potential of superconducting circuits for QIP has been recognised for more than twenty years \cite{Eikmans1990,Makhlin1999,MakhlinRMP2001}. The first experimental realisation of the simplest qubit, a Josephson-junction (JJ) based Cooper-pair Box (CPB) in the charge regime, was demonstrated in 1999 \cite{Nakamura1999}. Originally, the coherence time was very short, just a few nanoseconds, but it took only another few years to demonstrate a number of useful qubit concepts:  
the flux qubit \cite{Friedman2000,vdWal2000,Chiorescu2002}, the quantronium CPB \cite{Vion2002}, and the phase qubit \cite{Martinis2002}.
The next important step was to embed qubits in a superconducting microwave resonator, introducing circuit quantum electrodynamics (cQED)\cite{Yang2003,Yang2004,YouNori2003,Kis2004,Blais2004,DevWallMart2004}. The subsequent experiments using a 2D  superconducting coplanar microwave resonator \cite{Wallraff2004,Majer2007} demonstrated groundbreaking progress: 
\begin{enumerate}
\item Microwave qubit control, strong qubit-resonator coupling and dispersive readout \cite{Wallraff2004}; 
\item Coupling of CPB qubits and swapping excitations, in practice implementing a universal $\sqrt{iSWAP}$ gate \cite{Majer2007}.
\end{enumerate}
A basis for potentially scalable multi-qubit systems with useful long coherence times - the transmon version of the CPB - was published in 2007 \cite{Koch2007}. Moreover, in 2011 the invention of a transmon embedded in a 3D-cavity increased coherence times toward 100 $\mu$s \cite{Paik2011}. At present there is intense development of both 2D and 3D multi-qubit circuits with long-lived qubits and resonators, capable of performing a large number of high-fidelity quantum gates with control and readout operations. 

The purpose of this review is to provide a snapshot of current progress, and to outline some expectations for the future. We will focus on hardware and protocols actually implemented on current superconducting devices, with discussion of the most promising development to scale up superconducting circuits and systems. In fact, superconducting quantum circuits are now being scaled up experimentally to systems with several tens of qubits, to address real issues of quantum computing and simulation \cite{Chow2014,Corcoles2015,IBM_Takita2016,IBM_Takita2017,Versluis2016,Barends2014a,Saira2014,Riste2015,Kelly2015a,Barends2015a,Barends2016,OMalley2016,Asaad2015,Song2017,Kandala2017}. 

The aim is here to present a self-contained discussion for a broad QIP readership. To this end, time evolution and the construction and implementation of 1q and 2q gates in superconducting devices are treated in considerable detail to make recent experimental work more easily accessible. On the other hand, the more general discussion of theory, as well as much of the experimental work, only touches the surface and is covered by references to recent work.
Reviews and analyses of the QIP field are provided by \cite{SchoelGirv2008,ClarkWil2008,NatureInsight2008,Ladd2010,Siddiqi2011,LesHouches2014,DevSchoel2013,CiracZoller2010,Brown2010,Barreiro2011,BlattRoos2012,Georgescu2014}. For a comprehensive treatise on QIP we refer to Nielsen and Chuang \cite{NielsenChuang2010}. Extensive technical discussions of a broad range of superconducting circuits and qubits can be found in \cite{WeShu2007,Girvin2009,Girvin2014,Vool_Dev2016,Bylander2017,Gu2017}.

The present review also tries to look beyond the experimental state of the art, to anticipate what will be coming up in the near future in the way of applications. There is so much theoretical experience that waits to be implemented on superconducting platforms.
The ambition is to outline opportunities for addressing real-world problems in Physics, Chemistry and Materials Science.

\newpage

\section{Easy and hard problems}

Why are quantum computers and quantum simulators of such great interest? Quantum computers are certainly able to solve some problems much faster than classical computers. However, this does not say much about  solving computational problems that are hard for classical computers. Hard problems are not only a question of whether they take long time - the question is whether they can be solved at all with finite resources.

\begin{figure}[h]
\includegraphics[width=12cm]{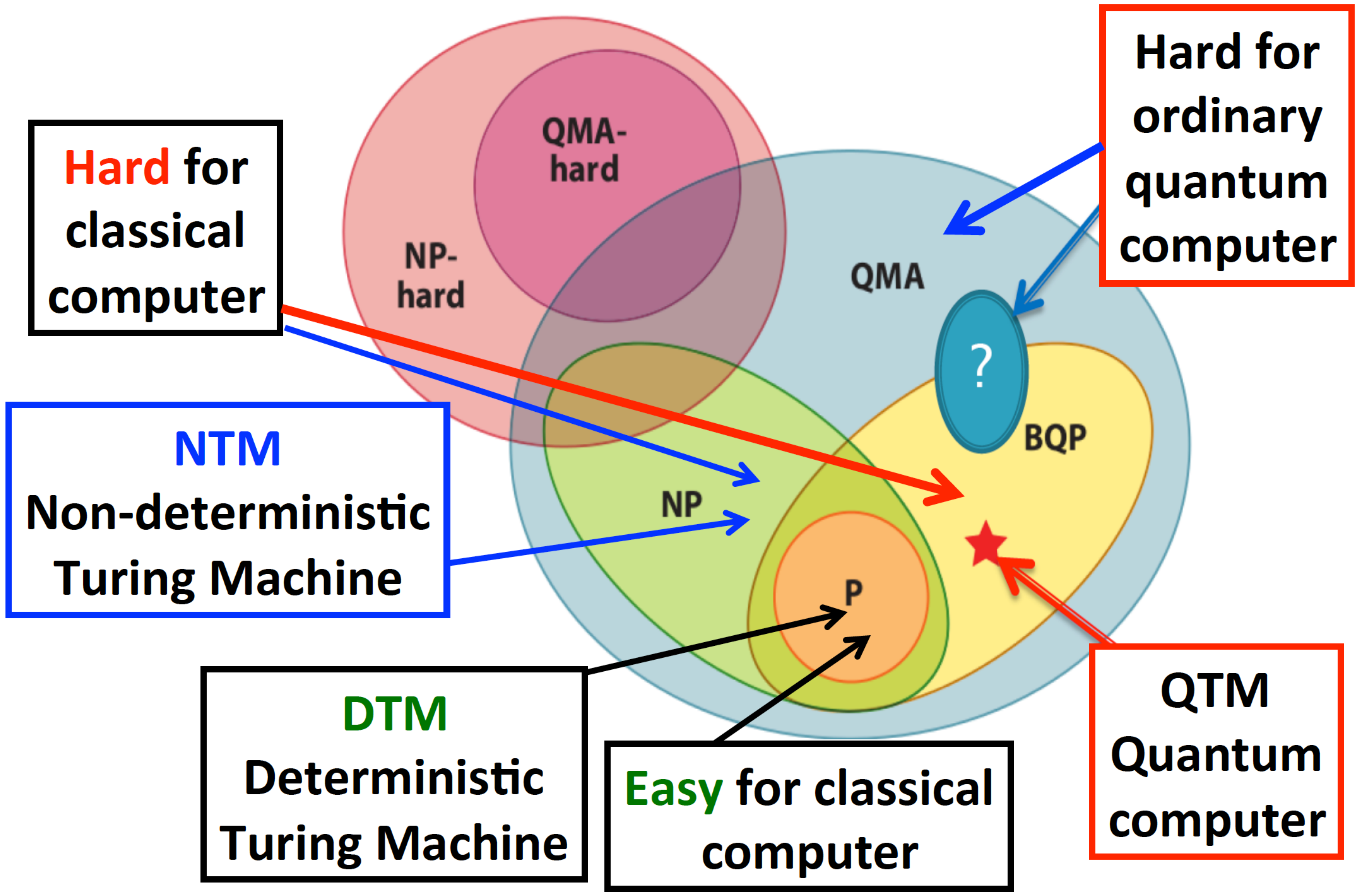}
\centering
\caption
{\small Computational complexity is defined by Turing machines (TM) providing digital models of computing  \cite {Preskill2012,Watrous2009,CplxZoo2014}: deterministic TM (DTM); quantum TM (QTM); classical non-deterministic TM (NTM).
Tractable problems are defined by polynomial time execution and define complexity classes: 
P denotes problems that are efficiently solvable with a classical computer; P is a subset of NP, the problems
efficiently checkable by a classical computer. QMA denotes the problems
efficiently checkable by a quantum computer. NP-hard problems are the problems at least
as hard as any NP problem, and QMA-hard problems are the problems at least as hard as any
QMA problem.  
For a nice tutorial on how to classify combinatorial problems, including games, see \cite{Aaronson2008}.
}
\label{cplxmap}
\end{figure}

The map of computational complexity in Fig.~\ref{cplxmap} classifies the hardness of computational (decision) problems. Quantum computation belongs to class BQP (bounded-error quantum polynomial time).
Figure~\ref{cplxmap} shows that the BQP only encompasses a rather limited space, basically not solving really hard problems. One may then ask what is the relation between problems of practical interest and really hard mathematical problems - what  is the usefulness of quantum computing, and which problems are hard even for quantum computers? 
 
\subsection{Computational complexity}

Computational complexity  \cite {Preskill2012,Watrous2009,CplxZoo2014,Montanaro2016} is defined in terms of different kinds of Turing machines (TM) providing digital models of computing. A universal TM (UTM) can simulate any other TM (including quantum computers) and defines what is computable in principle, without caring about  time and memory. Problems that can be solved by a deterministic TM (DTM) in polynomial time belong to class P (Fig.~\ref{cplxmap}), and are considered to be "easy", or at least tractable. A DTM is a model for ordinary classical computers - a finite state machine (FSM) reading and writing from a finite tape.

A probabilistic TM (PTM) makes random choices of the state of the FSM upon reading from the tape, and traverses all the states in a random sequence. This defines the class BPP (bounded-error probabilistic polynomial time). A PTM may be more powerful then a DTM since it avoids getting stuck away from the solution. Nevertheless, a PTM can be simulated by a DTM with only polynomial overhead, so the relation BPP=P is believed to be true.

A quantum TM (QTM) is a model for a quantum computer with a quantum processor and quantum tape (quantum memory). Problems that can be solved by a QTM in polynomial time belong to class BQP (Fig.~\ref{cplxmap}). There, outside P  (P $\subset$ BQP), we find problems like Shor's algorithm \cite {Shor1994} where a QTM provides exponential speedup. Nevertheless, Fig.~\ref{cplxmap} shows that BQP is a limited region of the complexity map, not including a large part of the NP-class containing many hard problems for a classical computer. It is unknown whether these are hard for a quantum computer. The class NP (non-deterministic polynomial) is defined by a non-deterministic TM (NTM), able to provide an answer that can be verified by a DTM in polynomial time. The NTM is not a real computer but rather works as an oracle, providing an answer. A subclass of NP is the MA (Merlin-Arthur) class where the all-mighty Merlin provides the answer which the classical Arthur can verify in polynomial time. Some of these problems are beyond a quantum computer to calculate, but it might be used to verify solutions in polynomial time. This is the large quantum Merlin-Arthur (QMA) complexity class shown in Fig.~\ref{cplxmap}. When not even a quantum computer can verify a solution in polynomial time, then that problem belongs to the NP-hard complexity class, where even a quantum computer is of no use.

\subsection{Hard problems}

Can a physical processes exist while being too hard to compute? Or does Nature actually not solve really hard instances of hard problems? Perhaps the results of Evolution are based on optimisation and compromises?

There is a long-standing notion that unconventional adaptive analogue computers can provide solutions to NP-hard problems that take exponential resources (time and/or memory) for classical digital machines to solve  \cite {Adleman1994,Lipton1995,Siegelmann1995,Copeland2011,Cabessa2011,Manea2007,Traversa2015}.
The key question therefore is: can unconventional computing provide solutions to NP-complete problems?   
The answer is in principle given by the Strong Church Thesis: Any finite analogue computer can be simulated  {\it efficiently} by a digital computer.
This is due to the time required by the digital computer to simulate the analogue computer being This is due to the time required by the digital computer to simulate the analogue computer being bounded by a polynomial function of the resources used by the analogue machine \cite{Vergis1986}. This suggests that physical systems (both digital and analogue) cannot provide solutions to NP-complete problems. NP-completeness is a worst case analysis, where at least one case requires exponential, rather than polynomial, resources in the form of time or memory.

Given the idea that "Nature is physical and does not solve NP-hard problems"  \cite{Aaronson2008,BroStepWen2017,Aaronson2005,Aaronson2013}, where does this place quantum computing? 
In principle, in a better position than classical computing, to compute the properties of physical quantum systems.
Tractable problems for quantum computers (BQP) are in principle hard for classical computers (P).
In 1982 Feynman  \cite{Feynman1982} introduced the concept of simulating one quantum system by another, emulating quantum physics by tailored quantum systems describing model quantum Hamiltonians (analogue quantum computers). The subsequent development went mostly in the direction of gate circuit models \cite{Barenco1995}, but two decades later the analogue/adiabatic approach was formally established as an equivalent universal approach \cite{Farhi2000,Farhi2001,Aharonov2004,Kempe2004,Cao2015}. Nevertheless, also for quantum computers the class of tractable problems (BQP)  is limited. There are many problems described by quantum Hamiltonians that are hard for quantum computers, residing in QMA, or worse  (QMA-hard, or NP-hard) \cite{Osborne2012,Gharibian2014,Rassolov2008,Schuch2009,Aaronson2009,Whitfield2013,Whitfield2014,WhitfieldZimboras2014,GossetNagaj2013,Farhi2014,Viglietta2012,Aloupis2012}. Although, the many-body problem is tractable
for quantum spin chains \cite{Verstraete2015,Landau2015}.
When quantum mechanics is not involved, e.g. in combinatorial problems, quantum computers may not have any advantage \cite{Aaronson2008}.

\subsection{Quantum speedup}

Quantum speedup is by definition connected with non-classical correlations \cite{JozsaLinden2003,Horodecki2009}.
{\it Entanglement} is a fundamental manifestation of quantum superposition and non-classical correlations for pure states  \cite{Gisin2014}.
An elementary example of classical behaviour is provided by a tensor product of  independent superpositions of two 2-level systems, $\ket{\psi_1} = \frac{1}{\sqrt{2}}(\ket{0} + \ket{1})$ and  $\ket{\psi_2} = \frac{1}{\sqrt{2}}(\ket{0} + \ket{1})$. The tensor product of $N$(= 2) states $\ket{\psi} = \ket{\psi_1} \otimes \ket{\psi_2}  = \frac{1}{\sqrt{2}}(\ket{0} + \ket{1}) \otimes \frac{1}{\sqrt{2}}(\ket{0} + \ket{1}) = \frac{1}{2}(\ket{00} + \ket{01} + \ket{10} + \ket{11})$ contains $2^N$ (= 4) superposed configurations: this is the basis for creating exponentially large superpositions with only a linear amount of physical resources (qubits). 

Highly entangled states are created by finite numbers of superpositions. In the present 2-qubit case, a maximally entangled state is the 2-qubit Bell state,  $\frac{1}{\sqrt{2}} (\ket{00} + \ket{11})$: it is not possible to assign a single state vector to any of the two subsystems, only to the total system. Entanglement allows us to construct maximally entangled superpositions with only a linear amount of physical resources, e.g. a large cat state: $\frac{1}{\sqrt{2}} (\ket{0......00} +  \ket{1.....11})$, entangling $N$ 2-level systems. This is what allows us to perform non-classical tasks and provide speedup \cite{JozsaLinden2003,Horodecki2009}.
Interestingly, just to characterise the entanglement can be a hard problem for a classical computer, because several  entanglement measures are NP-hard to compute \cite{Huang2014}. 
There is a large number of measures of entanglement, e.g. concurrence; entropy of entanglement (bipartite); entanglement of formation; negativity; quantum discord \cite{Horodecki2009,Huang2014,Wootters2001,EisertRMP2010,Eltschka2014,Girolami2014,XiLiFan2015,Ma2016}. Quantum discord is defined as the difference between two classically equivalent measures of information \cite{OllivierZurek2002}, and indicates the presence of correlations due to noncommutativity of quantum operators. For pure states it equals the entropy of entanglement \cite{EisertRMP2010}.
Quantum discord determines the interferometric power of quantum states \cite{Girolami2014}.
It provides a fundamental concept for computation with mixed quantum states in open systems, separable and lacking entanglement but still providing useful non-classical correlations.

Quantum speedup is achieved by definition if a quantum calculation is successful, as discussed by Dewes et al.  \cite{Dewes2012b} in the case of Grover search with a transmon 2-qubit system. It is then related to the expected known success probabilities of the classical and quantum systems.
In general, however, speedup of a computation is an asymptotic scaling property \cite{Ronnow2014}. 
Nevertheless, in practice there are so many different aspects involving setting up and solving different instances of various classes of problems that the time to solution (TTS) may be the most relevant measure \cite{Steiger2015}.

Polynomial or exponential speedup has not been much discussed in connection with digital QC because the systems are still small (5-10 qubits), and the limited coherence time does not allow very long calculations. 
In contrast, defining and detecting quantum speedup is presently a hot issue when assessing the performance of the D-Wave quantum annealing machines \cite{Ronnow2014,Steiger2015,Denchev2015,Zintchenko2015}.

\subsection{Quantum Supremacy}

Quantum Supremacy is a recent term for the old ambition of proving that quantum computers can outperform conventional classical computers \cite {Preskill2012}.
A simple implementation of Quantum supremacy is to create a physical quantum device 
that cannot be simulated by existing classical computers with available memory in any reasonable time. 
Currently, such a device would be a 50 qubit processor. It could model a large molecule that cannot be simulated by a classical computer. It would be an artificial physical piece of quantum matter that can be characterised by various quantum benchmarking methods in a limited time, but cannot be simulated by classical computers of today.  And by scaling up by a small number of qubits it will not be simulatable even by next generation classical computers. 
A recent example of this given by Boixo et al. \cite {Boixo2016b}, discussing how to characterise  Quantum Supremacy in near-term circuits and systems with superconducting devices.


\newpage

\section{Superconducting circuits and systems}

The last twenty years have witnessed a dramatic development of coherent nano- and microsystems. When the DiVincenzo criteria  were first formulated during 1996-2000 \cite {DiVincenzo1996,DiVincenzo2000} there were essentially no useful solid-state qubit devices around. Certainly there were a number of quantum devices: Josephson junctions (JJ), Cooper pair boxes (CPB), semiconductor quantum dots, implanted spins, etc. However, there was no technology for building coherent systems that could be kept isolated from the environment and controlled at will from the outside. These problems were addressed through a steady technological development during the subsequent ten years, and the most recent development is now resulting in practical approaches toward scalable systems.

\subsection{The DiVincenzo criteria (DV1-DV7)}
\label{DVcriteria}

The seven DiVincenzo criteria \cite{DiVincenzo2000} formulate necessary conditions for gate-driven (digital) QIP:\\
1. Qubits: fabrication of registers with several (many) qubits (DV1).\\
2. Initialisation: the qubit register must be possible to initialise to a known state (DV2).\\
3. Universal gate operations: high fidelity single and 2-qubit gate operations must be available (DV3).\\
4. Readout: the state of the qubit register must be possible to read out, typically via readout of individual qubits (DV4).\\
5. Long coherence times: a large number of single and 2-qubit gate operations must be performed within the coherence time of the qubit register, $T_2$ (DV5).\\\
6. Quantum interfaces for qubit interconversion: qubit interfaces must be possible for storage and on-chip communication between qubit registers (DV6).\\
7. Quantum interfaces to flying qubits for optical communication: qubit-photon interfaces must be available for long-distance transfer of entanglement and quantum information (DV7).

\subsection{Josephson quantum circuits}

\begin{figure}[h]
\center
\includegraphics[width=6cm]{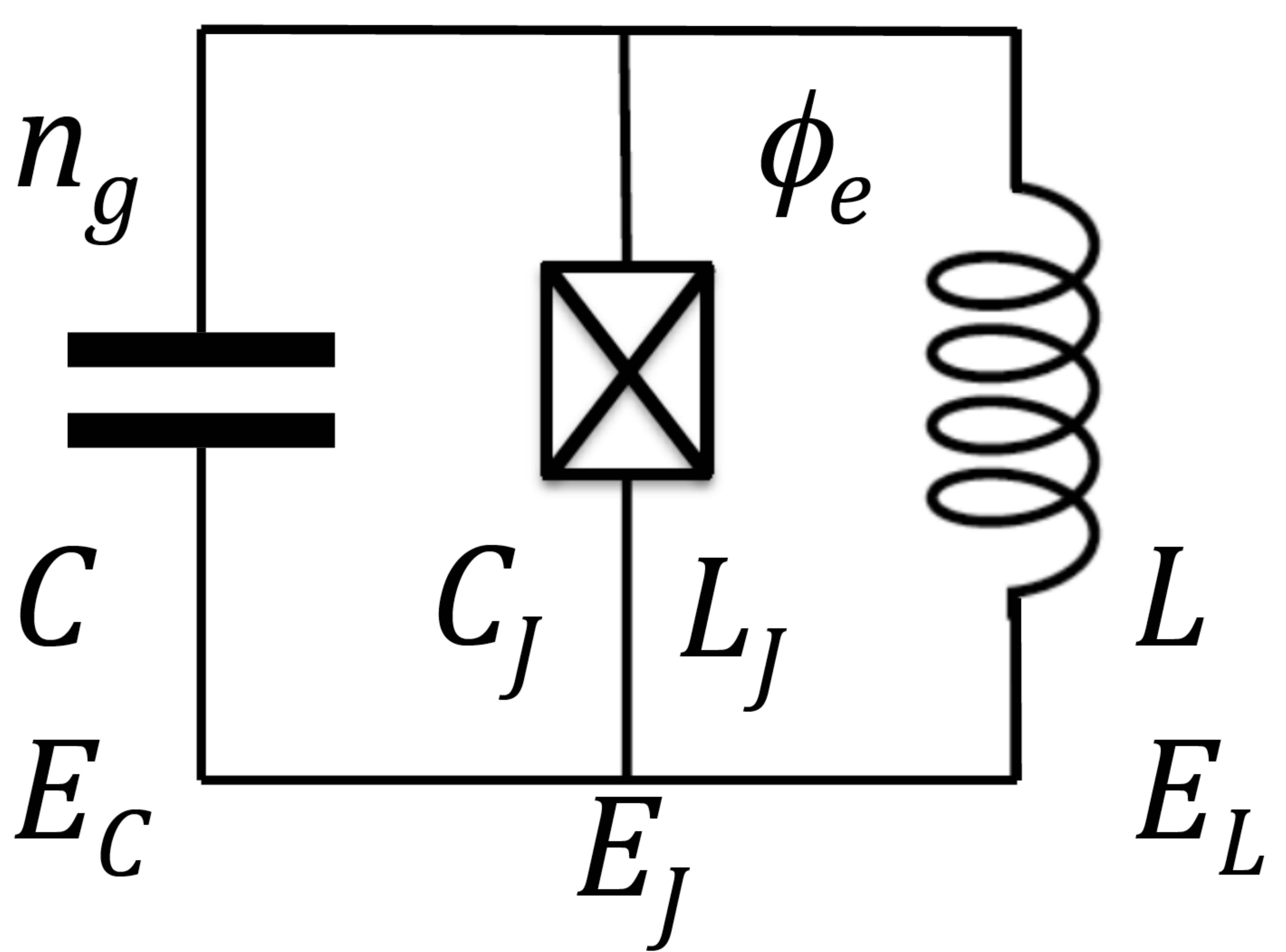}
\caption
{
\small   Basic equivalent circuit for all Josephson junction (JJ) based qubits.  $C$  represents a shunt capacitance and $L$ a shunt inductance. $C_J$ is the intrinsic capacitance of the JJ.
$E_C = {(2e)^2/2C}$,  $E_L = \hbar^2/(4e^2L)$ and  $E_{J0} = \hbar^2/(4e^2L_{J0})$. The Josephson inductance $L_{J0}$ is defined in the text. 
$n_g$ is the charge on the island induced by capacitive coupling to a voltage source (not shown), and $\phi_e$ is the phase across the JJ controlled by an external flux $\Phi$.
}
\label{CJL_fig}
\end{figure}

The recent systematic development of reliable transmon-based JJ-cQED circuits is now forming a basis for serious upscaling to 50 qubits in the near future, to develop system control, benchmarking, error correction and quantum simulation schemes. 

The qubits are based on the superconducting non-linear oscillator circuit shown in Fig. \ref{CJL_fig} (for an in-depth discussion, see Girvin \cite{Girvin2014}).
The Hamiltonian of the harmonic oscillator LC circuit alone is given by
\begin{equation}
\label{LCJ}
\hat H =  E_C\, \hat n^2  + E_L {\hat \phi^2\over 2},
\end{equation}
where $\hat n$ is the induced charge on the capacitor measured in units of 2e (Cooper pair), 
and $\hat \phi$ is the phase difference over the inductor. The charge $\hat n$ and phase $\hat\phi$ operators do not commute, $[\hat\phi,\hat n] = i$, which means that their expectation values cannot be measured simultaneously. 
$E_C = {(2e)^2/2C}$ (charging energy of one 2e Cooper pair),  $E_L = \hbar^2/(4e^2L)$, and the distance between the energy levels of the harmonic oscillator is given by $\hbar \omega = \hbar/ \sqrt{LC} = \sqrt{2 E_L E_C}$.

For a superconducting high-Q oscillator the energy levels are narrow and equidistant. However, in order to serve as a qubit, the oscillator must be anharmonic so that a specific pair of levels can be addressed. 
Adding the Josephson junction (JJ), the Hamiltonian of the  LCJ circuit becomes
\begin{equation}
\label{CJL_eq}
\hat H =  E_C\, (\hat n - n_g)^2 - E_{J0} \cos(\hat\phi) + E_L {(\hat \phi-\phi_e)^2\over 2}
\end{equation}
where $n_g$ is the voltage-induced charge on the capacitor C (qubit island), and $\phi_e$ is the flux-induced phase across the JJ. The Josephson energy $E_{J0}$ is given by $E_{J0} = \frac{\hbar}{2e} I_0$  in terms of the critical current $I_0$ of the junction \cite{WeShu2007}. Typically, the JJ is of SIS type (superconductor-insulator-superconductor) with fixed critical current. 

In order to introduce the Josephson nonlinear inductance, one starts from the fundamental Josephson relation
\begin{equation}
\label{JJ}
I_{J} = I_0$ sin $\phi
\end{equation}
Combined with Lenz' law:
\begin{equation}
\label{Lenz}
V= V= d\Phi/dt = \Phi_0/2\pi \; d\phi/dt, \;\;\;\Phi_0 = h/2e
\end{equation}
one finds that  
\begin{equation}
\label{Lenz}
 V= \Phi_0/2\pi \;(I_0 \;$cos$ \phi)^{-1}dI_J /dt. 
\end{equation}
Defining $dI_J /dt = V/L_J$, one finally obtains the Josephson inductance $L_{J0}$:
\begin{equation}
\label{Lenz}
L_{J} = \Phi_0/2\pi \; (I_0 $cos$ \phi)^{-1}  =  L_{J0} \;($cos$ \phi)^{-1} 
\end{equation}
This defines the Josephson inductance $L_{J0}$ of the isolated JJ circuit element in Fig. \ref{CJL_fig},
and allows us to express the Josephson energy as $E_{J0} = \hbar^2/(4e^2L_{J0})$.

In order to describe the energy-level structure of the quantum LCJ circuit in Fig. \ref{CJL_fig} one introduces $\hat n = -i\hbar  \frac {\partial }{\partial \phi}$ to get a Schr\"odinger equation for the circuit wave function $\psi$ in the phase variable $\phi$:
\begin{equation}
[E_C \; (-i\hbar   \frac {\partial}{\partial \phi} - n_g)^2+ U(\phi) ] \; \psi = E \; \psi
\label{PhaseSEq}
\end{equation}
\begin{equation}
U(\phi) =  - E_{J0} \cos(\phi) +  E_L{(\phi-\phi_e)^2\over 2}
\label{Upot}
\end{equation}
\begin{figure}[h]
\center
\includegraphics[width=12cm]{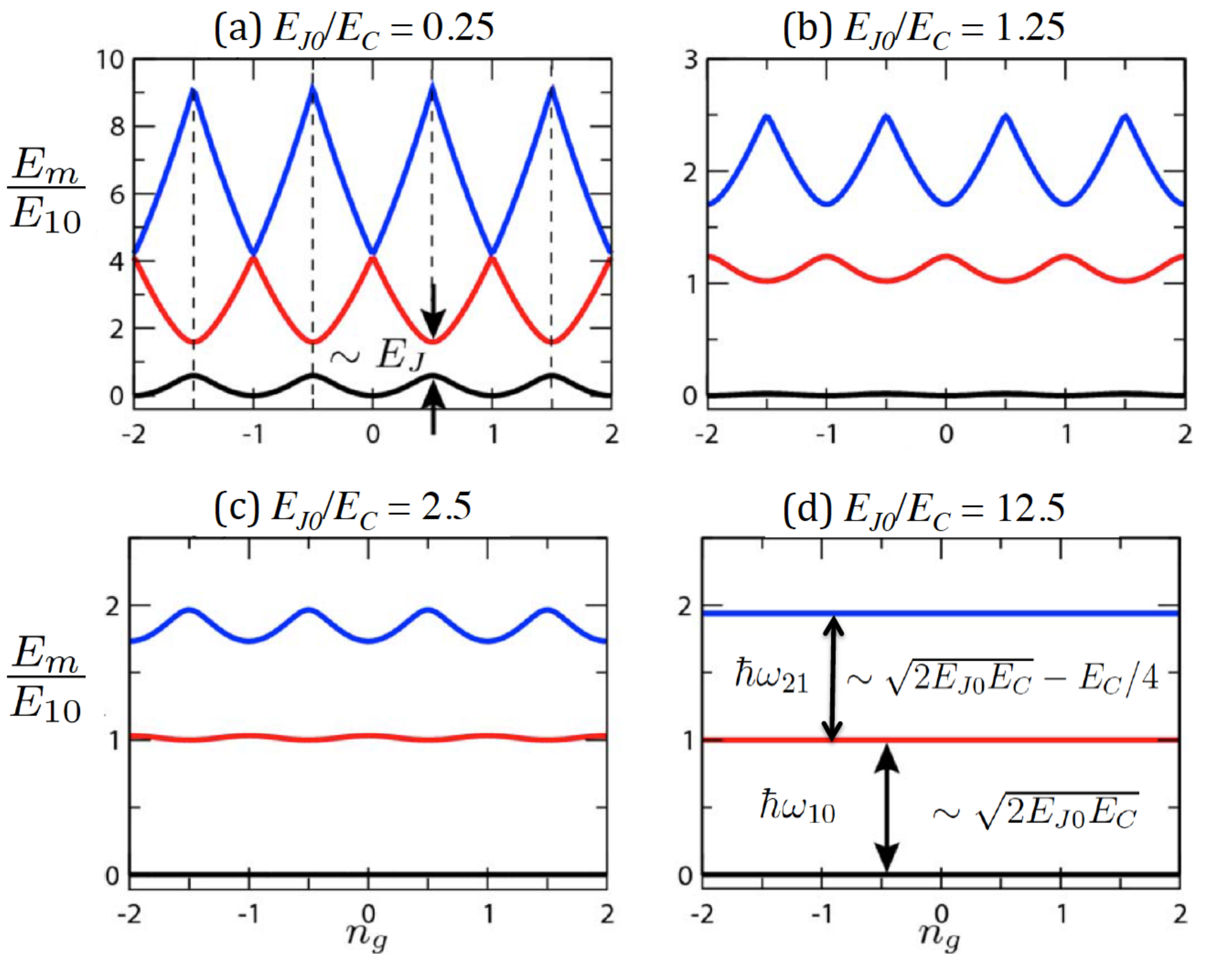}
\caption
{\small Level spectrum (band structure) of the Cooper pair box (CPB) as a function of the offset charge $n_g$ for different ratios $E_{J0}/E_C$  \cite{Koch2007}: (a) charge qubit \cite{Wallraff2004}; (b) Quantronium \cite{Vion2002};  (d)  Transmon  \cite{Koch2007}. 
"Historically", the CPB evolved from  the original charge qubit (1999) \cite{Nakamura1999} via the quantronium (2002) \cite{Vion2002,Cottet2002,Collin2004} and CPB-cQED (2004) \cite{Blais2004,Wallraff2004},  to the transmon (2007) \cite{Koch2007} and the Xmon (2013) \cite{Barends2013,MartinisMegrant2014}. The charge dispersion decreases exponentially with $E_{J0}/E_C$, while the anharmonicity only decreases algebraically with a slow power law in $E_{J0}/E_C$ \cite{Koch2007,Cottet2002} - 
this makes it possible to individually address selected transitions even for quite large ratios of $E_{J0}/E_C$.
Figure adapted from \cite{Koch2007}.
}
\label{transmon_Koch}
\end{figure}

With respect to Eq.~\ref{Upot} and Fig.~\ref{CJL_fig} there are two distinct cases:

 (1) $E_L$=0 ($L\sim\infty$)\;: $U(\phi)$ becomes a pure cosine periodic potential, and the wave function has the form $\psi = \psi(\phi,n_g) \: e^{in_g\phi}$, where $\psi(\phi,n_g)$ is a Mathieu function. The energy levels form bands $E(n_g)$ in the "momentum" direction \cite{DevWallMart2004,Koch2007}. The dispersion of these band depends on the ratio $E_{J0}/E_C$, as shown in Fig. \ref{transmon_Koch}. Of special interest is that a large capacitance $C$ results in flat low-lying bands, making the circuit insensitive to charge fluctuations (as well as to charge control via a DC gate voltage) \cite{Koch2007}.
  
(2) $E_L  \approx E_{J0}$\;: $U(\phi)$ is no longer periodic, but described by a parabola modulated by sinusoidal function. The shape of the potential toward the bottom of the parabola, and the associated qubit level structure, depend on the $E_L/E_{J0}$ ratio and on the external  flux $\phi_e$. This makes it possible to design a wide variety of qubits by tuning the circuit parameters in Fig.~\ref{CJL_fig}.

\subsection{Qubits (DV1)}

\begin{figure}[h]
\center
\includegraphics[width=10cm]{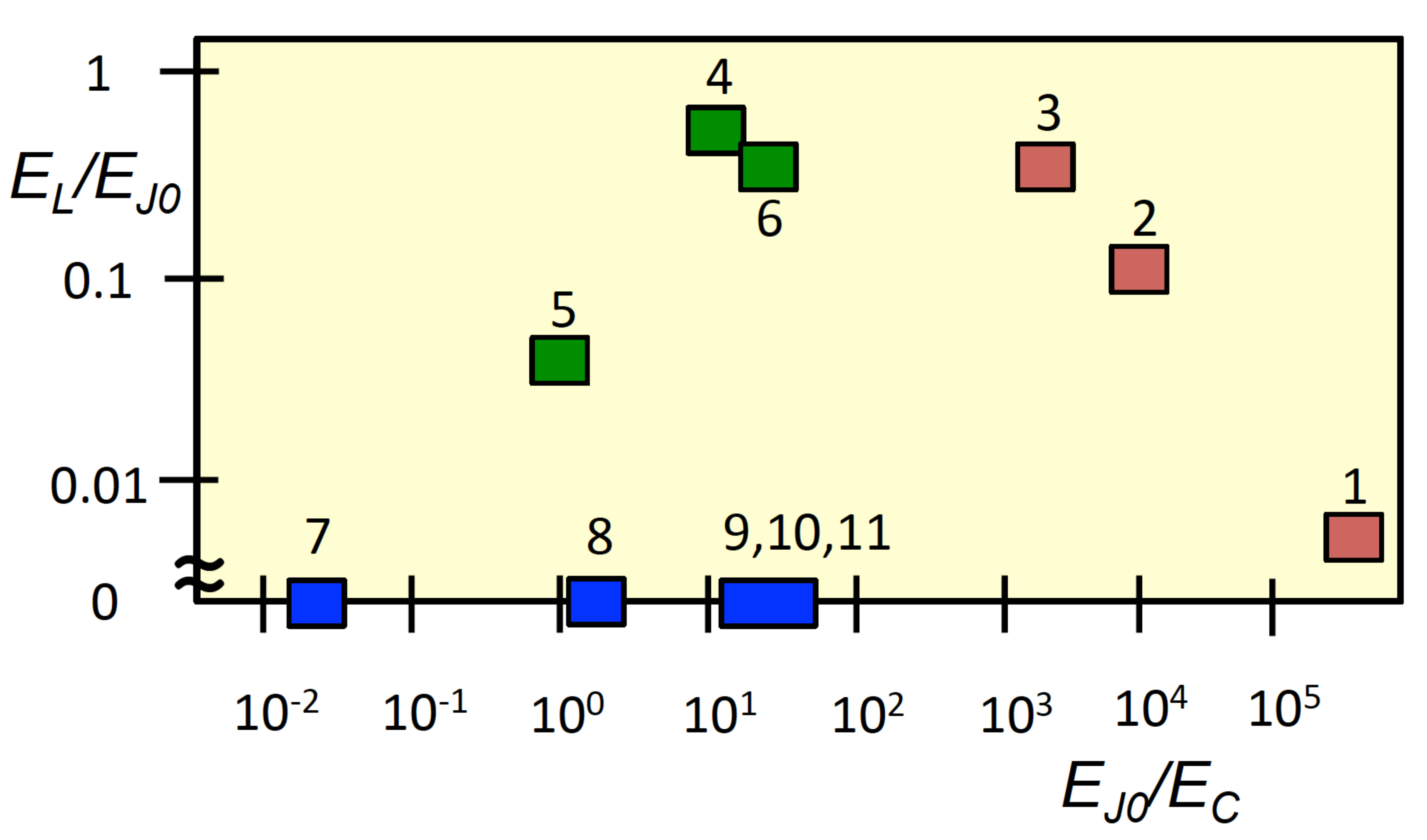}
\caption
{\small Graphic presentation of data from Table 1:
1.\;Phase qubit \cite{Martinis2002}; 2.\; Phase qubit  \cite{Steffen2006}; 3.\;rf-SQUID \cite{Friedman2000}; 4.\; Flux qubit \cite{Mooij1999}; 5.\; Fluxonium \cite{Pop2014}; 6.\;C-shunt \cite{Yan2016}; 7.\;Charge qubit \cite{Nakamura1999}; 8.\;Quantronium  \cite{Vion2002}; 9.\;Transmon  \cite{Koch2007,Paik2011}; 10.\;Xmon  \cite{Barends2013}; 11.\;Gatemon  \cite{Casparis2016}.
}
\label{qubit_graph}
\end{figure}

We will now briefly discuss the qubit families listed in Table~\ref{JJ-qubits} and shown in Fig.~\ref{qubit_graph}. In reality, all "qubits" are multi-level systems. However, since they are most often treated as quantum bits (binary logic), we will refer to them as qubits even if additional levels are used for gate operations.
The qubits can be defined in terms of different values of the circuit parameters through the $E_{J0}/E_C$ and $E_L/E_{J0}$ ratios, characterising a number of charge and flux types of devices. 

\begin{table}[h]
\caption{\label{JJ-qubits}
Main types of Josephson junction (JJ) based qubit circuits.} 
%
\begin{tabular}{@{}*{8}{l}}
\br                              
 & $C$ &$C_J$ & $L$  & $L_{J0}$ & $E_L/E_{J0}$ & $E_{J0}/E_C$  &  Z \cr 
& fF & fF & pH  &   pH &   &    & $\Omega$  \cr 
1. Phase qubit \cite{Martinis2002} & 0  & 6000 & 3300     & 16  &   0.005  & $\sim$10$^6$ & $\sim$ 1.5\cr
2. Phase qubit \cite{Steffen2006} & 800  & $\sim$0 & 720     & $\sim$ 80  &   0.11  & $\sim$10$^4$    &  $\sim$ 15 \cr
3. rf-SQUID \cite{Friedman2000} &  0  & 40 &  238   &  101  & 0.43  & 2000 &  48 \cr
4. Flux qubit \cite{Mooij1999} & 0  & 3 &   1200  & 600 & 0.5 & 10 & 450  \cr
5. Fluxonium \cite{Pop2014} & 0.15   & $\sim$0 &  3300  & 150 & 0.045  & 1 & 1400 \cr
6. C-shunt \cite{Yan2016} &  50  & $\sim$0 & 15000  &  4500  & 0.3 & 25 &  480  \cr
7. Charge qubit \cite{Nakamura1999} & 0.68  & $\sim$0 &   $\infty$   &  808 & 0    &  0.018  & $\sim$10$^4$  \cr
8. Quantronium \cite{Vion2002} & 2.8  & $\sim$0 &   $\infty$  &  1.1 10$^4$    & 0 &  1.27 &  1300 \cr
9. Transmon \cite{Koch2007,Paik2011} & 15-40  & $\sim$0 &   $\infty$  &   $\sim$10$^3$    & 0 &  10-50 &    $\sim$ 250 \cr
10. Xmon  \cite{Barends2013} & 100  & $\sim$0 &   $\infty$  &  $\sim$10$^4$  &  0   &  22-28 &   $\sim$ 500 \cr
11. Gatemon \cite{Casparis2016} & 100  & $\sim$0 &    $\infty$  &  $\sim$10$^4$  &   0  &  17-32 &    $\sim$ 500 \cr
\br
\end{tabular}
Charging energy of one Cooper pair (2e): $E_C = {(2e)^2/2C}$ \\
Inductive energy: $E_L = \hbar^2/(4e^2L)$ \\
Josephson energy: $E_{J0} = \hbar^2/(4e^2L_{J0})$ \\
Resistance "quantum": $R_Q = \hbar/(2e)^2 \approx  1.027059 \;k\Omega $ \\
Impedance: $Z \sim  \sqrt{L_{J0}/C} =  R_Q \sqrt{2} /\sqrt{E_{J0} / E_C}$ 
\end{table}

\subsubsection{Phase qubit}
The phase qubit is formed by the two lowest levels in the potential wells formed by a current-biased Josephson junction. 
In practice, current bias is achieved by flux-biasing an rf-SQUID ($\phi_e$ knob) (Fig.~\ref{CJL_fig}), placing the qubit in an anharmonic potential well on the slope of the parabola. 

The original phase qubit \cite{Martinis2002} was built on a large-area junction with large self-capacitance $C_J$ (Table~\ref{JJ-qubits}). However, the large area of the junction oxide gave rise to many defects, trapping two-level fluctuators (TLS) that severely limited the coherence time.
An improved phase qubit \cite{Steffen2006} was created by separating the device into a small-area (low $C_J$) JJ with the same critical current (and thus the same $E_{J0}$, $J_0$), and a large shunt capacitance $C$ with a dielectric with much fewer defects.
This phase qubit was the first one to be used for advanced and groundbreaking QIP applications with up to four qubits \cite{Ansmann2009,Barends2014b,Chen2014b,Roushan2014,Mariantoni2011,Lucero2012}. However, the coherence time has stayed rather short ($< 1 \mu s$); therefore, phase-qubit technology cannot be scaled up at the present time. See \cite{DevWallMart2004,WeShu2007} for detailed discussions.

\subsubsection{rf-SQUID flux qubit}

The rf-SQUID flux qubit  \cite{Friedman2000,Bennet2009} is a persistent-current qubit obtained by setting the biasing flux to $\Phi_0/2$ so that the Josephson part creates two potential wells separated by a barrier at the bottom of the parabola. This defines two low-lying "bonding-antibonding" qubit levels describing superpositions of left- and right-rotating supercurrents: $\ket{L} \pm \ket{R}$. Since the inductive SQUID loop is large, this flux qubit is sensitive to flux noise, and the relaxation and coherence times are quite short, $\sim 20 \;ns$ \cite{Bennet2009}, probably due to two-level fluctuators in the Nb/AlOx/Nb trilayer junction \cite{Bennet2009}. 
The D-Wave Systems' flux qubit is of this type \cite{Harris2010,Buynk2014}.

\subsubsection{Three-JJ flux qubit }

The three-JJ flux qubit  \cite{vdWal2000,Chiorescu2002,Mooij1999} consists of an rf-SQUID where the inductor $L$ has been replaced by two JJs to provide large inductance with a small SQUID ring. Since the added JJs also create an oscillating cosine potential, with the right parameters there appears a periodic potential with a double well at the bottom of each major well, defining two low-lying "bonding-antibonding" qubit levels. Tuning the flux bias with the $\phi_e$ knob makes it possible to vary the relative energies of the wells. Since the 3-JJ potential is periodic, it is associated with a band structure.

The three-JJ flux qubit has always been a major candidate for scaled-up multi-qubit systems, but the coherence time has not improved much, which has so far limited applications to cases making use of the SQUID properties and strong flux coupling  \cite{Niemczyk2010,Yoshihara2017,Forn-Diaz2017} for applications to  microwave technology \cite{Menzel2012,Zhong2013}, analog computing \cite{Haeberlein2015}, and metamaterials \cite{Jerger2012,Macha2014,Kakuyanagi2016}. A recent experiment has demonstrated somewhat longer coherence time of a flux qubit in a 3D cavity \cite{Stern2014}. Also, see the C-shunt flux qubit below.

\subsubsection{Fluxonium qubit}

The fluxonium \cite{Manucharyan2009,Pop2014} consists of a small JJ shunted by a very large inductance provided by a long array of large JJs. The resulting effective capacitance is very small (see Table~\ref{JJ-qubits}). 
This looks similar to the 3-JJ flux qubit in the sense that the two large JJs are replaced by a large JJ array. Approximately, the large array creates a wider parabola accommodating several potential wells. An important thing is that the capacitance C is so small that the there are practically no charge fluctuations (similar to the CPB charge qubit).
The relaxation time  $T_1 \approx 1 \;ms $  at 1/2 flux quantum bias is due to suppression of coupling to quasiparticles \cite{Pop2014}.

\subsubsection{C-shunt flux qubit} 
The C-shunt flux qubit \cite{YouHuAshabNori2007,Steffen2010} is usually viewed as a 3-JJ flux qubit shunted by a large capacitance. Viewed from another angle, it can be also be viewed as a transmon shunted by the effective large inductance of the two large flux qubits of the 3-JJ flux qubit. The effect is to flatten the bottom of the wells of the transmon cosine potential, making them quartic rather than quadratic. There is then no longer any double-well structure like in the flux qubit, but still strong anharmonicity (in contrast to the transmon)

Experimentally, presently the C-shunt flux qubit shows great promise \cite{Yan2016}, with broad frequency tunability, strong anharmonicity, high reproducibility, and coherence times in excess of 40 $\mu$s at its flux-insensitive point.

\subsubsection{2D Transmon qubit}
The transmon \cite{Koch2007} is a development of the CPB  toward a circuit with low sensitivity to charge noise,  and therefore much longer coherence times. This is achieved by radically flattening the bands in the charge direction by increasing the $E_{J0}/E_C$ ratio (Fig.~\ref{transmon_Koch}d). 
It should be noted that the transmon is really a flat-band multilevel system (qudit),  and the higher levels are often used for implemetation of 2-qubit gates. 
Since the influence of the charge offset $n_g$ will vanish (the situation in Fig. \ref{transmon_Koch}d), it follows that the energy levels can no longer be tuned statically by the charge gate - it can only be used for microwave excitation to drive transitions between energy levels. The driving is differential - the transmon is floating (not grounded). Tuning of the frequency of the transmon by varying the Josephson energy $E_{J0}$ can be achieved by replacing the JJ by a 2-JJ SQUID (which then also increases the sensitivity to flux noise).

The 2D transmon is now established as a central component of several scalable platforms \cite{Corcoles2015,Versluis2016,Riste2015}, with applications to a wide range of QIP problems.

\subsubsection{3D Transmon qubit} 

In the 3D transmon \cite{Paik2011}, the JJ qubit is coupled to the 3D cavity through a broadband planar dipole antenna.
Experiments with 3D devices and architectures are presently demonstrating important progress along two different lines: (i) like in 2D, a digital qubit approach with 1q and 2q gate operations controlled by microwave driving \cite{Paik2016}; (ii) a continuous-variable approach where the 3D cavities carry the information in multi-photon "cat-states", and the transmon qubits  mainly serve for creating and controlling the states of the cavities \cite{Mirrahimi2014,Brecht2016,Leghtas2015,Wang2016,Ofek2016,Narla2016,Heeres2016}.

\subsubsection{Xmon qubit}

The Xmon \cite{Barends2014a,Kelly2015a,Barends2015a,Barends2016,Barends2013,MartinisMegrant2014} is a transmon-type qubit developed for architectures with 2D arrays of nearest-neighbour capacitively coupled qubits. The large shunt capacitance $C$ (see Table~\ref{JJ-qubits}) has the shape of a cross and is grounded via the JJ-SQUID, allowing tunability of the qubit frequency.

The Xmon is established as a component for a major scalable platform. Circuits and systems with up to 9 Xmon qubits \cite{Kelly2015a,Barends2016} are presently being investigated with applications to a wide range of QIP problems. 

A variation is the gmon with direct tunable coupling between qubits \cite{Chen2014a}.

\subsubsection{Gatemon qubit}

The gatemon \cite{Larsen2015,Casparis2016} is a new type of transmon-like device, a semiconductor nanowire-based superconducting qubit.
The gatemon is of weak-link SNS type (superconductor-normal-metal-superconductor), and the Josephson energy is controlled by an electrostatic
gate that depletes carriers in a semiconducting weak link region, i.e. controls the critical current $I_c$ like in a superconducting transistor.
There is strong coupling to an on-chip microwave cavity, and coherent qubit control via gate voltage pulses. 
Experiments with a two-qubit gatemon circuit has demonstrated coherent capacitive coupling, swap operations and a two-qubit
controlled-phase gate \cite{Casparis2016}.

\subsubsection{Andreev level qubit}

The Andreev level qubit (ALQ) \cite{Zazunov2003,Zazunov2005} is a spin-degenerate, single-channel, SNS-type Josephson junction in an rf-SQUID loop. 
The ALQ can be strongly coupled to a coplanar resonator \cite{Romero2012}. Adding coupling to a spin degree of freedom in the junction makes it possible to manipulate the Andreev bound states (ABS) with a magnetic field \cite{Chtchelkatchev2003,Michelsen2008}.

Recently there has been significant experimental progress toward detecting and manipulating ABSs in atomic contacts (break junction point contacts) \cite{Bretheau2013,Janvier2015} and hybrid semiconductor-superconductor (Sm-S) nanostructures \cite{Lee2014,Woerkom2016}.  The device is typically an InAs nanowire (NW) between epitaxially grown superconducting Al electrodes (S). The resulting S-NW-S Josephson junction  \cite{Woerkom2016} is in fact a single-channel version of the gatemon \cite{Casparis2016}. The ALQ is potentially long-lived, but so far the coherence times are short - the ALQ remains a device for fundamental research.

\subsubsection{Majorana qubit} 

The ultimate system for quantum computing might be devices based on topological protection of information. One such system could be Majorana bound states (MBS) in Sm-S nanostructures that produce ABS at the interface between the normal  NW semiconductor (Sm) and the superconductor (S) \cite{Alicea2012,Mourik2012,DasSarma2015,Aasen2016,Deng2016}. 
By applying an axial magnetic field along the S-NW device, one can make the ABSs move to zero energy with increasing magnetic field and form mid gap states \cite{Lee2014,Deng2016}. If the states remain at zero energy in a long junction, a topological phase forms with MBSs at the endpoints of the nanowire \cite{Deng2016}.
The first experimental signatures of MBS in superconductor-semiconductor nanowires \cite{Mourik2012} have been confirmed \cite{Lee2014,Deng2016}, and extended to superconductor-atomic chain platforms \cite{Feldman2017}. 
A major issue is how to manipulate topologically protected qubits to allow universal quantum computation \cite{Clarke2016,Plugge2016}.

\subsection{Initialisation (DV2)} 

Qubit lifetimes are now so long that one cannot depend on natural relaxation time $T_1$ for initialization to the ground state. For fast initialisation on demand, qubits can be temporarily connected to strongly dissipative circuits, or to measurement devices \cite{Riste2012a,Johnson_Siddiqi2012,Bultink2016,McClure2016,Sank2016}.

\subsection{Universal gate operation (DV3)}

Universal high fidelity single- and two-qubit operations (Clifford + T gates; see Sect.~\ref{universal})  have been achieved for all major types of superconducting qubits. The shortest time needed for basic 1- and 2-qubit quantum operation is a few nanoseconds. Entangling gates with 99.4\% fidelity have recently been demonstrated experimentally \cite{Barends2014a}. However, it should be noted that high-fidelity gates may require carefully shaped control pulses with typically 10-40 $\mu$s duration.

\subsection{Readout (DV4)}

There are now well-established efficient methods for single-shot readout of individual qubits, typically performed via dispersive readout of a resonator circuit coupled to the qubit. 
A strong measurement  "collapses" the system to a specific state, and then repeated non-destructive measurements will give the same result. 
Single-shot measurements require extremely sensitive quantum-limited amplifiers, and it is the recent development of such amplifiers that has made single-shot readout of individual qubits possible \cite{DevSchoel2013,Riste2012a,Johnson_Siddiqi2012,Bultink2016,McClure2016,Sank2016,Vijay2009,Cast-Belt2007,Roch2012,Eichler2014,OBrien2014,Zhou2014,Schmitt2014,Sun2014,Jeffrey2014,Krantz2016,Walter2017}.

\subsection{Coherence times (DV5)} 

JJ-qubits are manufactured and therefore sensitive to imperfections. Nevertheless, there has been a remarkable improvement of the coherence times of both qubits and resonators during the last five years \cite{DevSchoel2013,MartinisMegrant2014,Riste2013,Bruno2015}.
Table~\ref{JJ-qubits_DV}  indicates the present state of the art.

\begin{table}[h]
\caption{\label{JJ-qubits_DV}
The DiVincenzo criteria \cite{DiVincenzo2000} and the status of the main types of superconducting JJ-based qubits (March  2017).
The figures in the table refer to the best published results, but may have limited significance - the coherence times in operational multi-qubit circuits are often considerably lower.} 
\begin{tabular}{@{}*{8}{l}}
\br                              
 & 2D Tmon & 3D Tmon & Xmon & Fluxm & C-shunt & Flux  &  Gatemon  \cr 
 & \cite{Koch2007} & \cite{Paik2011} &  \cite{Barends2013} & \cite{Pop2014} &\cite{Steffen2010,Yan2016}  & \cite{vdWal2000} & \cite{Casparis2016}   \cr 
&  &  &   &  &   &   \cr 
DV1, \#q  & 5 \cite{Riste2015,Rol2016} & 4 \cite{Paik2016} & 9 \cite{Barends2014a,Barends2016} & 1  & 2 \cite{Yan2016}  & 4 \cite{Grajcar2006} & 2 \cr
DV2 & $Yes$  & $Yes$   & $Yes$  & $Yes$  & $Yes$  &  $Yes$  &  $Yes$ \cr
DV3 & $Yes$ & $Yes$ & $Yes$ & $Yes$ & $Yes$ & $Yes$  & $Yes$  \cr
t$_{1q} (ns)$ & 10-20  &  30-40 &  10-20 & $-$ &  5-10  &  5-10   &  30  \cr
n$_{op,1q}$ & $> 10^3$  &  $> 10^3$ & $>10^3$   & $-$  & $\sim 10^3$  &  $\sim 10^3$   &  $\sim 10^2$ \cr
F$_{1q}$ & $\sim$ 0.999  &  $> 0.999$  & 0.9995 & $-$ & $-$  & $-$   &  $> 0.99$   \cr
t$_{2q}(ns)$ &10-40  & $\sim$ 450  & 5-30 &$-$  & $-$  & $-$  &  50  \cr
n$_{op,2q}$ & $\sim 10^3$ & $\sim10^2$ & $\sim10^3$ & $-$ & $-$  &$-$ & $\sim 10^2$  \cr
F$_{2q}$ & $>$ 0.99  & 0.96-0.98  & 0.9945 & $-$ & $-$  & $-$  & 0.91  \cr
DV4 & $Yes$ & $Yes$ & $Yes$ & $Yes$ & $Yes$ & $Yes$  & $Yes$ \cr
DV5 & $Yes$ & $Yes$ & $Yes$ & $Yes$ & $Yes$ & $Yes$  & $Yes$ \cr
$T_1 (\mu s)$ & $\sim$ 40  & 100  & 50  & 1000 & 55  & 20  \cite{Stern2014} & 5.3   \cr
$T^*_2 (\mu s)$ & $\sim$ 40  & $>$ 140   & 20 & $>10$ & 40  & $-$ & 3.7    \cr
$T^{echo}_2 (\mu s)$ & $\sim$ 40  &$>$ 140   & $-$ & $-$ & 85  & $-$ & 9.5    \cr
\br
\end{tabular}
-  The number of qubits ($DV1, \#q$) refers to operational circuits with all qubits connected. \\
-  $t_{1q}$ and $t_{2q}$ are gate times for 1q- and 2q-gates.\\
-  $n_{op,1q}$ and $n_{op,2q}$ are the number of 1q- and 2q-gate operations in the coherence time.\\
-  $F_{1q}$ and $F_{2q}$ are average fidelities of 1q- and 2q-gates, measured e.g. via randomised benchmarking (Sect.~\ref{characterisation}).\\
-  $T_1$ is the qubit energy relaxation time.\\
-  $T_2^*$ is the qubit coherence time measured in a Ramsey experiment.\\
-  $T_2^{echo}$ is the qubit coherence time measured in a spin-echo (refocusing) experiment. \\
-  Table entries marked with a hyphen (-) indicate present lack of data. \\
-  Note that average gate fidelities F$_{1q}$ and F$_{2q}$ do not necessarily correspond to thresholds for error correction \cite{Sanders2016}.\\
-  The t$_{2q}$ gate time for the 3D Tmon refers to a resonator-induced phase gate.\\
\end{table}

\subsection{Algorithms, protocols and software} 

A number of central quantum algorithms and protocols have been performed experimentally with multi-qubit circuits and platforms built from the main types of superconducting JJ-based qubits (see e.g. \cite{Barends2015a,Barends2016,Dewes2012b,Mariantoni2011,DiCarlo2009,Steffen2013,Salathe2015,Heinsoo2016,Oppliger2016}),
demonstrating proofs of principle and allowing several transmon-type systems to be scaled up.

In practice, a quantum computer (QC)  is always embedded in a classical computer (CC), surrounded by several classical shells of hardware (HW)  and software (SW) \cite{Versluis2016,Reilly2015,Bejanin2016}. Quantum computation and quantum simulation then involve a number of steps:

\begin{enumerate}
\item CC control and readout HW (shaped microwave pulses, bias voltages, bias fluxes).
\item CC control and readout SW for the HW ("machine language"; optimal control).
\item CC subroutines implementing gates (gate libraries).
\item High-level CC optimal control of quantum operations.
\item CC subroutines implementing quantum gate sequences (for benchmarking, QFT, time evolution, etc.).
\item High-level CC programming, compilation, and simulation of quantum algorithms and circuits  \cite{Wecker2014,Valiron2015,Haner2016a,Haner2016b}.
\item High-level CC programs solving problems \cite{Bauer2016,Reiher2016} 
\end{enumerate}

The only truly quantum part is step (v), explicitly performing quantum gates on quantum HW and quantum states.
This is where quantum speedup can be achieved, in principle.
Since the quantum gates have to be implemented by classical SW, it is necessary that the needed number of gates to describe a quantum circuit scales polynominally in the size of problem.
For the Clifford gates there are efficient (polynomial) representations.
However, to describe an arbitrary, universal quantum circuit needs T-gates and may take exponential resources \cite{BravyiGosset2016}.

If the quantum gates are executed in SW on representations of quantum states on a classical machine, then the quantum computer is emulated by the classical machine. 
Then to execute the gates scales exponentially, which means that a classical computer can only simulate a small quantum system. 
The present limit is around 50 qubits \cite{Boixo2016b,BravyiGosset2016} - beyond that is the realm of Quantum Supremacy \cite{Boixo2016b} .

\newpage

\section{Transmon quantum circuits}

The present development of quantum information processing with scalable Josephson Junction circuits and systems goes in the direction of coupling transmon-type qubits with quantum oscillators, for operation, readout and memory. 
In this section we will therefore focus on the transmon, and describe the components in some detail. For an in-depth  discussion, the reader is referred to the original article by Koch {\it et al.} \cite{Koch2007}.

\subsection{Transmon cQED}

\begin{figure}[h]
\center
\includegraphics[width=7cm]{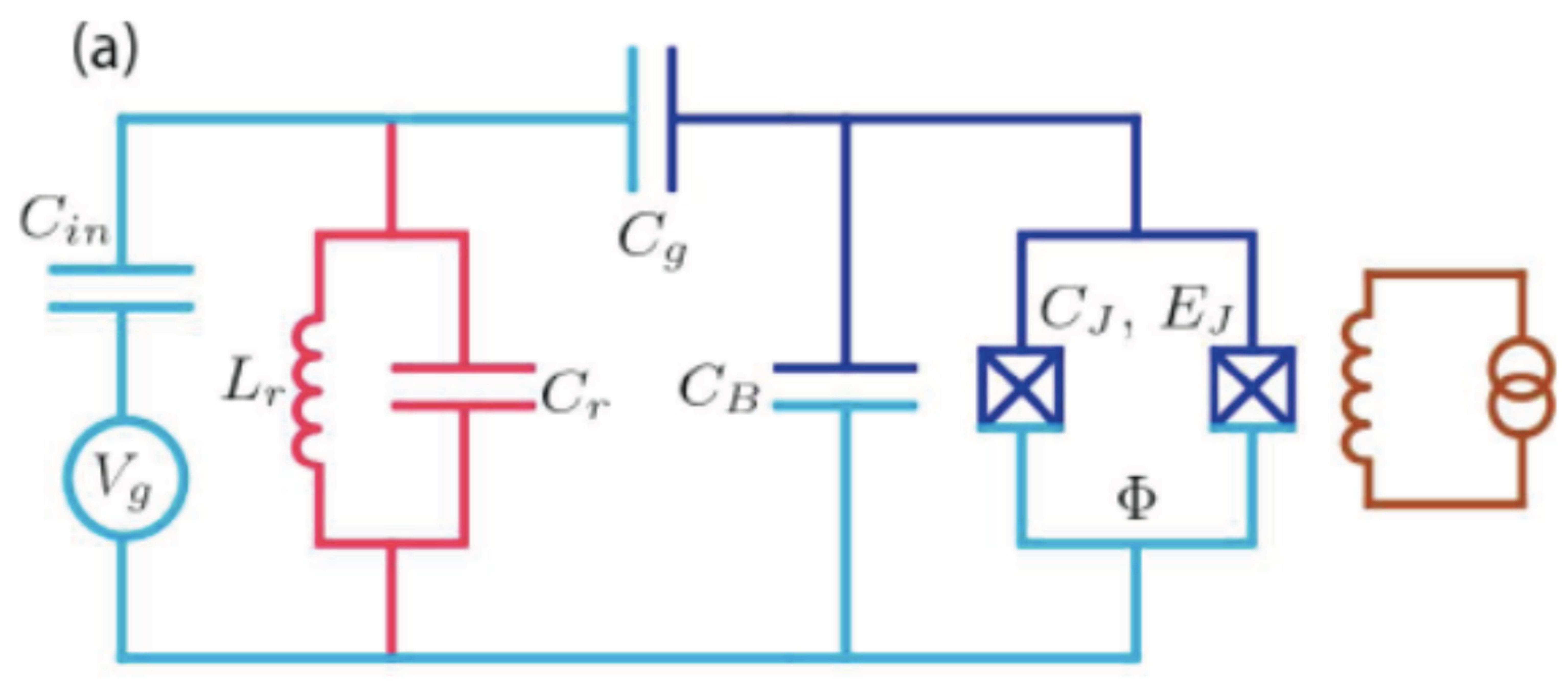}\;\;\; \;\;\; 
\includegraphics[width=6cm]{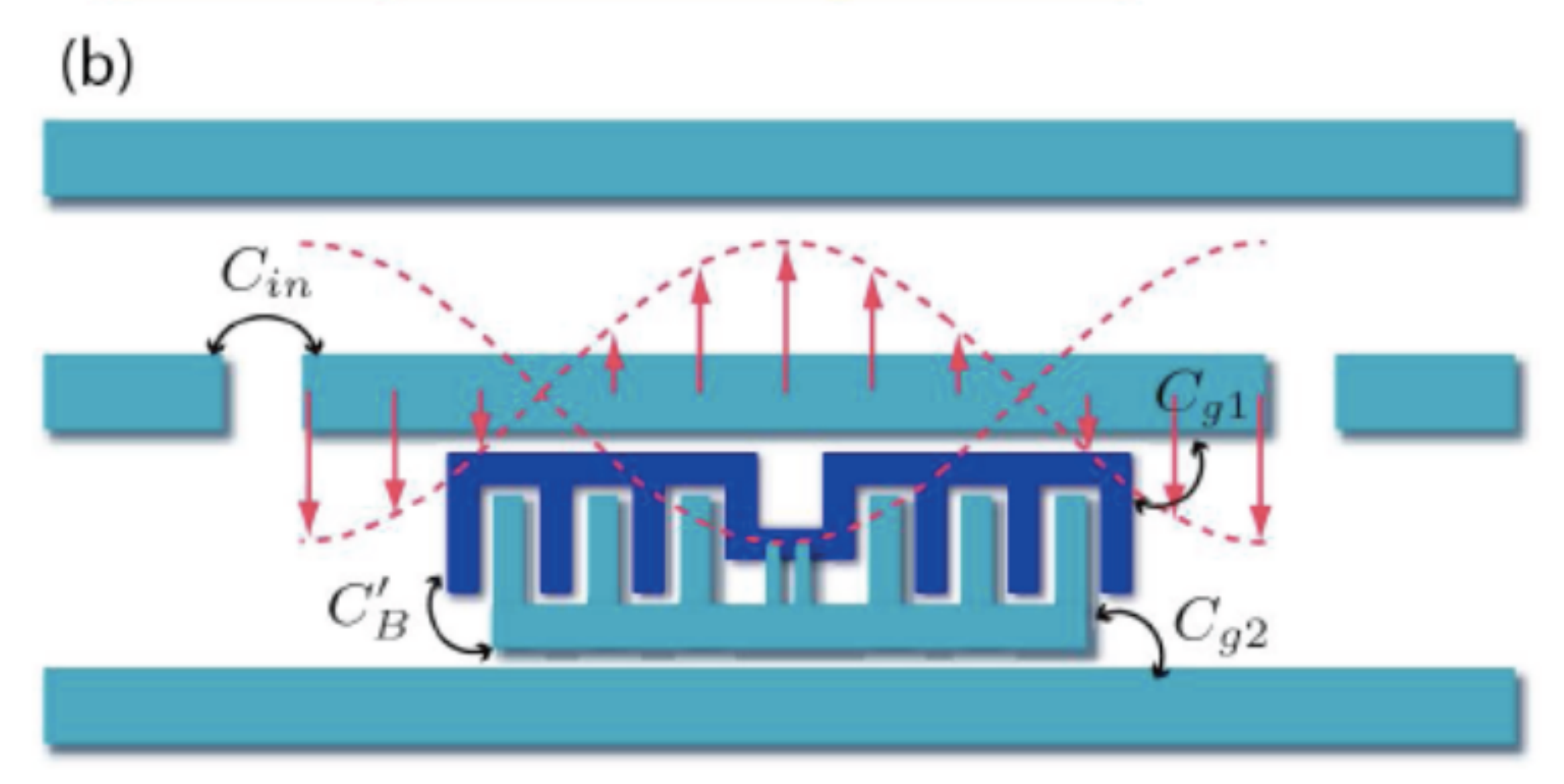}
\caption
{\small Transmon-cQED: (a) Equivalent circuit (see text); (b) Physical device. The 2-JJ SQUID is located at the centre of a large interdigitated shunt capacitor ($C_B$), and the entire transmon is capacitively coupled to a coplanar waveguide (CPW) resonator ($L_rC_r$). The transmon is not grounded - it is floating and driven differentially.
Adapted from \cite{Koch2007}.\\

}
\label{tmon_cQED}
\end{figure}

A generic compact circuit model for the device is shown in Fig.~\ref{tmon_cQED}a, and a hardware implementation is shown schematically in Fig.~\ref{tmon_cQED}b.

The transmon circuit in Fig.~\ref{tmon_cQED}a consists of a number of fundamental components: \\
-  A Cooper pair box (CPB) with one or two Josephson junctions (JJ) sitting in a closed circuit with a large shunting capacitance $C_B$ ("anharmonic oscillator"). The excitation energy of the two-level systems is $ \epsilon = E_1-E_0$\\
- A resonator circuit $L_rC_r$ (harmonic oscillator) with frequency $\omega = 1/\sqrt{L_rC_r}$.\\
- A capacitance $C_g$ coupling the transmon and the resonator with coupling constant $g$. \\
- A drive circuit (right) flux-coupled to the SQUID-type JJ circuit for tuning the qubit energy.\\
- A microwave drive circuit (left) capacitively coupled to the CPW for qubit operation. \\
- There is no explicit readout oscillator included in Fig.~\ref{tmon_cQED}a, but the $L_rC_r$ bus resonator can serve to illustrate a readout device..

Treating the transmon as an approximate two-level system with linear  coupling to a single-mode oscillator, the transmon-cQED Hamiltonian takes the form

\begin{equation}
\label{tmon-cQED1}
\hat H = \hat H_q + \hat H_{qr} + \hat H_r = - \frac{1}{2} \epsilon \; \sigma_{z} + g \; \sigma_{x} (a + a^+) +  \hbar \omega \;(a^+a + \frac{1}{2}) 
\end{equation}
where $\epsilon$ is the qubit excitation energy, $g$ is the qubit-oscillator coupling, and $\omega$ is the oscillator frequency.

Introducing the raising and lowering operators $\sigma_{\pm} = ( \sigma_x \pm i\; \sigma_y)/2$, the qubit-resonator coupling term $ \hat H_{qr}$ is split into two terms, Jaynes-Cummings (JC) and anti-Jaynes-Cummings (AJC):
\begin{equation}
\label{tmon-cQED2}
\hat H_{qr} = \hat H^{JC}_{qr} + \hat H^{AJC}_{qr} =  g \;  (\sigma^{+} a + \sigma^{-} a^+) + g  \; (\sigma^{+} a^+ + \sigma^{-} a)  
\end{equation}
This Hamiltonian describes the canonical quantum Rabi model (QRM) \cite{ShoreKnight1993,Casanova2010,Xie2016}. Equations~(\ref{tmon-cQED1},\ref{tmon-cQED2}) are completely general, applicable to any qubit-oscillator system.
Only keeping the Jaynes-Cummings (JC) terms corresponds to performing the rotating-wave approximation (RWA).

\subsection{Weak, strong, ultra-strong and deep-strong coupling}

There are five basic energy scales that determine the qubit-oscillator coupling strength: $g, \epsilon, \omega$, plus the oscillator decay rate $\kappa$ (resonance line width)  and the qubit decay rate $\gamma $ (transition line width) 
.

One typically distinguishes between four cases of qubit-oscillator coupling: \\
(i)  {\it Weak coupling}: $g  \ll \gamma, \kappa, \epsilon, \omega  $\;; RWA valid for $\epsilon - \omega \ll \omega$.\\ 
(ii)  {\it Strong coupling}: $  \gamma, \kappa \ll g \ll \epsilon, \omega  $\;; RWA valid; vacuum Rabi oscillations. \\
(iii)  {\it Ultra-strong coupling (USC)}: $g /\omega  \le 1$; RWA breaks down; $\hat H^{AJC}$ counter-rotating term  important. \\
(iv)  {\it Deep-strong coupling (DSC)}: $g /\omega  \ge 1$; RWA not valid at all; $\hat H^{AJC}$ essential; qubit-oscillator compound system.

In the two cases of {\it weak} and  {\it strong coupling}, one performs the rotating-wave approximation (RWA) and only keeps the first $ \hat H^{JC}_{qr}$ term, which gives the canonical Jaynes-Cummings model \cite{ShoreKnight1993,JaynesCummings1963,GerryKnight2004},
\begin{equation}
H \approx - \frac{1}{2}\epsilon \; \sigma_{z} + g \;  (\sigma^{+} a + \sigma^{-} a^+) + \hbar \omega \;(a^+a) 
\end{equation}
describing dipole coupling of a two-level system to an oscillator. 
In the non-resonant case, diagonalising the Jaynes-Cummings Hamiltonian to second order by a unitary transformation gives \cite{Blais2004,Koch2007,GerryKnight2004}
\begin{equation}
H = -{1\over 2}(\epsilon + {g^2\over\Delta})\;\sigma_z +
(\hbar\omega + {g^2\over\Delta}\;\sigma_z)\;a^+a 
\label{JC2ndorder}
\end{equation}
where $\Delta= \epsilon - \hbar\omega \gg g$ is the so-called detuning.
The result implies that (i) the qubit transition energy $\epsilon$ is Stark shifted (renormalized) by the coupling to the oscillator, and (ii) the oscillator energy $\hbar\omega$ is shifted by the qubit in different directions depending on the state of the qubit. This condition allows discriminating the two qubit states in dispersive readout measurement \cite{Blais2004,Wallraff2004}.

The {\it strong coupling} situation \cite{Yang2003,Yang2004,Blais2004}  was demonstrated experimentally already in 2004 with superconducting CPB-cQED \cite{Wallraff2004} by direct physical coupling of the CPB and the 2D resonator. 
The  {\it ultra-strong coupling} case \cite{Casanova2010,Ballester2012} is more difficult to achieve by direct statical physical coupling of a transmon qubit and a resonator, and but has recently been achieved experimentally using flux qubit cQED \cite{Niemczyk2010,Yoshihara2017,Forn-Diaz2017}.  
On the other hand, it is possible to simulate the QRM in the USC and DSC regimes by external time-dependent driving of the oscillator in analogue \cite{Ballester2012,Braumuller2017} or digital \cite{Mezzacapo2014b,Langford2016,Lamata2016} quantum simulation schemes.

\subsection{Multi-qubit Transmon Hamiltonians}
\label{mqtmonham}

In the following we will focus on transmon multi-qubit systems, and then the Hamiltonian takes the general form (omitting the harmonic oscillator term):
\begin{eqnarray}
\label{multiq}
\hat{H}  = \hat H_q + \hat H_{qr} + \hat H_{qq}  \nonumber \\ 
= - \frac{1}{2} \sum_i \epsilon_i \; \sigma_{zi} \;
+ \sum_i g_i  \; \sigma_{xi} (a + a^+) +\; \frac{1}{2} \sum_{i,j;\nu}
\lambda_{\nu,ij}\;\sigma_{\nu i} \; \sigma_{\nu j} 
\end{eqnarray}
For simplicity, in  Eq.~\ref{multiq} the qubit-resonator term $ \hat H_{qr}$ is considered only to refer to readout and bus operations, leaving indirect qubit-qubit interaction via the resonator to be included in $ \hat H_{qq}$ via the coupling constant $\lambda_{\nu,ij}$.

\subsubsection{Capacitive coupling}

\begin{figure}[h]
\center
\includegraphics[width=14cm]{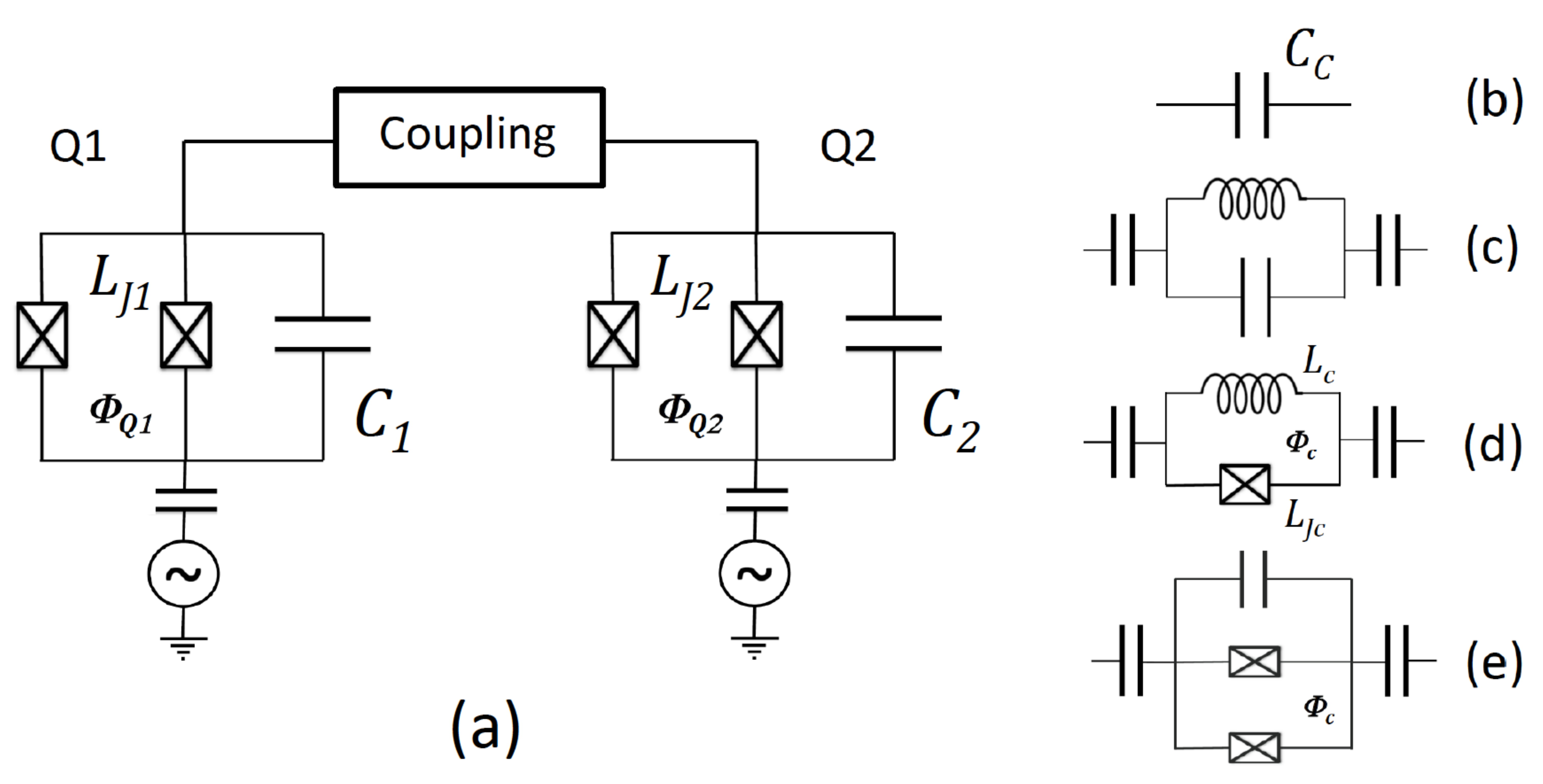}
\caption
{\small Two coupled transmon qubits flux-tunable energies $E_{10,1},E_{10,2}$. (a) Generic coupling scheme; (b) Capacitive coupling, $C_c$; (c) Resonator coupling; (d) Inductive coupling with tunable JJ \cite{Chen2014a}; (e) "Transmon-bus" coupling  \cite{McKay2016}.
}
\label{tmon_coupling}
\end{figure}

This case (Fig.~\ref{tmon_coupling}b) is described by an Ising-type model Hamiltonian with direct capacitive $(C_c)$ qubit-qubit charge coupling. 
For the transmon \cite{WeShu2007},
\begin{equation}
H  = -\sum_{i=1,2} \frac{\epsilon_i}{2}\sigma_{zi}\;+\; \lambda_{12} \; \sigma_{x1}\sigma_{x2}
\end{equation}
\begin{equation}
\lambda_{12} =  \frac{1}{2} \sqrt{E_{10,1} E_{10,2}} \; \frac{\sqrt{E_{C1} E_{C2}}} {E_{Cc}} =  \frac{1}{2} \sqrt{E_{10,1} E_{10,2}} \; \frac {C_c} {\sqrt{C_1 C_2}} \approx  \frac{1}{2} E_{10} \; \frac {C_c} {C}
\end{equation}
where the approximate result for $\lambda_{12}$ refers to identical qubits in resonance.
In the RWA one finally obtains the Jaynes-Cummings Hamiltonian
\begin{equation}
\label{JC}
H = -\sum_{i=1,2} \frac{\epsilon_i}{2}\sigma_{zi} \; +  \lambda_{12} (\sigma^+_1 \sigma^-_2 +\sigma^-_1 \sigma^+_2 )
\end{equation}

\subsubsection {Resonator coupling}
In this case (Fig.~\ref{tmon_coupling}c) the coupling is primarily indirect, via virtual excitation (polarisation) of the detuned bus resonator. Diagonalisation of the Hamiltonian gives the usual second-order qubit-qubit coupling \cite{Blais2004,Majer2007,WeShu2007}:
\begin{equation}
H  = -\sum_{i=1,2} \frac{\epsilon_i}{2}\sigma_{zi}\;+\; \lambda_{12} \; \sigma_{x1}\;\sigma_{x2}
\end{equation}
\begin{equation}
\lambda_{12} =  \frac{1}{2} \; g_1 g_2 \; ( \frac{1}{\Delta_1} +  \frac{1}{\Delta_2}) \equiv  g_1 g_2 \;  \frac{1}{\Delta}
\end{equation}
Here $\Delta_1= \epsilon_1 - \hbar\omega$ and $\Delta_2= \epsilon_2 - \hbar\omega$ are the detunings of the qubits, and $g_1 g_2 \ll \Delta_1, \Delta_2$.
Finally, in the RWA one again obtains the Jaynes-Cummings Hamiltonian
\begin{equation}
H = -\sum_{i=1,2} \frac{\epsilon_i}{2}\sigma_{zi}\; +\; \frac{g_1 g_2}{\Delta} \; (\sigma^+_1 \sigma^-_2 +\sigma^-_1 \sigma^+_2 )
\end{equation}

\subsubsection{Josephson junction coupling}

The transmon qubits can also be coupled via a Josephson junction circuit \cite{WeShu2007}, as illustrated in Fig.~\ref{tmon_coupling}d. A case of direct JJ-coupling (omitting the coupling capacitors in  Fig.~\ref{tmon_coupling}d) has recently been treated theoretically and implemented experimentally by Martinis and coworkers  \cite{Chen2014a,Geller2015b}. To a good approximation, the Hamiltonian is given by
\begin{equation}
H = -\sum_{i=1,2} \frac{\epsilon_i}{2}\sigma_{zi}\;+\; \lambda_{12} \; \sigma_{y1}\sigma_{y2}
\end{equation}
\begin{equation}
\lambda_{12} \approx  \frac{1}{2} \;  E_{10} \; \;\frac{L_c}{L_J} \; \frac{1}{2L_c + L_{Jc}/\mathrm{cos}\;\delta_c}
\end{equation}
where the approximate result for $\lambda_{12}$ refers to identical qubits in resonance,
and the RWA Hamiltonian is again given by Eq.~\ref{JC}. By varying the flux in the coupling loop, the Josephson inductance  can be varied between zero and strong coupling,  $0 \le 2\lambda_{12} \le 2\pi110 \;MHz$  \cite{Chen2014a}.

\subsubsection{Tunable  coupling}

Tunable qubit-qubit coupling can be achieved in a number of ways, for example (i)  by tuning two qubits directly into resonance with each other; (ii)  by tuning the qubits (sequentially) into resonance with the resonator; (iii) by tuning the resonator sequentially and adiabatically into resonance with the qubits \cite{Wallquist2006}; (iv) by driving the qubits with microwave radiation and coupling via sidebands; (v) by driving the qubit coupler with microwave radiation (e.g. \cite{McKay2016}); (vi) by flux-tunable inductive (transformer) coupling \cite{Kafri2016}. In particular, for JJ-coupling, the qubit-qubit coupling can be made tunable by current-biasing the coupling JJ \cite{Geller2015b,Lantz2004,Wallquist2005}.

\newpage

\section{Hybrid circuits and systems}

In this section we will discuss the status of the DiVincenzo criteria DV6 and DV7 listed in Sect.~\ref{DVcriteria}.

\begin{figure}[h]
\center
\includegraphics[width=10cm]{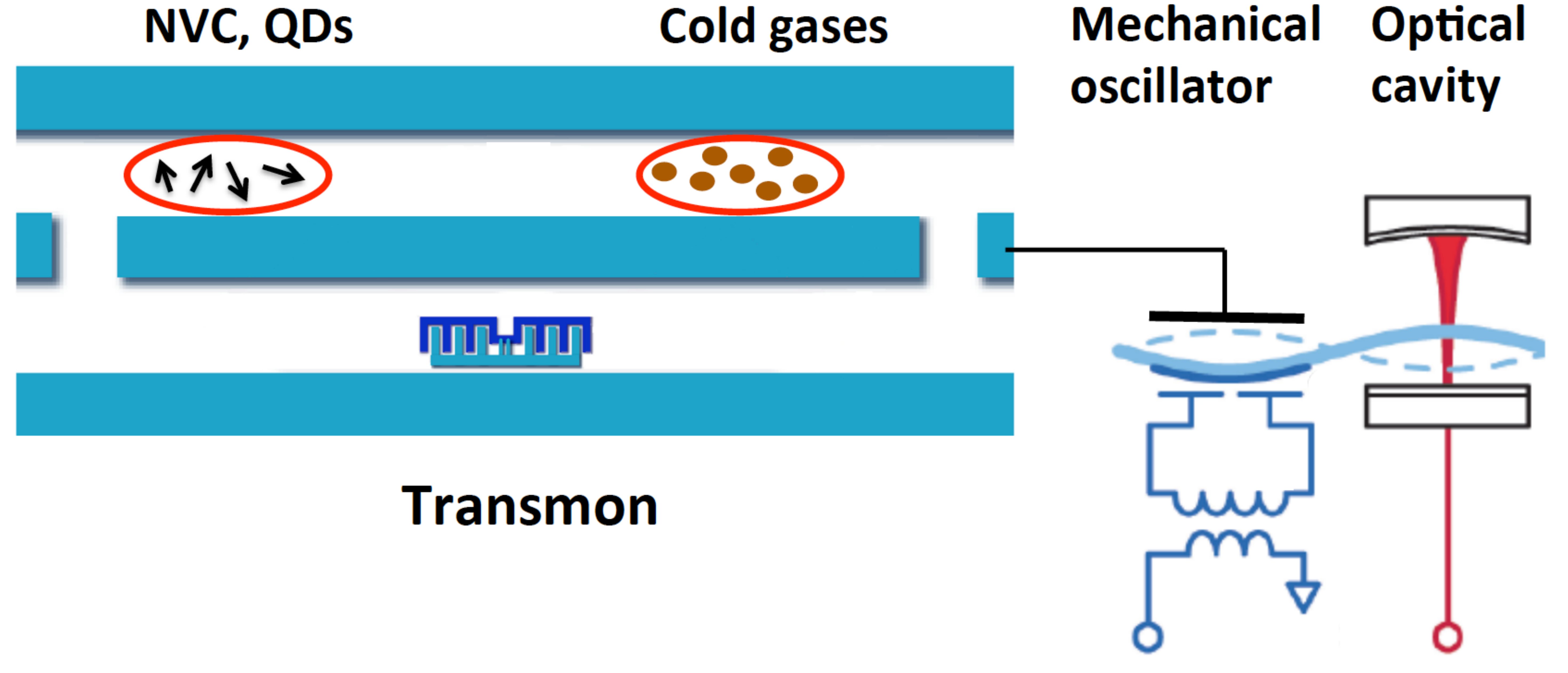}
\caption
{\small A conceptual view of a transmon hybrid system with "peripherals" serving as long-term memory and communication devices. 
}
\label{Hybrid}
\end{figure}

Even if a QIP system in principle can consist of a single large coherent register of qubits, practical systems will most likely be built as hybrid systems with different types of specialized quantum components: qubits, resonators, buses, memory, interfaces. The relatively short coherence time of JJ-qubits ($\mu s$) compared to spin qubits ($m s$) and trapped ions ($s$) has promoted visions of architectures with fast short-lived JJ-qubit processors coupled to long-lived memories and microwave-optical interfaces, as illustrated in Fig.~\ref{Hybrid}.

There are numerous demonstrations of coherent transfer between JJ-qubits and microwave resonators (both lumped circuits and microwave cavities, and mechanical resonators), as well as between JJ-qubits and spin ensembles. 
In principle, qubits coupled to microwave resonators (q-cQED) is already a hybrid technology.
An interesting aspect is that the development of long-lived transmon qubits and high-Q 2D and 3D resonators 
has changed the playground, and it is no longer clear what other kind of hybrid memory devices are needed for short-term quantum memory. Even for long-term memory, the issue is not clear: with emerging quantum error correction (QEC) techniques it may be possible to dynamically  "refresh" JJ-cQED systems and prolong coherence times at will. To achieve long-term "static" quantum memory, spin ensembles are still likely candidates, but much development remains.

The current situation for hybrid systems is described in three excellent review and research articles \cite{Aspelmeyer2014,Kurizki2015,Schuetz2015}. Here we will only briefly mention a few general aspects in order to connect to the DiVincenzo criteria DV6 and DV7, and to refer to some of the most recent work.

\subsection{Quantum interfaces for qubit interconversion (DV6)} 

The name of the game is to achieve strong coupling between elementary excitations (e.g. photons, phonons, spin waves, electrons) of two or more different components so that the mixing leads to pronounced sideband structures. This can then be used for entangling different types of excitations for information storage or conversion from localised to flying qubits.

\subsubsection{Transmon-spin-cQED}

Experimentally, strong coupling between an ensemble of electronic spins and a superconducting resonator (Fig.~\ref{Hybrid}) has
been demonstrated spectroscopically, using NV centres in diamond crystal \cite{Kubo2010,Schuster2010,Amsuss2011} and $Er^{3+}$ spins doped in a $Y_2SiO_5$  \cite{Bushev2011}.
 
Moreover, storage of a microwave field into multi-mode collective excitations of a spin ensemble has recently been achieved \cite{Kubo2012,Grezes2014}. This involved the active reset of the nitrogen-vacancy spins into their ground state by optical pumping and their refocusing by Hahn-echo
sequences. This made it possible to store multiple microwave pulses at the picowatt level and to retrieve them
after up to 35 $\mu s$, a three orders of magnitude improvement compared to previous experiments \cite{Grezes2014}.

The ultimate purpose is to connect qubits to the superconducting resonator bus, and to use the spin ensemble as a long-lived memory. Such experiments have been performed, entangling a transmon with a NV spin ensemble \cite{Kubo2011} via a frequency-tunable superconducting
resonator acting as a quantum bus, storing and retrieving the state of the qubit. Although these results constitute a proof of the concept of spin-ensemble-based quantum memory for superconducting qubits, the life-time, coherence and fidelity of spin ensembles are still far from what is needed.
Similar results were also achieved by directly coupling a flux qubit  to an ensemble of NV centers without a resonator bus \cite{Zhu2011}.

Finally, strong coupling between a transmon qubit and magnon modes in a ferromagnetic sphere has recently been achieved \cite{Tabuchi2015,Tabuchi2016}, demonstrating magnon-vacuum-induced Rabi splitting, as well as tunable magnon-qubit coupling utilising a parametric drive. 
The approach provides efficient means for quantum control and
measurement of the magnon excitations and  opens a new discipline of quantum magnonics.

\subsubsection{Transmon-micromechanical oscillator-cQED}

Mechanical oscillators (Fig.~\ref{Hybrid}) can be designed to have resonance frequencies in the microwave $GHz$ range and achieve strong coupling to superconducting qubits. Mechanical resonators therefore provide a new type of quantum mode - localised phonons. However, for this to be useful for quantum information processing one must be able to cool the mechanical oscillator to its ground state, to be able to create and control  single phonons \cite{OConnell2010,Lecocq2015}.
It is then possible to induce Rabi oscillations between the transmon and the oscillator by microwave driving via motional sidebands, resulting in periodic entanglement of the qubit and the micromechanical oscillator \cite{Pirkkalainen2013}.

\subsubsection{Transmon-SAW}

\begin{figure}[h]
\center
\includegraphics[width=12cm]{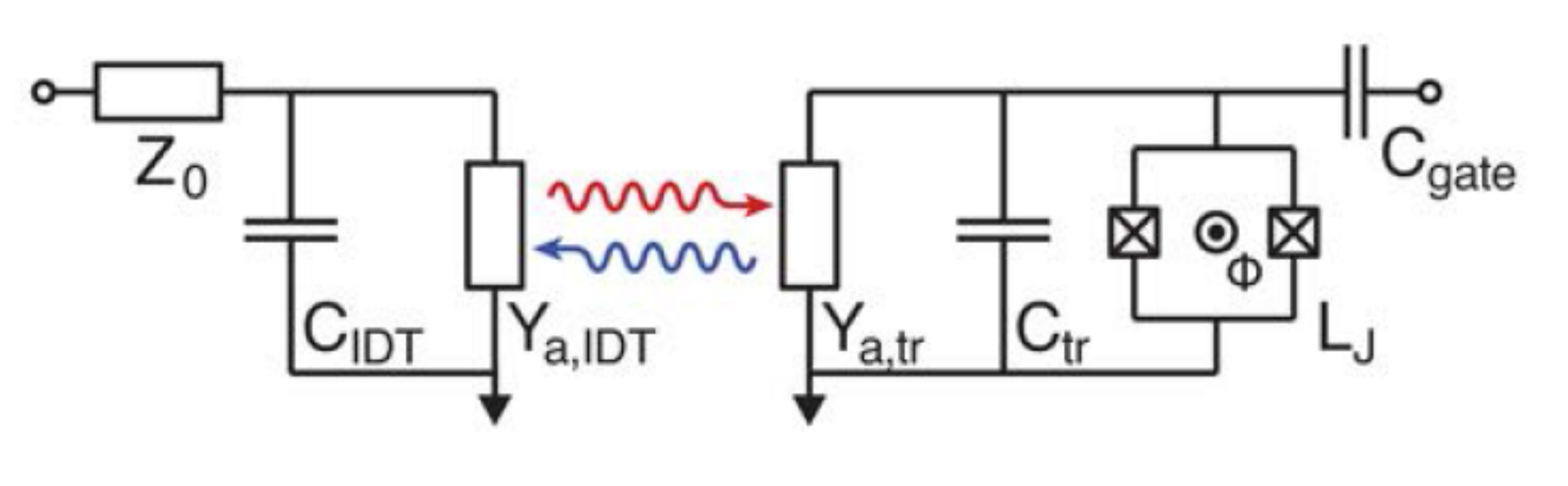}
\caption
{\small Propagating surface acoustic wave (SAW) phonons coupled to an artificial atom. Semi-classical circuit model for the qubit. The interdigital transducer (IDT) converts electrical signals to SAWs and vice versa. Adapted from \cite{Gustafsson2014}.}
\label{SAW}
\end{figure}

Surface acoustic waves (SAW) are propagating modes of surface vibrations  - sound waves. Recently, propagating SAW phonons on the surface of a piezoelectric crystal have been coupled to a transmon in the quantum regime (Fig.~\ref{SAW}), reproducing findings from quantum optics with sound taking over the role of light \cite{Gustafsson2014}. The results highlight the similarities between phonons and photons but also point to new opportunities arising from the unique features of quantum mechanical sound. The low propagation speed of phonons should enable new dynamic schemes for processing quantum information, and the short wavelength allows regimes of atomic physics to be explored that cannot be reached in photonic systems \cite{Kockum2014}.

The SAW-approach can be extended \cite{Manenti2017} to embedding a transmon qubit in a Fabry-Perot SAW cavity, piezoelectrically coupled to the acoustic field. This then realises a SAW version of cQED: circuit quantum acoustodynamics (cQAD) \cite{Shumeiko2016}.

\subsubsection{Transmon-HBAR} The recent development of bulk acoustic resonators \cite{OConnell2010,Han2016} has made possible the experimental demonstration \cite{Chu2017} of a high frequency, high-Q, bulk acoustic wave resonator (high-overtone bulk acoustic resonator) (HBAR) that is strongly coupled to a superconducting transmon qubit using piezoelectric transduction. The system was used to demonstrate basic quantum swap operations on the coupled qubit-phonon system. The $T_1$ relaxation time of the qubit was found to be 6 $\mu s$. Moreover,  $T_1$ and $T_2$ for the lowest phonon level in the resonator was found to be 17 $\mu s$ and 27 $\mu s$ resp. 
The analogy to 3D cQED is obvious, but the thickness of the device is only about 0.5 mm, so it looks more 2D than 3D.
It is expected \cite{Chu2017} that fairly straightforward improvements will make cavity quantum acoustodynamics (cQAD) a novel resource for building scalable hybrid quantum systems.

\subsection{Quantum interfaces to flying qubits (DV7)} 

The principle is that of good old radio technology:  from the transmitter side, one achieves low-frequency ($\omega$) modulation  of a strong high-frequency ($\Omega$) carrier (pump) beam by controlling the amplitude, frequency or phase of the carrier. The modulation is achieved by mixing the signals in a non-linear device, creating sidebands $\Omega \pm \omega$ around the carrier frequency.

In the present case, the mixers are different types of electro-optomechanical oscillators that influence the conditions for transmitting or reflecting the optical carrier beam. Typically three different oscillators are coupled in series, as illustrated in Fig.~\ref{Hybrid}: a microwave resonator ($\omega_{r}$), a micro/nanomechanical oscillator ($\omega_{m}$), and an optical cavity ($\Omega_{c}$), with coupling energies $g_{rm}$ and $g_{mc}$, resulting in the following standard Hamiltonian:
\begin{eqnarray}
\hat H =  \hbar \omega_{r} \;(a^+a + \frac{1}{2})  +  \hbar \omega_{m} \;(b^+b + \frac{1}{2}) +  \hbar \Omega_{c} \;(c^+c + \frac{1}{2}) \\  \nonumber
+  \; g_{rm}(a + a^+) (b + b^+) +  g_{mc} \;c^+c \; (b + b^+)
\end{eqnarray}
The mechanical oscillator changes the frequency of the optical cavity. This is the same principle as readout: the phase of the reflected carrier carries information about the state of the reflecting device. Here the phase of the reflected optical beam maps the state of the mechanical oscillator. Tuning the laser frequency $\Omega$ so that  $\Omega_{c} \approx \Omega \pm \omega$,  either sideband is now in resonance with the optical cavity. If the resonance linewidth of the optical cavity is smaller than $\omega$, then the sideband is resolved and will show a strong resonance.
Adding a (transmon) qubit coupled to the microwave resonator (Fig.~\ref{Hybrid}) one then has a chain of coupled devices that, if coherent, can entangle the localised qubit with the optical beam and the flying photon qubits. 

To create this entanglement is clearly a {\it major challenge}, and coherent coupling has so far only been achieved to varying degrees between various components. 
We will now briefly describe a few technical approaches to the central oscillator components: piezoelectric optomechanical oscillator \cite{Bochmann2013}, micromechanical membrane oscillator \cite{Andrews2014,Pirkkalainen2016,Bagci2014,Cernotik2016}, collective spin (magnon) oscillator\cite{Tabuchi2014,Osada2016,Hisatomi2016,Tabuchi2015,Tabuchi2016}, and SAW \cite{Shumeiko2016}.  

\subsubsection{Microwave-optical conversion: optomechanics}

\begin{figure}[h]
\center
\includegraphics[width=7cm]{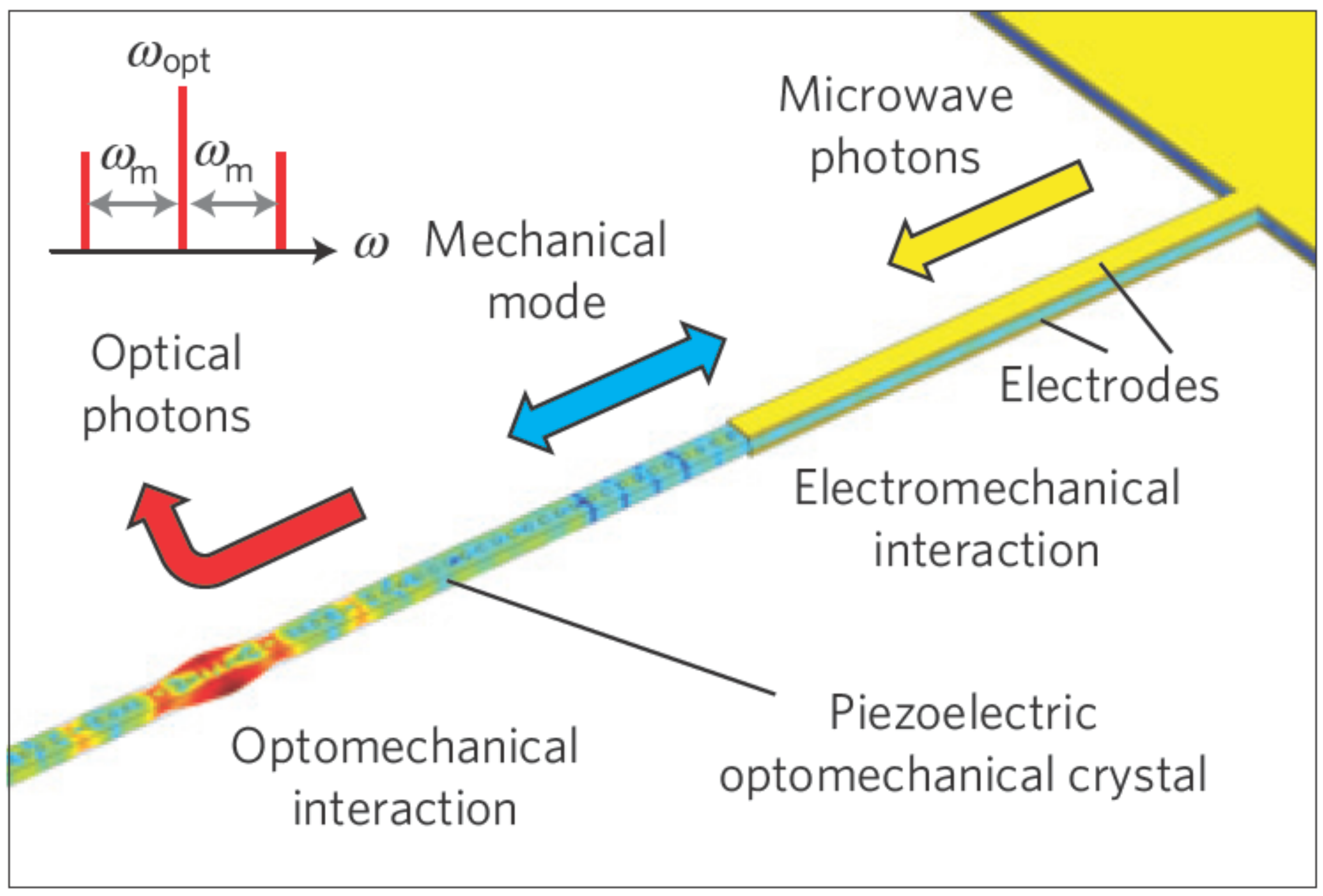} \;\;\;\;\;\;\;\;
\includegraphics[width=4cm]{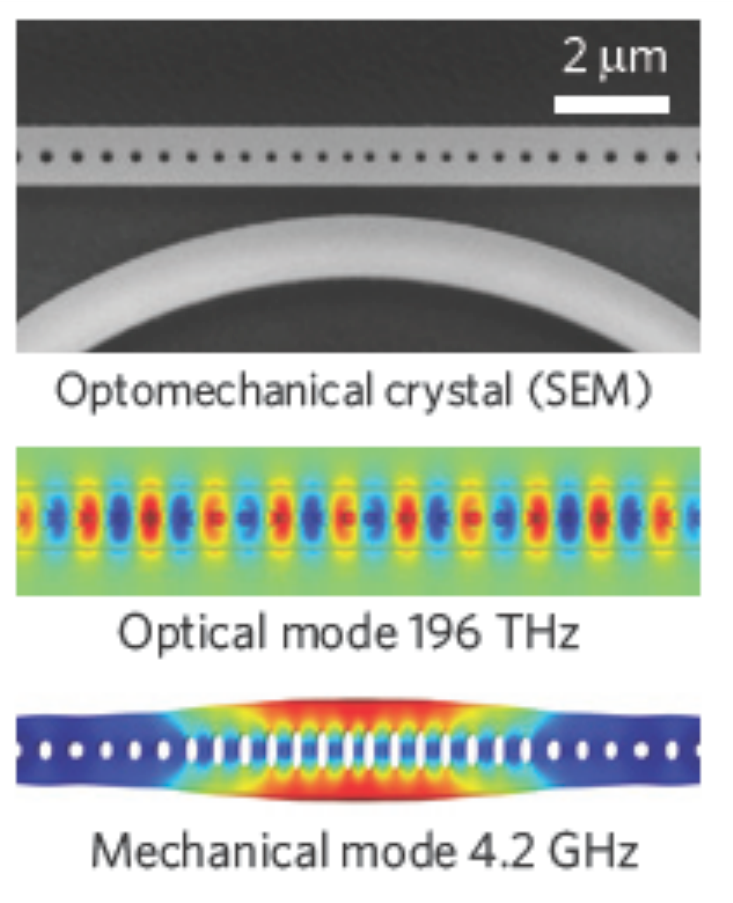}
\caption
{\small Layout and operation of microwave-to-optical converter using a piezoelectric optomechanical oscillator. Adapted from \cite{Bochmann2013}. }
\label{Bochmann}
\end{figure}
This approach (Fig.~\ref{Bochmann}) is based on the established optomechanical devices for modulating light \cite{Aspelmeyer2014}, and has been investigated experimentally \cite{Bochmann2013}.
A beam of piezoelectrical material is patterned to contain a nanophotonic (1D) crystal, localizing light in a region of enhanced vibrational amplitude.

\subsubsection{Microwave-optical conversion: micromechanics}

\begin{figure}[h]
\center
\includegraphics[width=14cm]{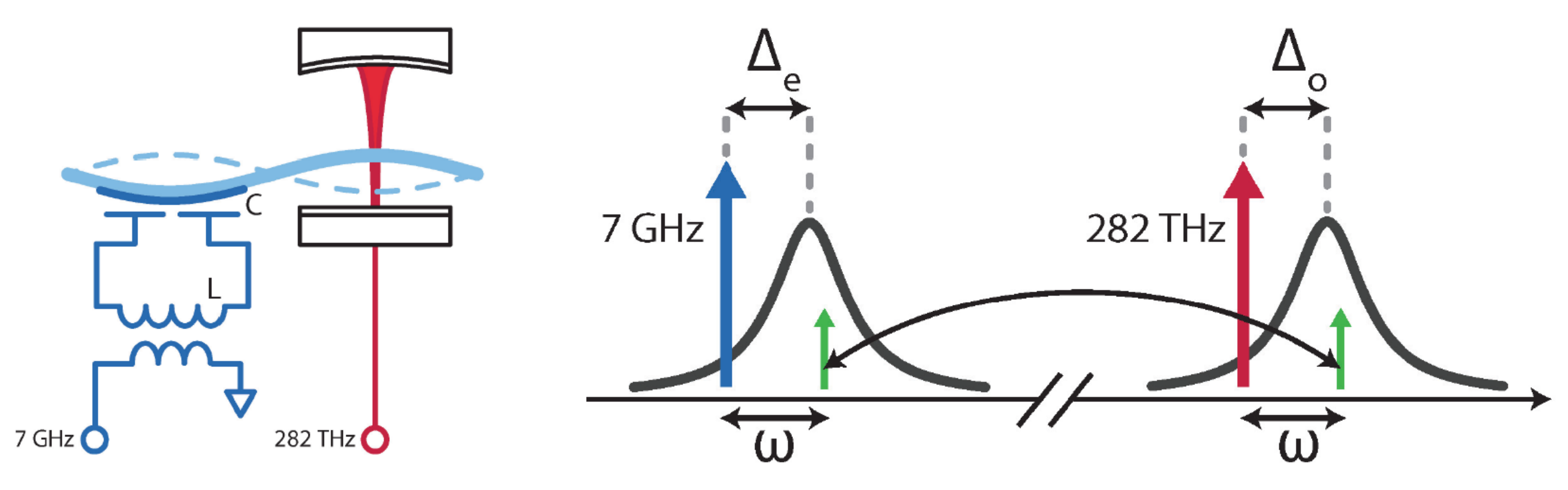}
\caption
{\small  Layout and operation of microwave-optical interface using an oscillating micromechanical membrane \cite{Andrews2014}. 
Microwave-to-optical conversion is achieved by pumping at optical frequency $\Omega$ with detuning so as to amplify the sidebands at  $\Omega \pm \omega$ inside the optical cavity resonance line. Optical-to-microwave conversion is achieved by pumping at MW frequency $\omega_p$ with detuning so as to amplify the sidebands at  $\omega_p \pm \omega$ inside the MW resonator resonance line. Adapted from \cite{Andrews2014}
}
\label{Andrews}
\end{figure}

This approach (Fig.~\ref{Andrews}) is based on the well-known technique of modulating reflected light, e.g. to determine the position of the tip of an AFM probe.
The radiation pressure (light intensity) exerts a ponderomotive force on the membrane (Fig.~\ref{Andrews}), coupling the mechanical oscillator and the optical cavity  \cite{Pirkkalainen2016}.
There are proof-of-concept experimental results showing coherent bi-directional efficient conversion of $GHz$ microwave photons and $THz$ optical photons \cite{Andrews2014}. Moreover, this technique was recently used for demonstrating optical detection of radiowaves \cite{Bagci2014}.

\subsubsection{Microwave-optical conversion: cavity optomagnonics} 
The Nakamura group has been investigating the coupling of microwave photons to collective spin excitations - magnons - in a macroscopic sphere of ferromagnetic insulator \cite{Tabuchi2014,Osada2016,Hisatomi2016,Tabuchi2015,Tabuchi2016}. They recently demonstrated strong coupling between single magnons in a magnetostatic mode in the sphere and a microwave cavity mode \cite{Tabuchi2014,Osada2016}, including bidirectional conversion \cite{Hisatomi2016}.

\subsubsection{Microwave-optical conversion: SAW}

\begin{figure}[h]
\center
\includegraphics[width=5cm]{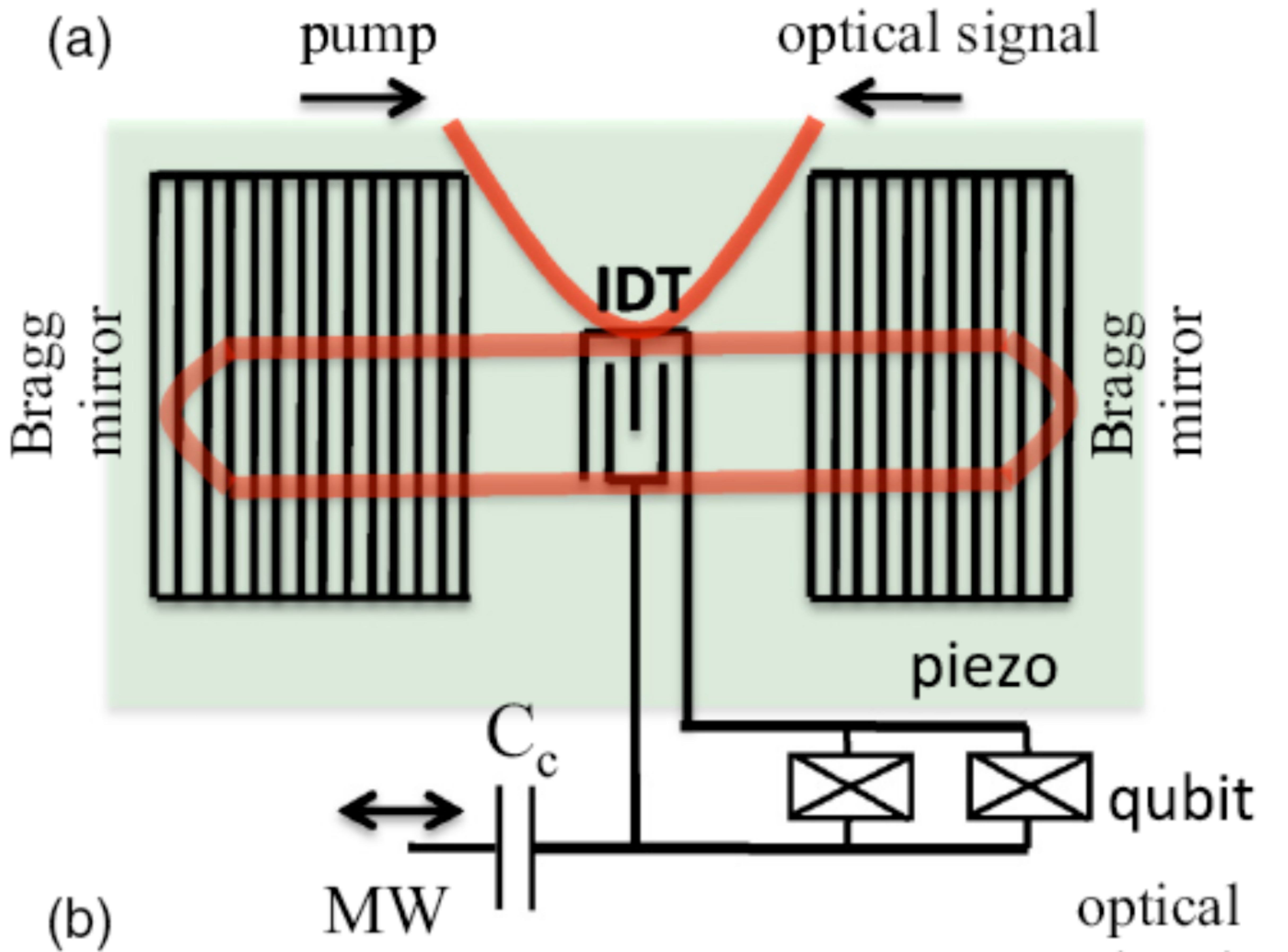} \;\;\;\;
\includegraphics[width=8cm]{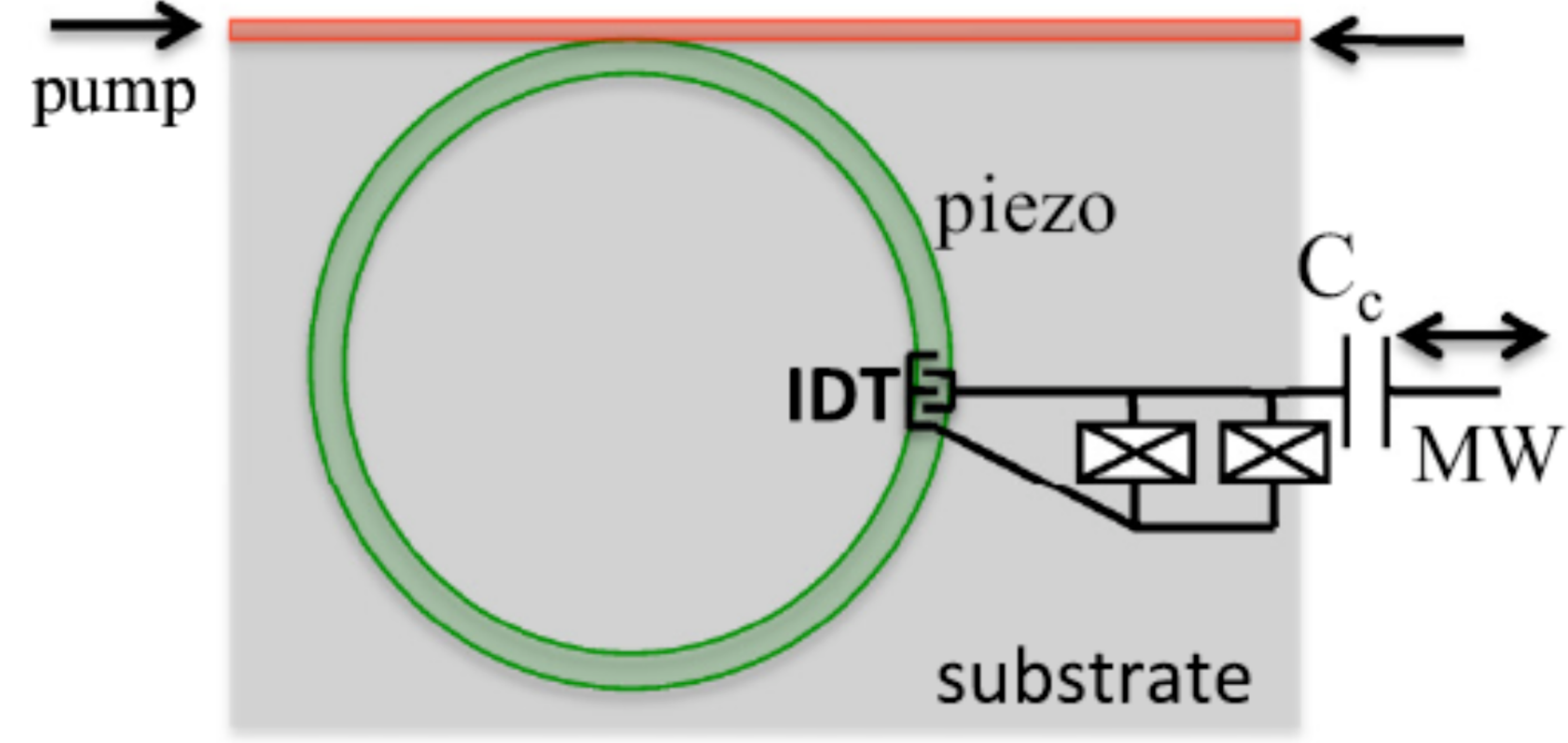}
\caption
{\small Layout and operation of microwave-to-optical converter using an SAW travelling wave. Adapted from \cite{Shumeiko2016}. }
\label{Shumeiko}
\end{figure}

Shumeiko \cite{Shumeiko2016} has presented a theory for a reversible quantum transducer (Fig.~\ref{Shumeiko}) connecting superconducting qubits and optical photons using acoustic waves in piezoelectrics. The approach employs stimulated Brillouin scattering for phonon-photon conversion,
and the piezoelectric effect for coupling of phonons to qubits. It is shown that full and faithful quantum conversion is feasible with state-of-the-art integrated acousto-optics.

\newpage

\section{Quantum gates}
\label{qgates}

\subsection{Quantum state time evolution}

In quantum information processing (QIP) one maps classical data on the Hilbert space of a given quantum circuit, studies the resulting time evolution of the quantum system, performs readout measurements of quantum registers, and analyses the classical output. At this level there is no difference between quantum computing (QC) and quantum simulation (QS).

\subsubsection*{Time-evolution operator}

The time evolution of a many-body system can be described by the Schr\"odinger equation for the state vector $\ket{\psi(t)}$, 
\begin{equation}
i \hbar \frac {\partial }{\partial t}\ket{\psi(t)} = \hat{H}(t) \ket{\psi(t)}.
\label{Seq}
\end{equation}
in terms of the time-evolution operator $\hat U(t,t_0)$ 
\begin{equation}
\ket{\psi(t)} = \hat U(t,t_0) \ket{\psi(t_0)}.
\label{psi}
\end{equation}
determined by the time-dependent many-body Hamiltonian $\hat{H}(t)$ of the system 
\begin{equation}
\hat{H}(t) =  \hat{H}_{syst} + \hat{H}_{ctrl}(t)
\end{equation}
describing the intrinsic system and the applied control operations. 
Gates are the results of applying specific control pulses to selected parts of a physical circuit. This affects the various terms in the Hamiltonian by making them time-dependent.

For simplicity,  $ \hat{H}_{syst}$ can be regarded as time-independent, and  $ \hat{H}_{ctrl}(t)$ taken to describe DC and microwave drives controlling the parameters of the total Hamiltonian. This involves e.g. tuning of qubits and resonators for coupling and readout, or setting up and evolving the Hamiltonian.
In addition, $ \hat{H}_{ctrl}(t)$ can introduce new driving terms with different symmetries. In general, the perturbing noise from the environment can be regarded as additional time dependence of the control parameters.

For the transmon, the system Hamiltonian takes the form in the RWA  (same notation as in Sect.~\ref{mqtmonham}): 
\begin{eqnarray}
\label{tmont}
\hat{H}_{syst} = - \; \frac{1}{2} \sum_{\nu i} \epsilon_i\; \sigma_{zi} \;
+ \sum_i g_i \;(\sigma^+_{i} a + \sigma^-_{i}a^+) \; + \; \hbar \omega \; a^+a  \\ 
+ \; \frac{1}{2} \sum_{i,j;\nu} \lambda_{\nu,ij}\;(\sigma^+_i \sigma^-_j +\sigma^-_i \sigma^+_j )  
\end{eqnarray}
and the control term can be written as
\begin{eqnarray}
\label{drivet}
\hat{H}_{ctrl}(t) =  \sum_{i;\nu} f_{\nu i}(t)\;\sigma_{\nu i} \; +\; \frac{1}{2} \sum_{i,j;\nu} h_{\nu,ij}(t)\;\sigma_{\nu i} \; \sigma_{\nu j}
+ \; k(t)\; a^+a 
\end{eqnarray}
The time dependent $ \hat{H}_{ctrl}(t)$  allows switching on and off or modulating  the various terms in  $ \hat{H}_{syst}$, as well as introducing pulse shapes for optimal control. In Eq.~(\ref{drivet}), the first term provides general types of single-qubit gates, the second term describes qubit-qubit coupling explicitly introduced by external driving, and the third term tuning of the frequency of the oscillator.
Moreover, in Eq.~(\ref{drivet}), the first term allows tuning of the qubit energies in and out of resonance with the oscillator, making it possible to switch on and off the qubit-oscillator coupling as well as creating oscillator-mediated qubit-qubit coupling. In the same way, the third term makes it possible to tune the oscillator itself in and out of resonance with the qubits.

The solution of Schr\"odinger equation for $\hat U(t,t_0)$
may be written  as
\begin{equation}
\label{Tevol}
\hat U(t,t_0) = \hat U(t_0,t_0) + \int_{t_0}^{t} \hat{H}(t') \hat U(t',t_0)dt'
\end{equation}
and in terms of the time-ordering operator $\hat{T}$:
\begin{equation}
\label{Tevol}
\hat U(t,t_0) = \hat{T} \;e^{-\frac{i}{\hbar} \int_{t_0}^{t} \hat{H}(t') dt'} \;,
\end{equation}
describing the time evolution of the entire many-particle state in the interval $[t_0,t]$.  
$\hat U(t,t_0)$ in Eq.~\ref{Tevol} is the basis for describing all kinds of quantum information processing, from the gate model for quantum computing to adiabatic quantum simulation. 
If the total Hamiltonian commutes with itself at different times, the time ordering can be omitted,
\begin{equation}
\label{tevol}
\hat U(t,t_0) = \;e^{-\frac{i}{\hbar} \int_{t_0}^{t} \hat{H}(t') dt'} \;.
\end{equation}
This describes the time-evolution controlled by a homogeneous time-dependent potential or electromagnetic field, e.g. dc or ac pulses with finite rise times, or more or less complicated pulse shapes, but having no space-dependence. Moreover, if the Hamiltonian is constant  in the interval $[t_0,t]$, then the evolution operator takes the simple form
\begin{equation}\label{UHconst}
\hat U(t,t_0) = e^{-\frac{i}{\hbar}\hat{H}(t-t_0)} \;,
\end{equation}
describing stepwise time-evolution.

Computation is achieved by sequentially turning on and off 1q and 2q gates, in parallel on different groups of qubits, inducing effective $N$-qubit gates. 

\subsection{Gate operations}

The time-development will depend on how many terms are switched on in the Hamiltonian during a given time interval. In the ideal case all terms are switched off except for those selected for the specific computational step. A single qubit gate operation then involves turning on a particular term in the Hamiltonian for a specific qubit, while a two-qubit gate involves turning on an interaction term between two specific qubits. In principle one can perform direct $N$-qubit gate operations by turning on interactions among all $N$ qubits.

\subsection{1q rotation gates}
1q gates are associated with the time-dependent 1q term of the control Hamiltonian:
$\hat{H}_{ctrl}(t) =  \sum_{i;\nu} f_{\nu i}(t)\;\sigma_{\nu i}$.
Expanding the state vector $\ket{\psi(t)}$ in a computational 1q basis, one obtains for a given single qubit,
\begin{eqnarray}
\ket{\psi(t)} 
= \sum_{m} a_{m} \sum_{k}\ket{k}\bra{k} e^{-\frac{i}{\hbar} \int_{t_0}^{t}  \sum_{\nu} f_{\nu}(t') dt' \sigma_{\nu}} \ket{m} 
\end{eqnarray}
For a general control Hamiltonian the $\sigma_{\nu}$-operators do not commute, and the exponential cannot be factorised in terms of products of  $\sigma_x$, $\sigma_y$ and $\sigma_z$ terms. To get a product we must apply the operators sequentially, acting in different time slots. In that case, for a given $\sigma_{\nu}$-operator we get
\begin{eqnarray}
\ket{\psi(t)}
= \sum_{m} a_{m} \sum_{k}\ket{k}\bra{k} e^{-\theta(t) \sigma_{\nu}} \ket{m} 
\end{eqnarray}
where $\theta$ = $\theta(t)$ = $\frac{i}{\hbar} \int_{t_0}^{t}  f_{\nu}(t') dt'$.

Expanding the exponential, calculating the $\bra{k} \sigma_{\nu} \ket{m}$ matrix elements, and resumming, one obtains the time evolution in terms of rotation operators $R_{\nu}(\theta)$:
\begin{eqnarray}
\ket{\psi(t)}
= \sum_{m} a_{m} \sum_{k}\ket{k}\bra{k} R_{\nu}(\theta)_{km} \ket{m} 
\end{eqnarray}
where 
\begin{equation}
\label{Urx}
R_x(\theta) = 
\left(\begin{array}{cc}
cos(\theta/2)  &  -i \;sin(\theta/2) \\
 -i \;sin(\theta/2)  & cos(\theta/2)  
\end{array}\right)  \;\;\;\; 
\end{equation}
\begin{equation}
\label{Ury}
R_y(\theta) = 
\left(\begin{array}{cc}
cos(\theta/2)  &  - \;sin(\theta/2) \\
 \;sin(\theta/2)  & cos(\theta/2)
\end{array}\right) 
\end{equation}
\begin{equation}
\label{Urz}
R_z(\theta) = 
\left(\begin{array}{cc}
\exp(-i\theta/2)  &  0 \\
 0  & \exp(i\theta/2)
\end{array}\right) 
\end{equation}
describing single qubit rotations around the x-, y-, and z-axes.

\subsection{2q resonance gates}

\subsubsection {iSWAP} 
\label{iSWAP}
The 2q iSWAP gate can be implemented by using $ \hat H_{ctrl}$ for tuning the energy of one of the qubits onto resonance with the other qubit, thereby effectively turning on the $ \hat H_{12}$ qubit-qubit interaction in $ \hat H_{syst}$.

Expanding the state vector $\ket{\psi(t)}$ in a computational 2q basis, one obtains
\begin{eqnarray}
\ket{\psi(t)} = \sum_{m,n} a_{mn} \sum_{k,l}\ket{kl}\bra{kl}e^{-i \hat H_{12} \; t} \ket{mn} 
\end{eqnarray}
If the qubits are on resonance ($\epsilon_1 =  \epsilon_2 $), then the matrix elements of the 2-qubit interaction part of the time evolution operator take the form
\begin{eqnarray}
\bra{kl} \hat H_{12} \ket{mn} = \lambda\;  \bra{kl}\sigma^+_1 \sigma^-_2 +\sigma^-_1 \sigma^+_2 \ket{mn} \\ \nonumber
=   \lambda\;  ( \delta_{k,m-1} \delta_{l,n+1}  +  \delta_{k,m+1} \delta_{l,n-1} )
\end{eqnarray}
Expanding the exponential function, introducing the matrix elements and resumming yields

\begin{equation}
\label{H12}
\hat U(t) =   \bra{kl}e^{-\frac{i}{\hbar}  \hat H_{12} \;t}\ket{mn} =   \bra{kl}  iSWAP \ket{mn},
\end{equation}
\begin{equation}
iSWAP = 
\left(\begin{array}{cccc}
1 &  0  & 0  & 0 \\
0  &  cos(\lambda t) & -i\;sin(\lambda t) &  0 \\
0  &  -i\;sin(\lambda  t) & cos(\lambda t) &  0  \\
0 &  0  & 0  & 1 
\end{array}\right),
\end{equation}
referred to as the iSWAP gate.

Creating an excited $\ket{01}$ state from $\ket{00}$ by a $\pi$ pulse, the iSWAP gate describes how the system oscillates between the $\ket{01}$ and $\ket{10}$ states.
The $\sqrt {iSWAP}$ gate is obtained by choosing $\lambda t = \pi/2$, 
\begin{equation}
\sqrt{iSWAP} =
\left(\begin{array}{cccc}
1 &  0  & 0  & 0 \\
0  &  1 & -i &  0 \\
0  &  -i  & 1&  0  \\
0 &  0  & 0  & 1 
\end{array}\right),
\end{equation}
putting the system in a Bell-state type of superposition $\ket{\psi} = \frac{1}{\sqrt{2}}(\ket{01} + i \ket{10})$.

\subsubsection {CPHASE} 

\begin{figure}[h]
\center
\includegraphics[width=14cm]{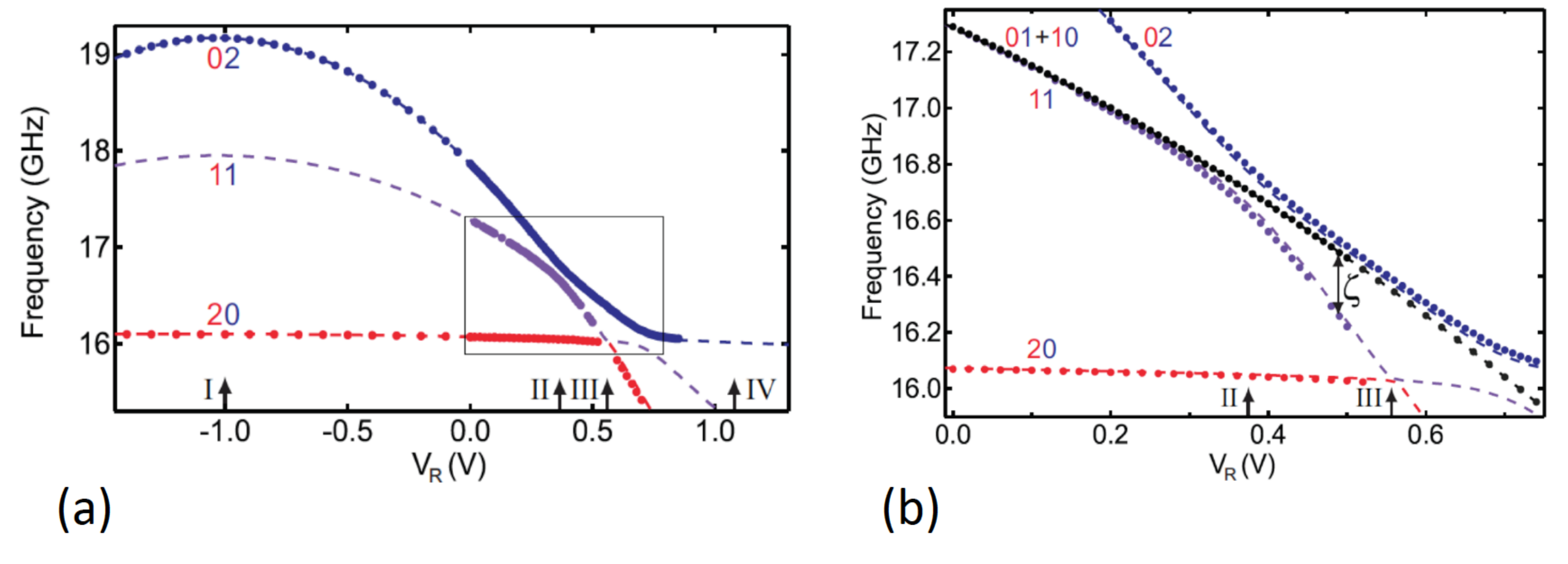}
\caption
{\small Energy level spectra explaining the CPHASE two-transmon resonance gate. The frequencies of the transmons are controlled by voltages $V_L$ and $V_R$ applied to CPWs controlling the flux in the transmon 2-JJ loops (Fig.~\ref{tmon_coupling}). Keeping $V_L$ constant and varying $V_R$ produces the energy level dispersion. The CPHASE gate is produced by moving from point  (I) to the curve crossing point (II), staying for a prescribed time ($\ket{11} \leftrightarrow  \ket{02}$), and then moving back to (I).  The frequency shift  $\zeta = f_{10} + f_{01} - f_{11}$ (Fig.~\ref{cphase}b) makes the phase of the  $\ket{11}$ level evolve more slowly that of  $\ket{01+10}$, producing a controlled phase gate.   Adapted from \cite{DiCarlo2009}. }
\label{cphase}
\end{figure}

The CPHASE gate can be implemented by making use of the spectral repulsion from the third level $\ket{2}$ of the transmon, as shown in Fig.~\ref{cphase}.  One first uses two-colour $\pi$-pulses to drive both qubits from $\ket{0}$ to $\ket{1}$, inducing a $\ket{00} \rightarrow \ket{11}$ transition (point I). Then tuning one of the qubits rapidly into (near) resonance with the other one, staying at the crossing of the $\ket{11}$ and $\ket{02}$ levels for a certain time (point II), produces an interaction-dependent shift $\zeta$ of the $\ket{11}$ level relative to $\ket{01+10}$. This induces an iSWAP gate between $\ket{11}$ and $\ket{02}$.

To see this, one expands the state vector in an extended computational basis:
\begin{eqnarray}
\ket{\psi(0)} = \sum_{m,n} a_{mn}\ket{mn} \nonumber \\
= a_{00}\ket{00} + a_{01}\ket{01} + a_{10}\ket{10}  + a_{20}\ket{20}  + a_{11}\ket{11} + a_{02}\ket{02}  
\end{eqnarray}
and calculates the matrix elements $\bra{kl}\hat H\ket{mn}$ (like in Sect. \ref{iSWAP}). The resulting energy level spectra in Fig.~\ref{cphase} show an avoided level crossing and a frequency shift $\zeta$ of the $\ket{11}$ level due to repulsion from the $\ket{02}$ level, as shown in detail in Fig.~\ref{cphase}b.

In this representation the evolution operator is diagonal, with the result that

\begin{equation}
\label{}
\hat U(t) = 
\left(\begin{array}{cccc}
1  &  0  & 0  & 0 \\
0  &  e^{\phi_{01} }   & 0 &  0 \\
0  &  0  & e^{\phi_{10}}    &  0 \\
0  &  0  & 0  &  e^{\phi_{11} (t)}  
\end{array}\right)
\end{equation}
with 

\begin{equation}
\phi_{01} = - \bra{01}\hat H\ket{01} \;t = - \frac{\epsilon_1}{2}  \;t \;; \;\;\phi_{10} = - \bra{10}\hat H\ket{10} \;t= - \frac{\epsilon_2}{2}  \;t
\end{equation}
\begin{equation}
\phi_{11} (t) = - \bra{11}\hat H\ket{11} \;t = -\frac{\epsilon_1+\epsilon_2}{2} \; t + h \int_{0}^{t}  \zeta(t)) dt
\end{equation}
In the experiment, the 11-02 splitting is determined by the time-dependent bias tuning voltage $V_R(t)$ in Fig.~\ref{cphase}.
If $\epsilon_1 = \epsilon_2$ (point II), then 
\begin{equation}
\phi_{01} = \phi_{10} = - \frac{\epsilon}{2} \; t \;; \;\;\; \phi_{11} (t) = -\epsilon \; t + h \int_{0}^{t}  \zeta(V_R(t)) dt
\end{equation}
After time t such that 
\begin{equation}
\phi_{01} = \phi_{10} = - \frac{\epsilon}{2}  \; t = 2\pi
\end{equation}
then the 11 state has rotated twice, and the phase is given by  $- 4\pi + h \int_{0}^{t} \zeta(V_R(t)) dt$.
\begin{equation}
\label{CPHASE}
\hat U(t)  = 
\left(\begin{array}{cccc}
1  &  0  & 0  & 0 \\
0  & 1 & 0 &  0 \\
0  &  0  & 1 &  0 \\
0  &  0  & 0  & e^{i \phi_{11} (t)}
\end{array}\right)
;\;\;\;\;\phi_{11}(t) = h \int_{0}^{t} \zeta(V_R(t)) dt
\end{equation}
At this point, the excursion of the bias voltage  will decide the integrated strength needed for achieving  
$\phi_{11}(t) = \pi$, providing the CPHASE gate (Fig.~\ref{2q_gates}a):
\begin{equation}
\label{CPHASE}
CPHASE = CZ =  Ctrl\;R_z(\pi) =
\left(\begin{array}{cccc}
1  &  0  & 0  & 0 \\
0  & 1 & 0 &  0 \\
0  &  0  & 1 &  0 \\
0  &  0  & 0  & -1
\end{array}\right)
\end{equation}

\subsubsection {CNOT}
The CNOT gate can be expressed in terms of CPHASE and two Hadamard gates, as commonly implemented in transmon circuits (Fig.~\ref{2q_gates}b): 
\begin{eqnarray}
\label{CNOT}
CNOT = CX = Ctrl\;R_y(\pi) = 
\left(\begin{array}{cccc}
1  &  0  & 0  & 0 \\
0  & 1 & 0 &  0 \\
0  &  0  & 0 &  1 \\
0  &  0  & 1  & 0
\end{array}\right)
\end{eqnarray}
The first H-gate changes from the z- to the x-basis, and the second H-gate transforms back.

\subsubsection{Controlled rotation}
CPHASE is a special example of the general controlled Z-rotation - Ctrl-Z($\theta$) - gate in Eq.~(\ref{CPHASE}) and Fig.~\ref{2q_gates}c, allowing one to control time evolution and (Fig.~\ref{2q_gates}d) to map states to ancillas for phase estimation,

\subsubsection{2q time evolution}
\begin{figure}[h]
\center
\includegraphics[width=12cm]{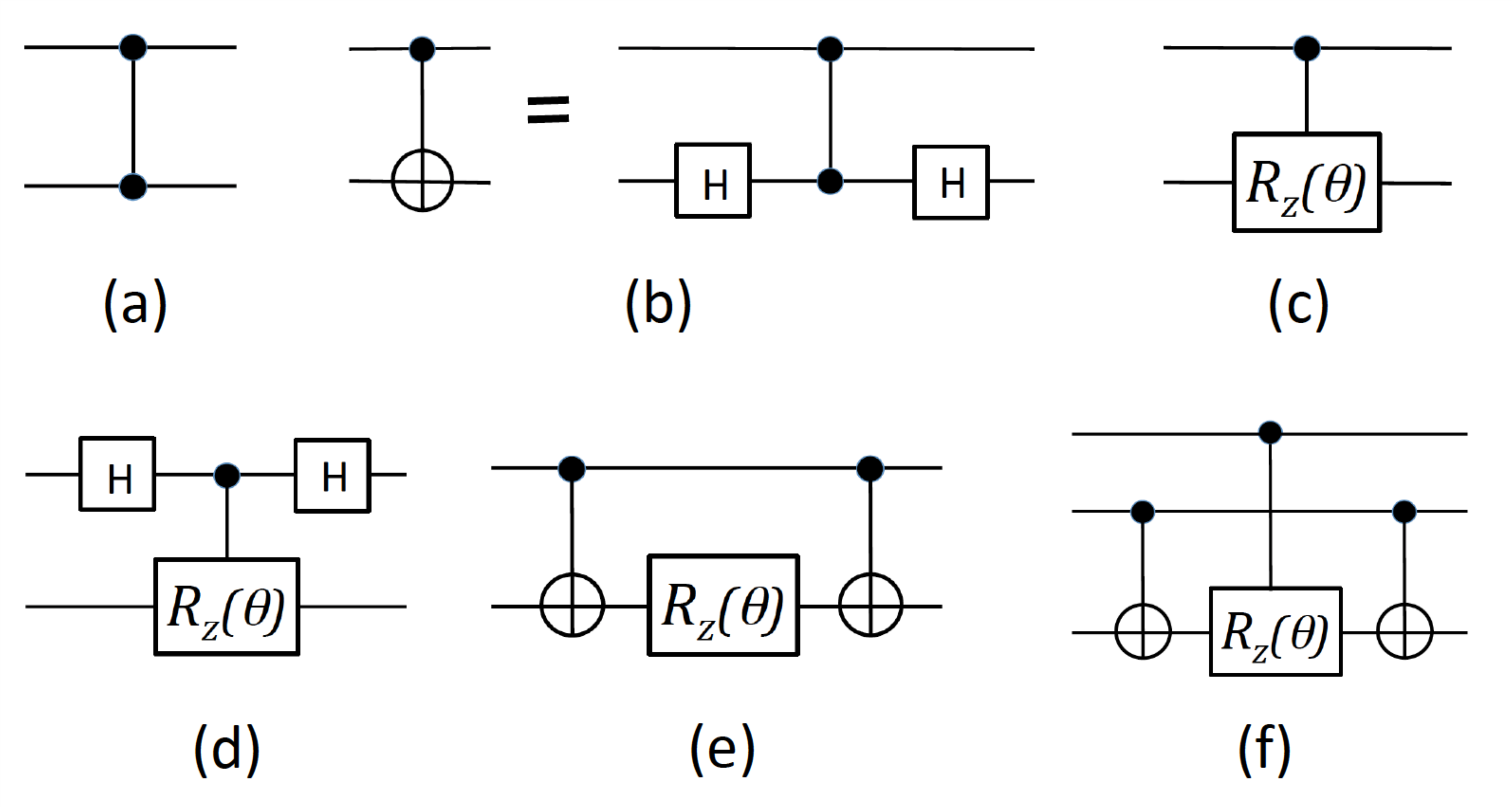}
\caption
{\small Circuits for implementation of (a) CPHASE; (b) CNOT; (c) Ctrl-Z($\theta$), $\theta$ arbitrary; (d) basic circuit for phase estimation using an ancilla (top qubit); (e) the $U = exp[{-i\frac{\theta}{2}\sigma_{z} \otimes \sigma_{z}}]$ operator; (f) a controlled version of (e) for controlled time evolution and phase estimation (top qubit).
 }
\label{2q_gates}
\end{figure}

We now have the tools to describe the time evolution operator corresponding to 2-qubit interaction terms. The parts of the Hamiltonian with 
$\sigma_{z} \otimes \sigma_{z}$ products, $U = exp[{-i\frac{\theta}{2}\sigma_{z} \otimes \sigma_{z}}]$ can be implemented by a quantum circuit of the form shown in Fig.~\ref{2q_gates}e \cite{NielsenChuang2010}. 
Operators like $exp[{-i\frac{\theta}{2}\sigma_{x} \otimes \sigma_{x}}]$ and $exp[{-i\frac{\theta}{2}\sigma_{z} \otimes \sigma_{x}}]$ can be generated by adding a number of 1q-rotation gates. Moreover, Fig.~\ref{2q_gates}f represents a controlled version of Fig.~\ref{2q_gates}e for controlled time evolution and phase estimation.

\subsection {2q gates induced by microwave driving}
\label{MWdrive}

The flux-tunability of transmons make them sensitive to flux noise, resulting in decoherence. One approach is therefore to use fixed-frequency transmon qubits, replacing the frequency-tuning squid by a single JJ. This is also an important design for arrays of 3D transmon qubits where direct access for tuning individual qubits may be difficult or impossible.

The generic approach for coupling non-linear oscillators is to use electromagnetic driving  fields to induce parametric coupling with tunable strength by creating a spectrum of sidebands bridging frequency gaps. 
In this way it is possible to entangle superconducting qubits with different frequencies using (i) fixed linear couplings, (ii) only microwave control signals, and (iii) tunable effective interaction strengths. 
Recently these methods have been applied experimentally through a variety of schemes based on two different principles: (i) driving qubits, and (ii) driving coupling resonators, e.g. a tunable bus.

\subsubsection {Driving qubits} 

\subsubsection* {Cross resonance (CR) 2q gates} 

The CR scheme \cite{Paraoanu2006,Rigetti2010,Chow2011,Corcoles2013,Sheldon2016} exploits already present nonlinearities to achieve tunable coupling, circumventing the need for nonlinear coupling elements.
The CR two-qubit gate scheme irradiates one of the qubits at the transition frequency of the other qubit. 
In the presence of this cross-resonant microwave drive, an effective coupling emerges between the two qubits whose strength increases linearly with the ratio (drive amplitude)/(difference frequency).

The CR-coupling of two qubits, Q1 and Q2, can be understood in the dressed state picture of quantum optics \cite{Paraoanu2006,Rigetti2010}.
Under CR driving, the central transition at the irradiation frequency of the driven dressed Q1 system is matched to the bare transition of the undriven Q2. One thus creates a resonance between the central feature of the Mollow triplet on Q1 and the bare transition of Q2. The tunability of the effective coupling strength $G$ results from the evolution of the dressed Q1 eigenstates as the field amplitude $F$  is adjusted \cite{Paraoanu2006,Rigetti2010}:
\begin{eqnarray}
\hat{H}_{eff} =  g(F) \; \sigma_{z1}\;\sigma_{x2},
 \end{eqnarray}
which is related to the CNOT gate by one additional local $\pi/2$ rotation of each qubit.  

In addition to the CR scheme, one approach is to create a microwave-activated conditional-phase gate (MAP) \cite{Chow2013} based on driving the $\ket{03}$ and  $\ket{12}$  transmon states into resonance. 
A general problem with driving qubits is that the couplings may depend sensitively on the qubit level structure. For transmon qubits the CR scheme is limited by the weak anharmonicity of the transmon, and the MAP scheme employs specific  higher excited states of the transmon. These schemes may therefore be challenging to scale up to many qubits.

\subsubsection {Driving a tunable bus} 

Attaching a SQUID  to the end of a coplanar wave-guide resonator (CPW) makes it possible to vary the boundary condition (effective length) and create a flux-tunable resonator   \cite{Sandberg2008,Palacios-Laloy2008} and to couple qubits \cite{Walllquist2006,Sandberg2009}.  In \cite{Walllquist2006}, fixed-frequency qubits with different frequencies were coupled by successively bringing each qubit quasi-statically in and out of resonance with the tunable CPW, effectively creating multi-qubit gates. 
In \cite{Sandberg2008,Sandberg2009}, the CPW was rapidly tuned (chirped) to create interference and beating of microwave emission, which in principle could dynamically couple qubits \cite{Sandberg2009}.
Alternatively, one can drive the resonator at high frequency to create sideband structure and dynamic parametric coupling between qubits. This is presently at the focus of extensive and promising research  \cite{Paik2016,McKay2016,Cross2015,Andersen2015,Felicetti2014}, potentially providing multi-qubit gate architectures for scaled-up systems. A recent proposal is based on the  {\em Dynamical Casimir Effect} \cite{Felicetti2014}: A SQUID is then connected to the midpoint of a CPW resonator that is connected to transmon qubits at both ends, varying the coupling between the two halves by flux tuning. Driving the SQUID at microwave frequencies emits pair of photons that can entangle the qubits \cite{Felicetti2014}.

\subsubsection* {Resonator-induced phase gate (RIP)} 
  
In the resonator-induced phase gate (RIP) scheme \cite{Paik2016,McKay2016,Cross2015}  fixed-frequency transmons are statically coupled to the same bus resonator driven at the difference frequency of two qubits. 

In a two 2D-transmon setup \cite{McKay2016}, parametrically oscillating a flux-tunable "bus qubit" (similar to a combination of the qubit-qubit couplings in Figs.~\ref{tmon_coupling}b,c) at the qubit-qubit detuning enables a $\sigma_+ \sigma_- +\sigma_- \sigma_+$ resonant exchange (XX+YY) interaction. The interaction is said to implement a 183 $ns$ two-qubit iSWAP gate between qubits separated in frequency by 854 $MHz$ with a measured average fidelity of 0.98 from interleaved randomized benchmarking. This gate may be an enabling technology for surface code circuits and for analog quantum simulation \cite{McKay2016}.

In a 3D-transmon-cQED setup with four superconducting qubits \cite{Paik2016}, RIP gates are experimentally implemented between pairs of qubits, demonstrating high-fidelity CZ gates between all possible pairs of qubits. The qubits are arranged within a wide range of frequency detunings, up to as large as 1.8 $GHz$. This setup was used to generate a four-qubit Greenberger-Horne-Zeilinger (GHZ) type of state \cite{Paik2016}.

\subsubsection* {M\o lmer-S\o rensen (MS) 2q gate}

In ion traps,  the qubits (ions) are naturally coupled by collective vibrational modes generating a sideband structure \cite{Monz2011}.
The M\o lmer-Sorensen (MS) gate \cite{MoSo1_1999,MoSo2_1999,SoMo_2000}  is a single-step 2-qubit entangling gate $\ket{00} \rightarrow \ket{11}$  driven by 2-tone 2-photon excitation assisted by collective (vibrational)  modes, providing resonant intermediate states with sideband structure. 
It looks like Rabi driving of a single qubit, coupling the $\ket{0, 1ph} $ and  $\ket{1, 0ph} $ levels, $\ket{0,1ph} \pm \ket{1,0ph} \; \leftrightarrow  \; \lambda \; \sigma_{x1}$ extended to resonant 2-photon direct driving of two qubits, coupling the $\ket{00, 2ph} $ and  $\ket{11, 0ph} $ levels, $\ket{00,2ph} \pm \ket{11,0ph} $ giving rise to an effective 2-qubit interaction $ \lambda \; \sigma_{x1}\;\sigma_{x2}$.

\subsubsection*{Multiqubit gates}
can be implemented in several ways: (i) sequentially, by series of 1q and 2q gates to yield CCNOT (Toffoli) or CCZ  \cite{Mariantoni2011,Reed2012,Fedorov2012}; (ii) by a "single-shot" optimised pulse sequence \cite{Zahedinejad2015,Zahedinejad2016}; or (iii) by  single-shot collective excitation via bus dynamics. The state of the art of (iii)  is currently defined by an ion-trap experiment entangling 14 qubits  via a M\o lmer-S\o rensen (MS)  gate \cite{Monz2011}.
The MS gate can be generalised to direct Rabi-like driving of N qubits, coupling the $\ket{0...00, Nph} $ and  $\ket{1...11, 0ph} $ levels, leading to an effective N-qubit interaction $ \lambda \; \sigma_{x1}\;\sigma_{x2} ....... \;\sigma_{xN}$.
There is a recent proposal for implementing MS gates in transmon-cQED  \cite {Andersen2015} by driving a SQUID to create sideband structure, as discussed above, and simultaneously driving the qubits with two microwave tones. 
Also, there are recent proposals how to implement controlled multi-qubit gates by multi-tone microwave driving of the bus, the control qubit, and the target qubits, in cQED \cite{Yang2005,Yang2010,Yang2014,Li2017}.

Finally we mention two very different recent approaches to gate construction and time evolution:  a genetic algorithm (GA) approach to decomposing the time evolution operator in a series of high-fidelity gates adapted to the system under study  \cite{LasHeras2016}; and a supervised-learning approach to tuning a 4-spin circuit to performing a 3q-Toffoli gate in a feed-forward style without external control \cite{Banchi2016}.

\subsection {Gate synthesis and universal sets of gates}
\label{universal}
An important problem in quantum computing is how to decompose an arbitrary unitary operation in a sequence of standard elementary gates from a gate library \cite{GilesSelinger2013}. The Clifford stabiliser group can be generated by three elementary gates:  two single-qubit gates, the Hadamard gate H, and the S-gate (phase gate), 
\begin{equation}
\label{U}
H = \frac{1}{\sqrt{2}}
\left(\begin{array}{cc}
1  &  1 \\
1  & -1
\end{array}\right); \;\;\;\;\;\;\;\;
\label{U}
S  = 
\left(\begin{array}{cc}
1  &  0 \\
0  & i
\end{array}\right) \;\;\;\;\;\;\;\;
\end{equation}
and a two-qubit gate, e.g. CNOT (Eq.~(\ref{CNOT})) or CPHASE (Eq.~(\ref{CPHASE})). This constitutes a standard set of Clifford gates {\it \{H, S, CNOT\}}.
Adding a non-Clifford gate, such as the T gate: 
\begin{equation}
\label{U}
T  = 
\left(\begin{array}{cc}
1  &  0 \\
0  & e^{\pi/4}
\end{array}\right) \;\;\;\;\;\;\;\;
\end{equation}
forms a universal set {\it \{H, S, CNOT, T\}}, and makes it possible to generate all quantum circuits, i.e. in principle to perform universal, but not necessarily efficient (poly-time),  quantum computing. 

Any quantum circuit of operations can be described by a sequence of gates from this finite universal set. For the specific case of single qubit unitaries the Solovay-Kitaev theorem \cite{DiMatteoMosca2016} guarantees that this can be done efficiently, involving only a polynomial number of gates. For a general quantum circuit, it may take an exponential number of gates  to synthesise the operation. Recently Bravyi and Gosset \cite{BravyiGosset2016} presented a new algorithm for classical simulation of quantum circuits over the universal Clifford + T gate set which is polynomial in the number of qubits and the number of Clifford gates but exponential in the number of T gates. However, the exponential scaling was sufficiently mild that it was possible to use the algorithm to simulate classically a medium-sized quantum circuit dominated by Clifford gates, namely a hidden-shift quantum algorithm with 40 qubits, a few hundred Clifford gates, and nearly 50 T gates \cite{BravyiGosset2016}.

\newpage

\section{Quantum state preparation and characterisation}

\subsection{Quantum state characterisation}
\label{characterisation}
Quantum state characterisation and quantum computing are two sides of the same coin. Both involve application of long series of gates. The difference lies in the purpose of the protocol: characterisation of a quantum state, or solution of a computational problem.   Tomography is all about visualisation of quantum states and processes, which means measurement and presentation of  large quantities of quantum information.

In the general case, the N-qubit density matrix is defined by $(2^{N})^2$ matrix elements, requiring $2^{2N} -1$ independent measurements for characterization, based on an ensemble of replicas of the state in question. In the 2-qubit case, 15 matrix elements have to be determined via a set of 15 measurements at different angles. In solid-stated devices the integrated detectors cannot be rotated. Therefore, measurements are typically performed by rotation gates applied to the various qubits before measurement. The data are then used to calculate the elements of the density matrix.
The discussion is of course not limited to two-level systems: the transmon is a multi-level system and a single qutrit 3-level density matrix was investigated in\cite{Bianchetti2010}.

For large-scale quantum circuits and systems to be trustworthy, it is all-important to be able to verify the functionality and performance at various levels by appropriate measurements and data analysis. Full characterisation via quantum state tomography (QST) and quantum process tomography (QPT) \cite{ChuangNielsen1997,Mosheni2008,Yuen-Zhou2014,ChristandlRenner2012} scales exponentially with the number of qubits - an NP-hard problem - and is only possible for small systems. For up to three qubits, QST and QPT are well-established tools for characterising quantum states, gates and processes, as demonstrated by a large number of recent applications involving superconducting circuits
(see e.g. \cite{Fedorov2012,Yamamoto2010,Baur2012,Chow2012}). 

For larger systems, however, full QPT becomes impractical, and methods have to be developed for reducing the information needed, e.g. via {\em randomised benchmarking} (RB) (a randomised QPT procedure, applying a random sequence of gates) \cite{Knill2008,Wallman2014,Ball2016}, compressed sensing QPT (CSQPT) (exploiting sparsity of matrices) \cite{Kosut2008,Gross2010,Shabani2011,Flammia2012,Rodionov2014}, and adaptive Bayesian quantum tomography \cite{Huszar2012,Kravtsov2013,Mahler2013} 

The rapid scaling up of transmon systems will make reduced-information methods indispensable, and there has already been a number of recent applications of RB and twirling protocols  \cite{Chow2009,Magesan2011,Kimmel2014,Ryan2015,Johnson2015,Cross2016} and  CSQPT  \cite{Rodionov2014} to superconducting circuits. While standard QPT provides information about a single gate, RB gives a measure of the accumulated error over a long sequence of gates. Standard QPT is limited by errors in state preparation, measurement and one-qubit gates and suffers from inefficient scaling with number of qubits. RB yields estimates of the computationally relevant errors without relying on accurate state preparation and measurement. Since it involves long sequences of randomly chosen gates, it also verifies that error behaviour is stable when used in long computations. 
{\em Interleaved} RB (IRB)  is a scalable protocol for estimating the average error of individual quantum
computational gates.  IRB involves sequentially mixing Clifford and Pauli gates.
The technique takes into account both state preparation and measurement errors and is scalable in the number of qubits
\cite{Barends2014a,Magesan2012,Sheldon2015,Chasseur2016}.
{\em Twirling} provides useful partial information of a quantum channel by averaging the channel under the action of a set of unitaries \cite{Moussa2012,Lu2015}, pre-multiplying the input state by an operation, running the original process, post-multiplying by the inverse operation, and finally averaging over a set of operations. Randomly applying the Pauli operators with uniform probability to any density operator - Pauli twirling - gives the maximally mixed state. Twirling is often a part of the abovementioned RB protocols to introduce averaging. 

\subsection{Quantum Supremacy characterisation}

In just a few years, quantum computers without error correction are expected to be able to approximately sample the output
of random quantum circuits that state-of-the-art classical computers cannot simulate \cite {FarhiHarrow2016}. 
In this spirit, Boixo et al. \cite{Boixo2016b} study the computational task of sampling from the output distribution of pseudo-random
quantum circuits composed from a universal gate set, a typical benchmarking problem, and introduce the concept of Cross Entropy (CE)  to characterise Quantum Supremacy.
The Shannon entropy $S(A) = -\sum_a P(A_a) \; ln P(A_a)$ is a measure of the inherent uncertainty of a single random variable $A$.
The Cross Entropy $CE(A;B) = -\sum_x P(A_x) \; ln P(B_x)$  is a {\em measure of inaccuracy} \cite{Kerridge1961} and gives the average number of bits needed to identify events that occur with probability
$P(A_x)$, if a coding scheme is used that is optimal for the probability distribution $P(B_x)$ \cite{Crooks2016}.

Boixo et al. \cite{Boixo2016b} consider a 7 x 6 qubit 2D lattice with gate depth 25, close to the limit of present classical computers (7 x 7 qubits).
They show how to estimate the cross entropy between the sampled output distribution $P(A)$ of an experimental implementation of a random quantum circuit providing quantum chaos, and the ideal output distribution $P(B)$ simulated by a supercomputer, and argue that the cross entropy is closely related to the circuit fidelity. 
If the experimental quantum device achieves a cross entropy surpassing the performance of the
state-of-the-art classical competition, this will be a first demonstration of Quantum Supremacy \cite{Boixo2016b,Crooks2016} . 

A crucial aspect of a near-term Quantum Supremacy proposal is that the computational task can only be performed
classically through  direct simulations with cost that is exponential in the number of qubits. Quantum Supremacy can be claimed if the theoretical estimates are in good agreement with the experimental extrapolations.

\subsection{Multi-qubit state preparation}

\subsubsection {Bell states}

Bell state preparation and characterisation represent pioneering experiments \cite{Ansmann2009,DiCarlo2009}.
By now these experiments are "routine", and focus is rather on multi-qubit entangled states for QEC and QIP. 
\begin{figure}[h]
\includegraphics[width=15.5cm]{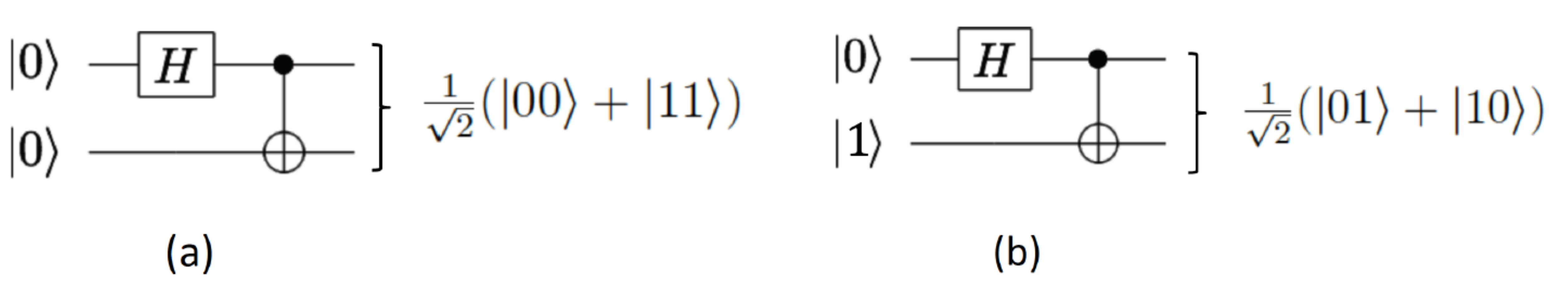}
\caption
{\small Bell states. The Hadamard gate creates superposition states  $ \ket{0} \rightarrow   \ket{0} +  \ket{1}$ and $ \ket{1} \rightarrow   \ket{0} -  \ket{1}$, which allows the CNOT gate to create 2q-superposition Bell states: (a)  $\frac{1}{\sqrt{2}} (\ket{00} +  \ket{11})$; (b) $ \frac{1}{\sqrt{2}} (\ket{01} +  \ket{10}$).
}
\label{Bell}
\end{figure}

\begin{figure}[h]
\center
\includegraphics[width=15cm]{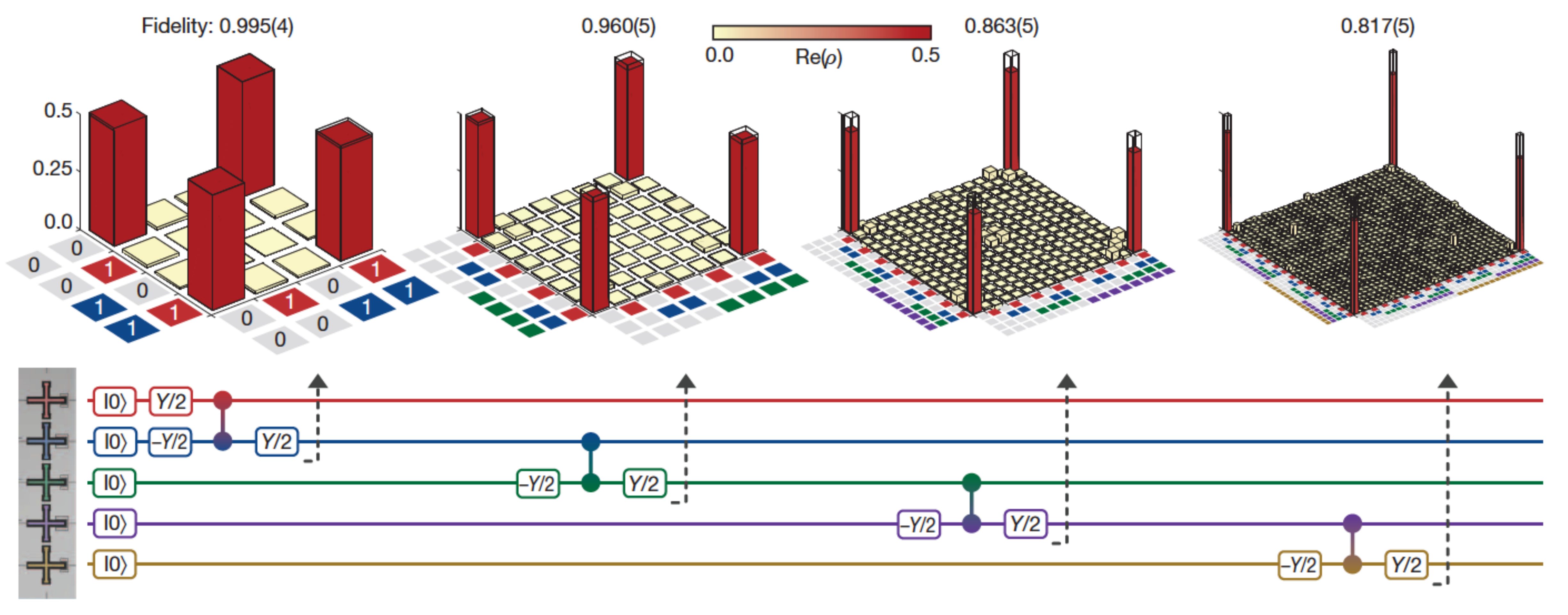}
\caption
{\small Quantum state tomography (QST) and generation of GHZ type of  states. Top row: Real part of the density matrix for the N = 2 Bell state and the N = 3, 4 and 5 GHZ states, measured by quantum state tomography. Ideal density matrix elements are transparent, with value 0.5 at the four corners. Bottom row: Algorithm used to construct the states. Adapted from \cite{Barends2014a}.
 }
\label{UCSB_5q_GHZ}
\end{figure}

For the double-excitation Bell state $\ket{\psi}  = \frac{1}{\sqrt{2}}(\ket{00} + \ket{11}) $,  the density matrix  is given by
\begin{equation}
\label{rhomatrix}
\hat\rho  = \frac{1}{2}(\ket{00} + \ket{11})(\bra{00} + \bra{11}) =  \frac{1}{2}
\left(\begin{array}{cccc}
1  & 0  &  0 &  1 \\
0  & 0  &  0 &  0 \\
0  & 0  &  0 &  0\\
1  &  0 &  0 & 1
\end{array}\right)
\end{equation}
Figure~\ref{UCSB_5q_GHZ} shows characterisation of maximally entangled GHZ-type states with five capacitively coupled Xmon qubits \cite{Barends2014a} via quantum state tomography (QST).
The Bell state density matrix is experimentally demonstrated in the leftmost panel of Fig.~\ref{UCSB_5q_GHZ}, showing the characteristic four corner pillars.
(Note that Song et al. \cite{Song2017} recently published tomographic results for a 10-qubit GHZ state).

For the single-excitation Bell state $\ket{\psi}  = \frac{1}{\sqrt{2}}(\ket{01} + \ket{10}) $, the density matrix  is given by
\begin{equation}
\label{rhomatrix}
\hat\rho  = \frac{1}{2}(\ket{01} + \ket{10})(\bra{01} + \bra{10}) =  \frac{1}{2}
\left(\begin{array}{cccc}
0  & 0  &  0 &  0 \\
0  & 1  &  1 &  0 \\
0  & 1  &  1 &  0\\
0  &  0 &  0 & 0
\end{array}\right)
\end{equation}
spanning a different part of Hilbert space than the double-excitation Bell state.

\subsubsection {GHZ states}

For the triple-excitation GHZ state $\ket{\psi}  = \frac{1}{\sqrt{2}}(\ket{000} + \ket{111}) $,  the density matrix  is given by
\begin{equation}
\label{rhomatrix}
\hat\rho  = \frac{1}{2}(\ket{000} + \ket{111})(\bra{000} + \bra{111}) 
\end{equation}
corresponding to the corner pillars in the second panel in Fig.~\ref{UCSB_5q_GHZ}.

\subsubsection {W-states}

\begin{figure}[h]
\center
\includegraphics[width=15.5cm]{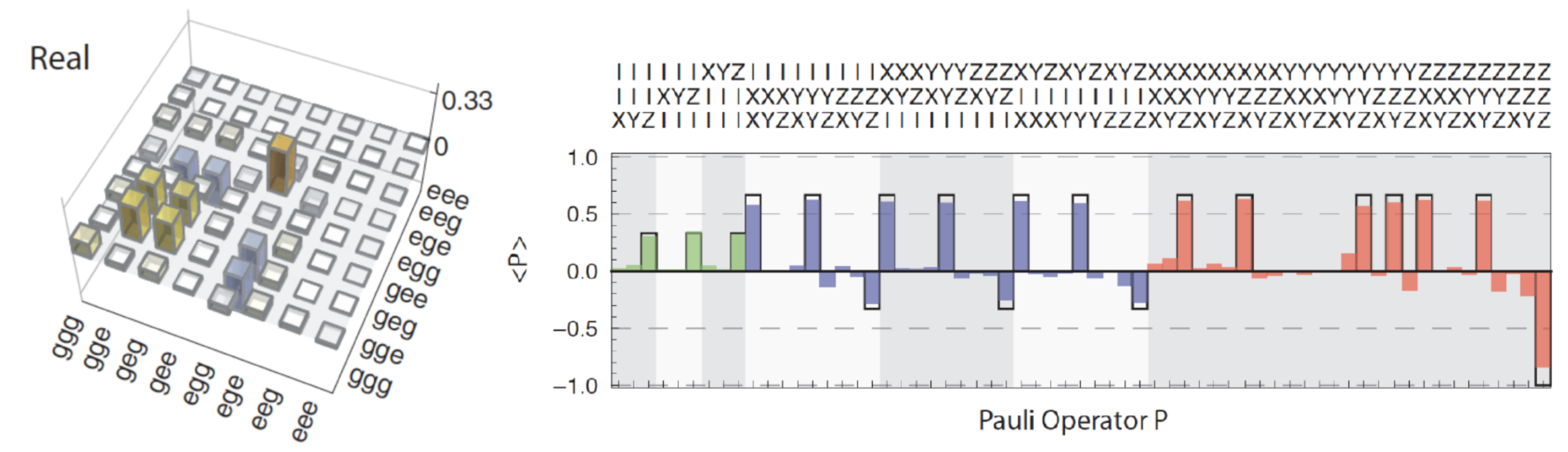} 
\caption
{\small Tomography of the W-state \cite{Mlynek2012}. Left: Real part of the density matrix. Right: Pauli set from quantum process tomography. Adapted from \cite{Mlynek2012}.
 }
\label{ETHZ_W}
\end{figure}

For the 3-qubit single-excitation Werner W-state, $\ket{\psi}  =  \frac{1}{\sqrt{3}}(\ket{001} + \ket{010} + \ket{100}) $,  the density matrix  is given by
\begin{eqnarray}
\label{rhomatrix}
\hat\rho  = \frac{1}{3}(\ket{001} + \ket{010} + \ket{100})(\bra{001} + \bra{010}  + \bra{100}) 
\end{eqnarray}
The form of the density matrix is illustrated by the experimental results shown in Fig.~\ref{ETHZ_W} \cite{Mlynek2012} (cf. the N=3 GHZ state in Fig.~\ref{UCSB_5q_GHZ}).

\subsubsection{Generating Bell states by parity measurement}

\begin{figure}[h]
\center
\includegraphics[width=6cm]{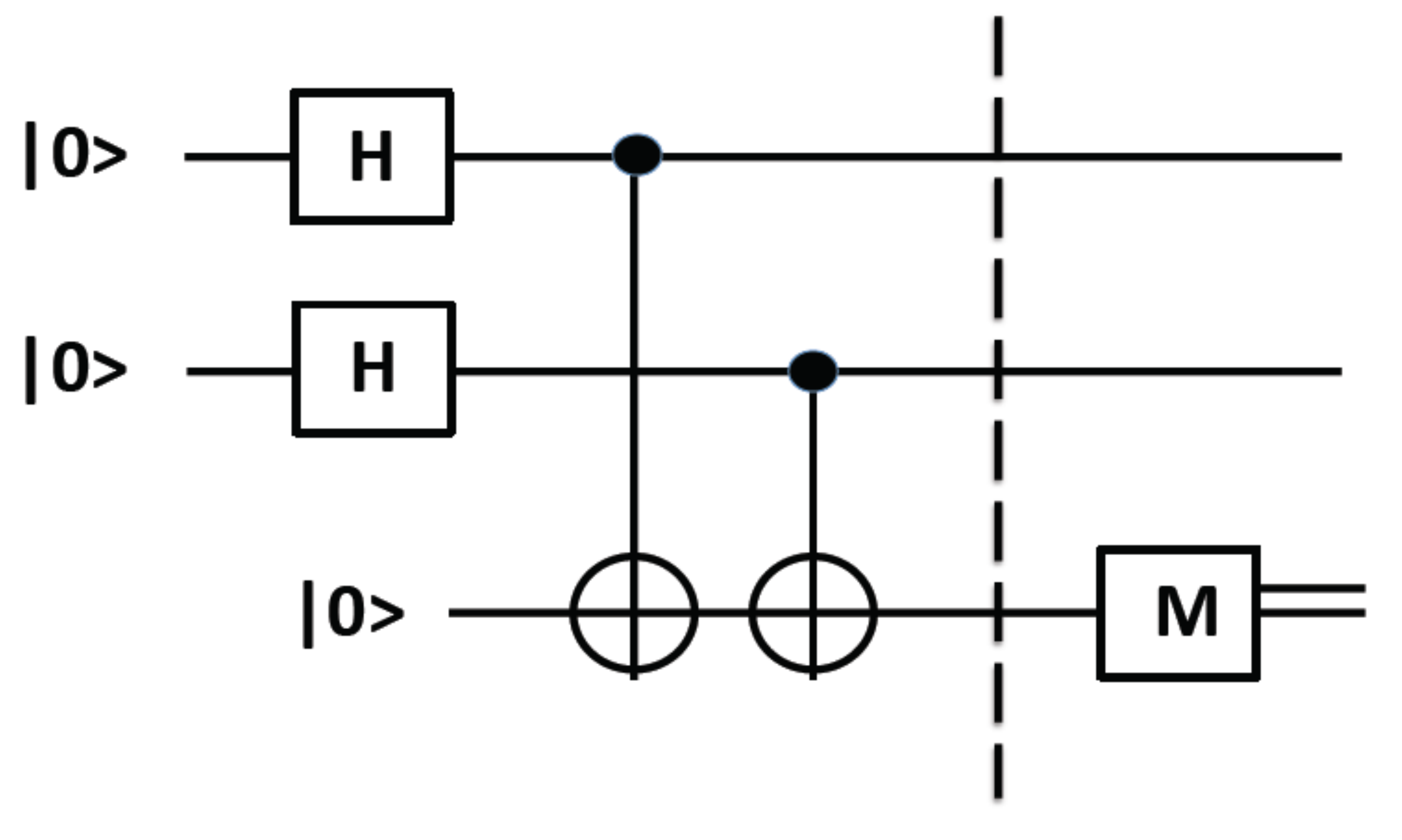}
\caption
{\small Entanglement by parity measurement. The state at the dashed line is given by $ [\ket{00} + \ket{11}] \ket{0} + [\ket{01} + \ket{10}] \ket{1}$. Projective measurement of the ancilla collapses the state into one of the Bell states \cite{Saira2014}. The information about the state of the ancilla can be used for deterministic entanglement.
 }
\label{parity}
\end{figure}

Measurement provides an important way to prepare quantum states. 
Figure~\ref{parity} shows a way to prepare entangled states by parity measurement \cite{Saira2014}. 
The Hadamard gates generate the two-qubit product state $ \ket{00} + \ket{01} + \ket{10} +\ket{11}$. Adding the ancilla, the three-qubit state becomes  $ \ket{000} + \ket{010} + \ket{100} +\ket{110}$. After applying the CNOT gates, the state (at the dashed line) is given by $ \ket{000} + \ket{011} + \ket{101} +\ket{110} = [\ket{00} + \ket{11}] \ket{0} + [\ket{01} + \ket{10}] \ket{1}$, representing a sum of Bell pairs with opposite parities. Measurement of the state (parity) of the ancilla collapses the state into one of the Bell states. 
The entanglement process can be made deterministic by feedback control, inducing a bit flip on demand if the unwanted parity is detected, as demonstrated by Saira {\it et al.}  \cite{Saira2014}. 

\subsection {Teleportation}

\subsubsection {Teleportation of states}

\begin{figure}[h]
\center
\includegraphics[width=10cm]{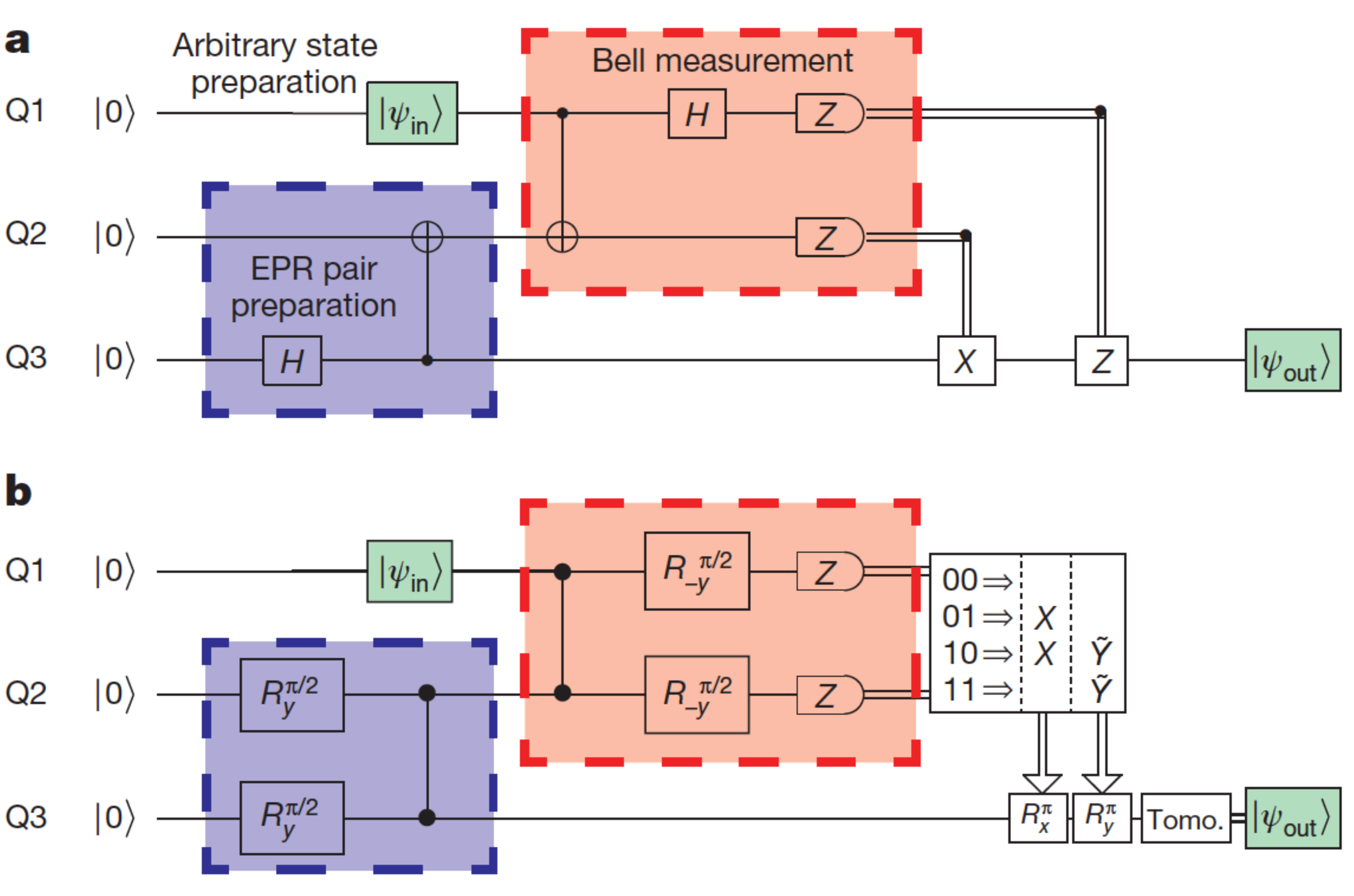} \;\;\;
\includegraphics[width=5cm]{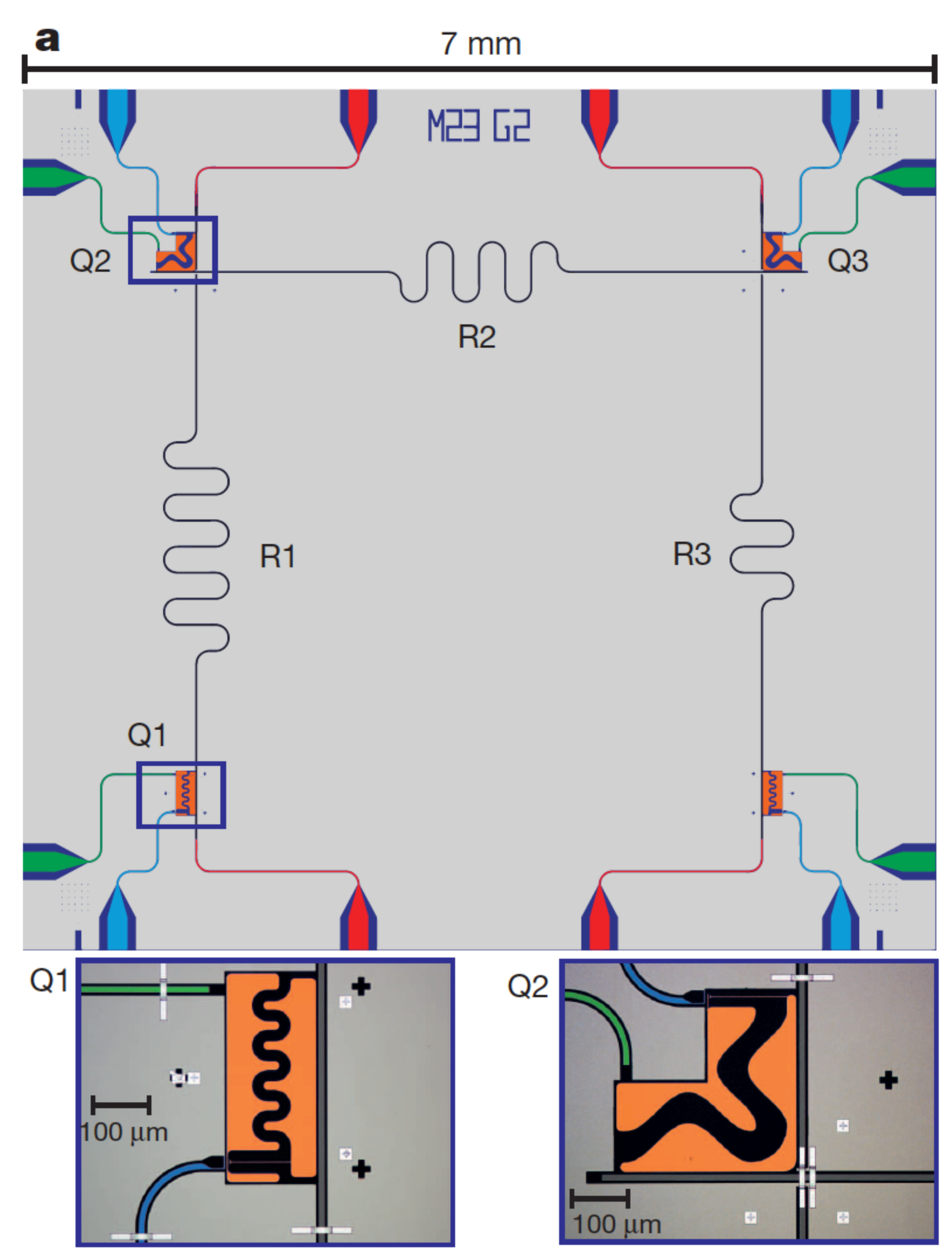}
\caption
{\small Quantum circuits for teleportation. Left: (a) Generic standard circuit. (b) The first application with superconducting 2D transmon-cQED and FPGA control of readout and feedforward gates \cite{Steffen2013}. Right: The hardware circuit used in the experiment. Adapted from \cite{Steffen2013}.
}
\label{ETHZ_teleport}
\end{figure}

Teleportation is a fundamental QIPC protocol - it allows quantum states to be reconstructed in distant places given a coherent quantum communication channel and additional classical channels for sending control information. Teleportation is a fundamental milestone that has to be passed by any competitive technology \cite{Riebe2004,Barrett2004}. 

The successful demonstration of teleportation with a transmon circuit (Fig.~\ref{ETHZ_teleport}) \cite{Steffen2013}  therefore represents a very important step for superconducting circuits, even though it "only" involved transfer over a distance of a millimeter between two qubits on the same chip, limited by the length of the microwave bus resonator. It remains to perform communication between distant qubits (cf. Ref.~\cite{Pfaff2014}), requiring long microwave transmission links, or optical links made possible by MW-optical interfaces.

\subsubsection {Teleportation of entanglement}
\label{entswap}
\begin{figure}[h]
\center
\includegraphics[width=9cm]{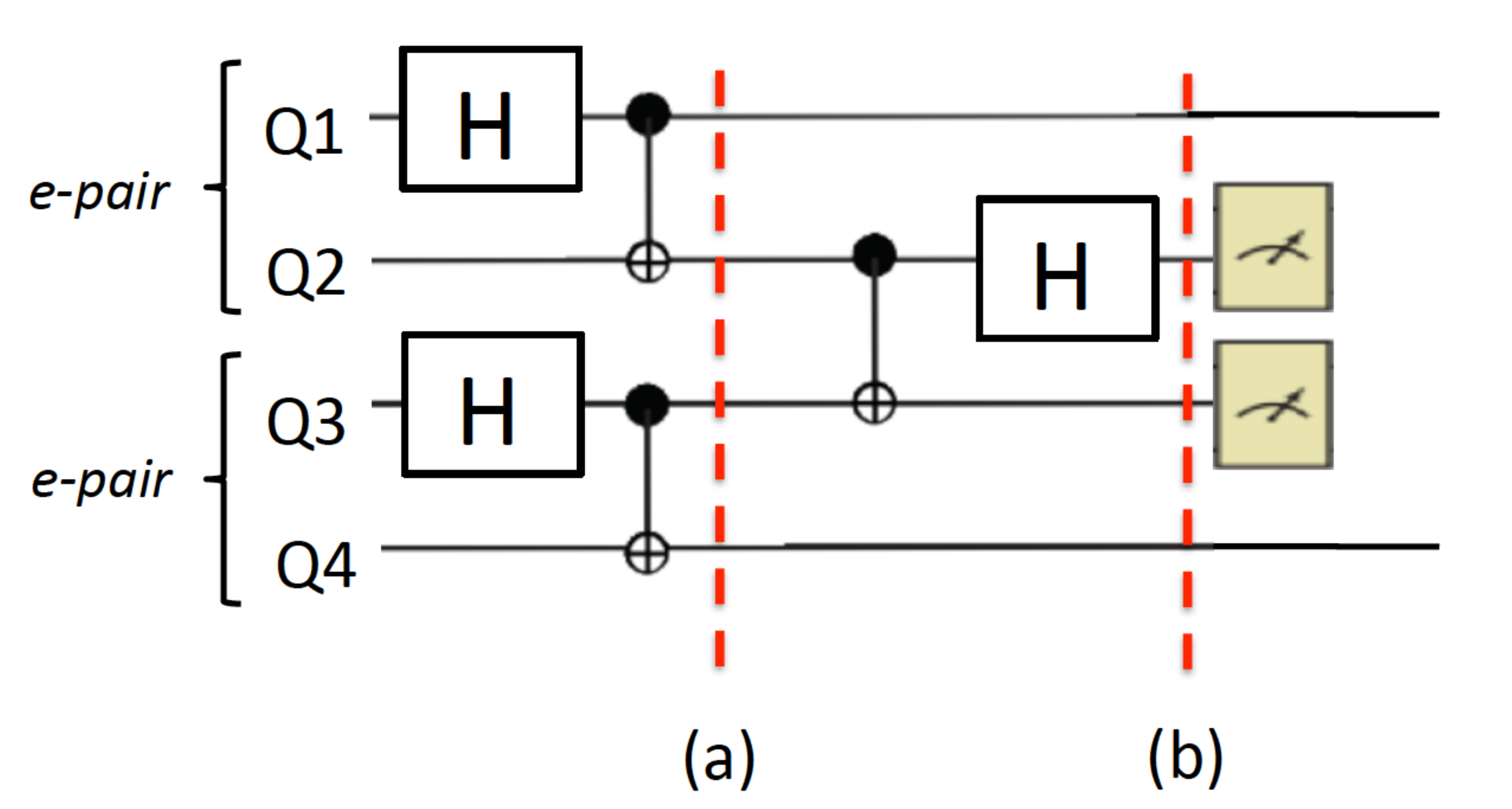}
\caption
{\small Entanglement swapping. The protocol first creates two independent entangled pairs, applying CNOT gates to (Q1,Q2) and (Q3,Q4), and then further applies a CNOT gate to entangle Q2 and Q3. This results in a special 4-qubit entangled state, and projective measurement of Q2 and Q3 provides classical information how to create an entangled pair (Q1,Q4) at a distant location, given that the states of qubits Q1 and Q4 are available.
}
\label{swap}
\end{figure}

This is an extension of the teleportation protocol - called entanglement swapping -  entangling two independent qubits that never interacted in the past \cite{Zukowski1993,Herbst2015}. 
The quantum circuit is shown in Fig.~\ref{swap}. 

As a simple example, consider the input state $\ket{0000}$. The two first Hadamard and CNOT gates then create a product of two Bell states at (a): $(\ket{00} +  \ket{11})(\ket{00} +  \ket{11}) = \ket{0000} + \ket{0011} +\ket{1100} +\ket{1111}$. Applying a CNOT gate to Q2, Q3 and a Hadamard gate to Q2 results in (b):  $ (\ket{0000} + \ket{1001}) + (\ket{0011} +\ket{1010}) + (\ket{0100} - \ket{1101}) + (\ket{0111} - \ket{1110})$. Performing a Bell measurement of Q2 and Q3 then projects the state with probability $0.25$ to one of the following entangled (Q1,Q4) Bell pairs: 
\begin{eqnarray}
\label{xx}
00 \rightarrow \ket{00} +  \ket{11}\\
01 \rightarrow \ket{01} +  \ket{10}\\
10 \rightarrow \ket{00} -  \ket{11}\\
11 \rightarrow \ket{01} -  \ket{10} 
\end{eqnarray}
The measurement entangles the remaining coherent qubits, and the classical 2-bit information tells exactly what Bell state was created. This at the core of a repeater protocol. Heinsoo {\it et al.} \cite{Heinsoo2016} have implemented the circuit in Fig.~\ref{swap} with a 4-transmon circuit (similar to Fig.~\ref{ETHZ_teleport}), measuring Q2, Q3  and identifying the four Bell states via quantum state tomography of Q1, Q4.

\subsection{Distillation of entanglement}

The purpose of entanglement distillation is  to extract a maximally entangled state from a collection of less entangled states. This can be used as an entanglement  resource,  e.g. for repeaters in quantum communication \cite{Waeldchen2016}. The distillation concept applies to both pure and mixed states. 

\begin{figure}[h]
\center
\includegraphics[width=10cm]{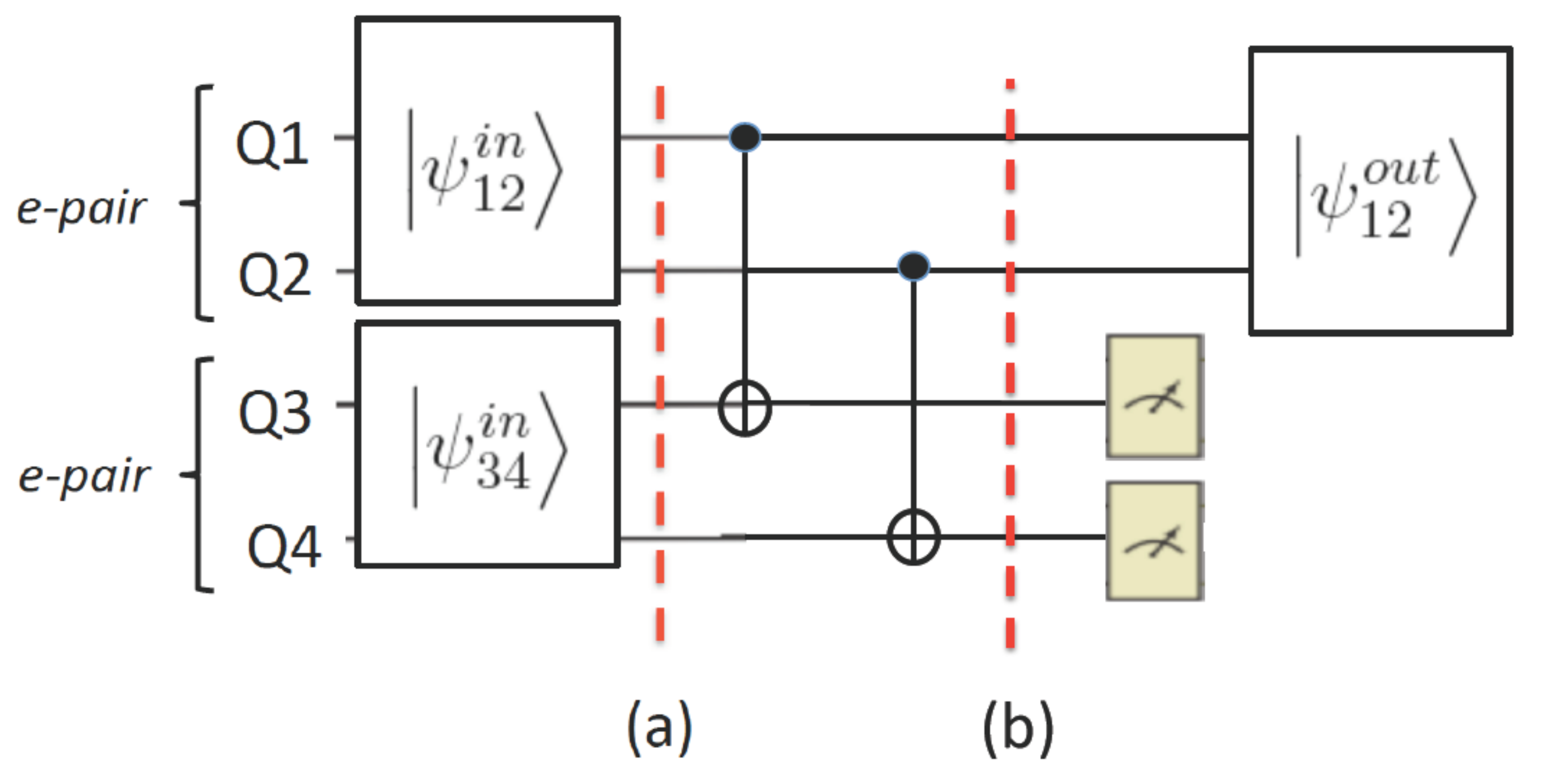}
\caption
{\small Distillation of entanglement of two partially entangled pairs, $\psi^{in}_{12}$  and $\psi^{in}_{34} $. The 4-qubit state at (a) is a product of two independent pairs. In contrast, the 4-qubit state $\psi_{1234}$ at (b) is not factorisable at all, in general. Measurement of Q3,Q4 projects  out a 2-qubit state, with two different outcomes: (Q3,Q4) = (0,0)  $\rightarrow \psi^{out1}_{12}$ and  (Q3,Q4) = (1,1)  $\rightarrow \psi^{out2}_{12}$. Both represent superpositions of Bell states. The instance of identical input pairs is very special, (0,0) producing a state with low concurrence $C_1 < C_{in}$, and (1,1) producing a Bell state state with $C_2 = 1$.
}
\label{distill}
\end{figure}

A general two-qubit state can be written as $\psi = a_0 \ket{00} + a_1 \ket{01} + a_2 \ket{10} + a_3 \ket{11}$. A measure of entanglement is the Concurrence: $C = 2 |a_0 a_3 - a_1 a_2| $. For a product state, $C=0$, which can be achieved in a number of ways. For a Bell state, on the other hand, $C=1$.

An example of a distillation quantum circuit is shown in Fig.~\ref{distill}. 
Two partially entangled and independent pairs of input states $\psi_{12}$ and $\psi_{34}$:
\begin{eqnarray}
\label{xx}
\ket{\psi^{in}_{12}} = [ \cos(\epsilon_{12}\frac{\pi}{4}) \ket{00} + \sin(\epsilon_{12}\frac{\pi}{4})  \ket{11} ]; \;\;\;C=\sin(\epsilon_{12}\frac{\pi}{2})\\
\ket{\psi^{in}_{34}} =  [\cos(\epsilon_{34}\frac{\pi}{4}) \ket{00} + \sin(\epsilon_{34}\frac{\pi}{4})  \ket{11} ]; \;\;\;C=\sin(\epsilon_{34}\frac{\pi}{2}) 
\end{eqnarray}
are entangled by two CNOT gates, creating a 4-qubit entangled state $\psi_{1234}$ at (b). Introducing $\epsilon_{12}+\epsilon_{34} = 2\epsilon$ and $\epsilon_{12}-\epsilon_{34} = \Delta\epsilon$ one obtains
\begin{eqnarray}
\label{xx}
\ket{\psi_{1234}} =  [\cos(\Delta\epsilon \frac{\pi}{4}) + \cos(\epsilon \frac{\pi}{2})] \ket{0000} + [\sin(\epsilon \frac{\pi}{2}) - \sin(\Delta\epsilon \frac{\pi}{4})] \ket{0011} \nonumber   \\
+ [\sin(\epsilon \frac{\pi}{2}) + \sin(\Delta\epsilon \frac{\pi}{4})] \ket{1111} +   [\cos(\Delta\epsilon \frac{\pi}{4}) - \cos(\epsilon \frac{\pi}{2})] \ket{1100}
\end{eqnarray}

Performing a Bell measurement of Q3 and Q4 results in (0,0) or (1,1) outcomes with equal probability,  and projects the 4q-qubit state to one of two different entangled 2-qubit (Q1,Q2) Bell pairs (not normalised):  
\begin{eqnarray}
\label{xx}
(0,0)  \rightarrow \ket{\psi^{out1}_{12}} \nonumber   \\
=  [\cos(\Delta\epsilon \frac{\pi}{4}) + \cos(\epsilon \frac{\pi}{2})] \ket{00} +  [\cos(\Delta\epsilon \frac{\pi}{4}) - \cos(\epsilon \frac{\pi}{2})] \ket{11} \\
C_1 = |[ \cos^2(\Delta\epsilon \frac{\pi}{4}) -  \cos^2(\epsilon \frac{\pi}{2})] / [\cos^2(\Delta\epsilon \frac{\pi}{4}) +  \cos^2(\epsilon \frac{\pi}{2})]
\end{eqnarray}
\begin{eqnarray}
\label{xx}
(1,1)  \rightarrow \ket{\psi^{out2}_{12}}  \nonumber  \\   
=  [ \sin(\epsilon \frac{\pi}{2}) -\sin(\Delta\epsilon \frac{\pi}{4})] \ket{00} +  [\sin(\epsilon \frac{\pi}{2}) + \sin(\Delta\epsilon \frac{\pi}{4})] \ket{11} \\  
 C_2 = | [\sin^2(\epsilon \frac{\pi}{2}) - \sin^2(\Delta\epsilon \frac{\pi}{4})] / [\sin^2(\epsilon \frac{\pi}{2}) + \sin^2(\Delta\epsilon \frac{\pi}{4})] |
\end{eqnarray}

In the case of identical input states ($\Delta\epsilon = 0$),  investigated experimentally by Oppliger {\it et al.} \cite{Oppliger2016},
\begin{eqnarray}
\label{xx}
 C_1 = | \sin^2(\epsilon \frac{\pi}{2}) / [1 +  \cos^2(\epsilon \frac{\pi}{2})] |  \\  
 C_2 =1, \;\;\; \epsilon \neq 0
 \end{eqnarray}
Of the two resulting entangled pairs, the (0,0) outcome produces a less entangled $\psi^{out1}_{12}$ output state than the input  $\psi_{12}$ state, while the (1,1) outcome produces a fully entangled $\psi^{out2}_{12}$ Bell state with probability $P_S = 0.5 \sin^2(\epsilon \frac{\pi}{2})$, as demonstrated  experimentally \cite{Oppliger2016}.

If the input states are not identical ($\Delta\epsilon \neq 0$), there is a new dimension, and one must evaluate the concurrence as a functions of $\epsilon_1$ and $\epsilon_2$, $C(\epsilon_1,\epsilon_2)$. \\

\newpage

\section{Quantum state protection}

\subsection{Quantum  control}

\subsubsection*{Quantum optimal control}   \cite{WhaleyMilburn2015} is essentially a question of controlling qubit driving and time evolution via pulse shaping. 
The quantum system to be controlled is modeled by unperturbed and control Hamiltonians $\hat H$ and $\hat H_{ci}$ with $u_i(t)$ the control fields (e.g. microwave driving fields) to be shaped:

\begin{equation}
\hat H  :=  \hat H +  \sum_{i} u_{i}(t) \hat H_{ci}
\end{equation}
Pulse shaping \cite{Motzoi2009,Chow2010,Gambetta2011} can be used to reduce single-qubit gate errors arising from the weak anharmonicity of transmon superconducting qubits \cite{Bianchetti2010,Chen2016,Vesterinen2014,Kelly2014a}.

Motzoi et al.  \cite{Motzoi2011} developed optimal control methods for rapidly time-varying Hamiltonians, in the form of a numerical method to find optimal control pulses that accou nt for the separation of timescales between the variation of the input control fields and the applied Hamiltonian. The  simulation of the quantum evolution is accurate on the timescale of the fast variation in the applied Hamiltonian.

Egger and Wilhelm \cite{Egger2014a} have recently developed adaptive hybrid optimal quantum control for imprecisely characterized systems (Ad-HOC). The method combines open- and closed-loop optimal control by first performing a gradient search towards a near-optimal control pulse and then an experimental fidelity estimation with a gradient-free method. For typical settings in solid-state quantum information processing, Ad-HOC enhances gate fidelities by an order of magnitude, making optimal control theory applicable and useful.

\subsection{Feedforward control}

Feedforward control means reading out information from a qubit, or group of qubits, and sending the classical information at a later time (forward) to a device that controls another group of (distant) qubits. 
Teleportation is a typical example of strong (projective) measurement and digital feedforward control (Fig.~\ref{ETHZ_teleport}) \cite{Steffen2013}.

\subsection{Feedback control}

Feedback control differs from feedforward control in that the measured classical control information is fed "back" to the same group of qubits via a classical feedback loop (of course still forward in time).

\subsubsection{Digital feedback control} 
involves strong (projective) measurement on the device to be controlled, classical processing  using fast electronics (FPGA), and finally communication of classical signals back to the device to operate digital quantum gates  \cite{Riste2015b}, e.g. to reset a qubit \cite{Riste2012a,Riste2012b}, or to create deterministic entanglement by parity measurement and feedback \cite{Riste2013a}. \\

\subsubsection{Analogue feedback control} 
involves weak (non-projective) measurement on the device to be controlled, classical processing using fast electronics (FPGA), and finally communication of classical signals back to the device to counteract the disturbance, e.g. in order to stabilise Rabi oscillations  \cite{RuskovKorotkov2002,Korotkov2005,Vijay2012} or  quantum trajectories \cite{Murch2013,Campagne-Ibarcq2013,Weber2014,deLange2014,Roch2014,Tan2015,Huang2013}.

\subsubsection{Measurement and back-action} 

 is a continuous process with speed controlled by the interaction strength between the system and the measurement device. This means that the "collapse of the wave function", i.e. the separation into distinguishable projections,  has a time scale, from weak (slow) to strong (fast) measurement \cite{Gillett2010,Katz2008}. There is a connection between information gain and back action, and the effects of back action can be undone under the right circumstances \cite{Tornberg2010,Tornberg2014,Kockum2012,Groen2013}.

Groen {\it et al.} \cite{Groen2013} performed a two-step indirect measurement of a transmon qubit with 
tunable measurement strength by partially entangling the qubit with an ancilla qubit (weak measurement), 
followed by a projective ancilla measurement using a dedicated resonator (strong, projective measurement). This revealed the back-action
of the measurement on the qubit as a function of qubit-ancilla interaction strength and ancilla measurement basis.
Nonclassical weak values were observed upon conditioning ancilla measurements on the
outcome of a projective measurement of the qubit. 

Monitoring a quantum state by weak measurements also makes it possible to "uncollapse" quantum states and achieve decoherence suppression by quantum measurement reversal \cite{Korotkov2010,Sherman2013,Keane2012,Zhong2014}

\subsection{Quantum networks and machine learning}

Artificial intelligence (AI) and machine learning (ML) represent frontlines in classical computer science. Quantum machine learning (QML)  \cite{Biamonte2016}
is now of great and increasing  interest in QIP \cite{Zahedinejad2016,Biamonte2016,Lamata2017,Alvarez-Rodriguez2016,Wiebe2015a,Wiebe2015b,Wiebe2015c,Amin2016}, in addition to classical ML optimisation processes of quantum hardware  \cite{Banchi2016} and software \cite{Haner2016b,CarleoTroyer2017}.
Machine learning is an adaptive dynamical data-driven process where a computer code is modified by training - learning how to provide an optimal solution to a given task according to certain criteria. The main processes are: supervised learning, unsupervised learning, and reinforcement learning. In all cases, the program code is modified in response to the training. 

In {\it supervised learning}, the machine adapts in response to sets of already classified (labeled) training data and feedback. 
In {\it unsupervised learning}, the machine develops the ability to classify unlabeled data without feedback, grouping data into different clusters (without necessarily "understanding" what the clusters represent).
In {\it reinforcement learning} the machine selects a series of actions depending on feedback (reward) from the environment.  To achieve an optimal transition from a given initial state to a desired final state the system needs feedback. Typically, the teacher only responds by telling the machine that its behaviour was inappropriate in terms of the distance in an appropriate fitness measure. 
The "machine" can be a piece of real hardware (e.g. a neural net with adjustable synaptic weights), or it can be an adaptive dynamical  program, or a combination of both. 

The topic is of central importance because AI and QML (or even  classical ML) might provide optimal solutions to the single most important issue in quantum computing and simulation - how to design a time evolution operator that evolves a system from A to B in an efficient manner, i.e. in polynomial time as a function of the size of the input? 
Effectively this might mean to avoid expressing the evolution operator  in series of 1q and 2q gates, or even multi-qubit gates, and instead find efficient paths through the high-dimensional Hilbert space from the initial to the final state of, e.g., a large molecule. Carleo and Troyer \cite{CarleoTroyer2017} recently showed that systematic reinforcement machine learning with a variational representation of the wave function  based on artificial neural networks (restricted Boltzmann machine, RBM) can reduce the complexity to a tractable computational form for some  interacting  quantum spins models in one and two dimensions.

This then puts focus on machine learning for optimal control and searching for optimal paths in complex energy landscapes. 
Recently, a quantum algorithm for solving a system of linear equations $Ax=b$  (HHL) \cite{HHL2009,Childs2009} has attracted great interest and been used for constructing algorithms for both supervised and unsupervised machine learning  \cite{Lloyd2013,Lloyd2014,Rebentrost2013}, e.g. to classify data or to identify patterns in large data sets ("Big Data"). 
This type of work is touching some core questions regarding speedup of quantum algorithms: the HHL algorithm in itself can achieve quantum speedup when allowed to work on suitably conditioned quantum data, but this does not  necessarily admit exponential speedup of finding solutions to given real-world problems - there are a number of caveats \cite{Aaronson2015}. 

Finally, Las Heras et al. \cite{LasHeras2016} have used a genetic algorithm (GA) to synthesize high-fidelity 2q gates by adding ancillas to increase the fidelity and optimise the resource requirements of digital quantum simulation protocols, while adapting naturally to the experimental constraints.
Moreover, Zhang and Kim \cite{ZhangKim2017,Carrasquilla2017} devised machine learning for a neural network to analyse topological phase transitions.

\subsection{Error correction codes and stabilisers}

Quantum error correction (QEC) presents one of the greatest experimental challenges. QEC is quite well developed theoretically,  but experimentally it is just at the beginning.
In this section we will describe a few simple examples of experimental applications, including pieces of the surface code on the way to the complete scheme \cite{Corcoles2015,IBM_Takita2016,IBM_Takita2017,Versluis2016,Kelly2015a}.

An error in a single qubit cannot be corrected - if an error changes the state of the qubit, a measurement of that state says nothing about the original state - the information is lost. Expanding the space of a qubit changes this situation because the information about the error can be stored for later correction - this is essentially the same as discussed earlier in terms of feedback undoing measurement back action.

Expanding the space means coding a qubit $ \alpha \ket{0} + \beta \ket{1} $ by representing it as a cluster of qubits - a logical qubit. 
For a given code, there are operators that commute with the code operators and have the same eigenstates - these are called stabilisers  (see e.g. \cite{Martinis2015}).
A measurement of a stabiliser operator then results in an eigenstate of the logical qubit, with no knowledge of the individual qubits.

A common measure of the size of the error is the Hamming distance, stating how many bits differ between the correct and corrupted codewords.  Classically the simplest form of error correction is redundancy at a level corresponding to the Hamming distance. If one expands $0$ and $1$ into bit-string code words $00$ and $11$ with distinct parity +1, defined by the bit sum modulo 2 checked by XOR gates, then a bit-flip error will lead to $01$ or $10$, and can be detected as a parity change to -1. Parity checks represent fundamental steps in both classical and quantum error detection schemes.

\begin{figure}[h]
\center
\includegraphics[width=8cm]{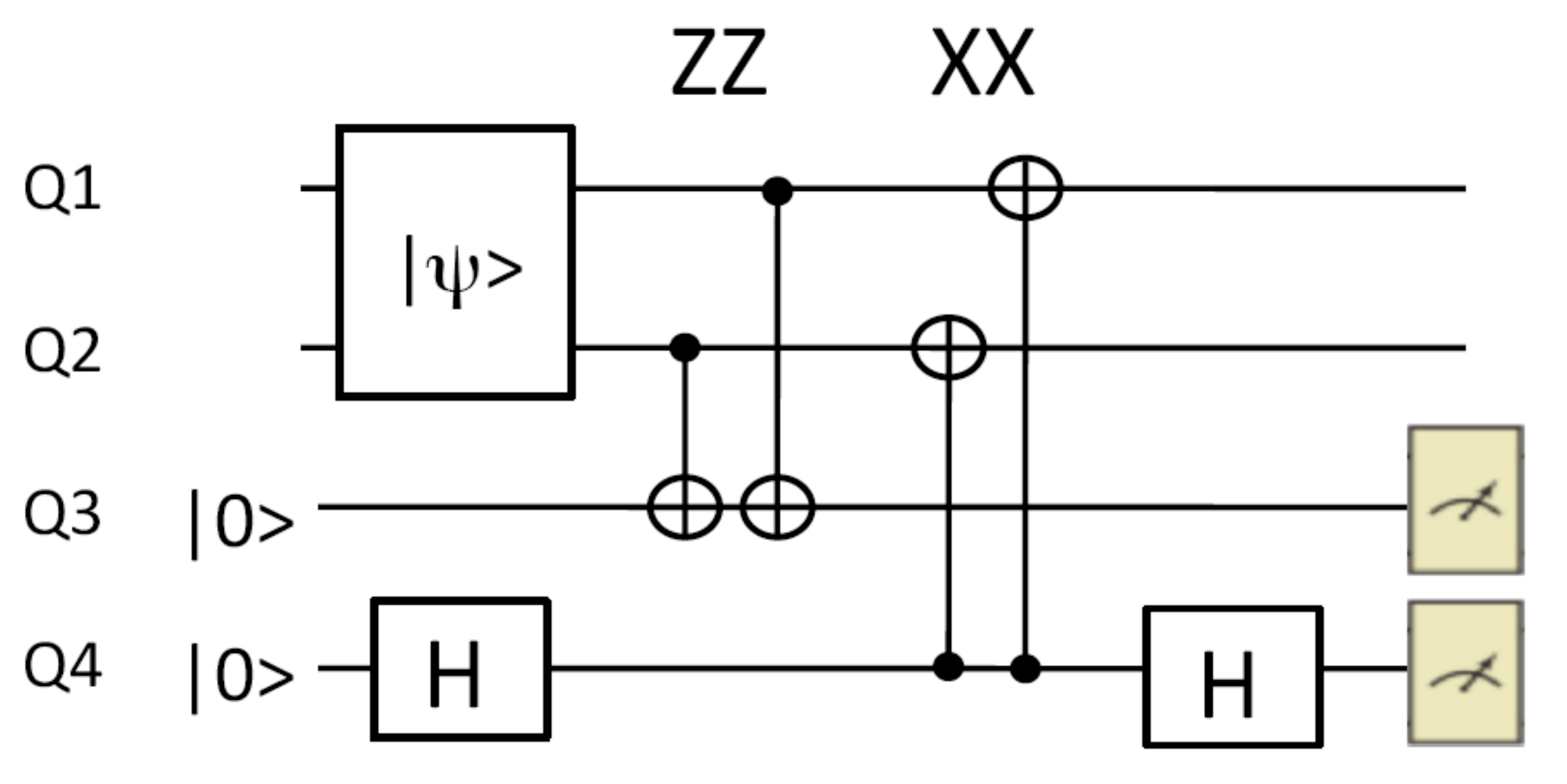}
\caption
{\small Quantum circuit for checking the ZZ (bit-flip) and XX (phase-flip) parities of a  2-qubit state $\ket{\psi}$. Since ZZ and XX commute, bit-flip and phase-flip parities can be checked independently. The Hadamard gates switch to the  $\ket{+}$ X-basis
 }
\label{parity2}
\end{figure}

In the quantum version in Fig.~\ref{parity2}, the qubit $\ket{\psi} = \alpha \ket{0} + \beta \ket{1}$ is coded as $ \alpha \ket{00} + \beta \ket{11}$. One then adds a third and a fourth qubit for checking and storing the ZZ and XX parities of the codeword. The parity checks are then performed via CNOT (XOR) gates between each of the ancillas and the two qubits. 

As a specific example, consider checking for bit-flips with the ZZ $(\sigma_{z1}\sigma_{z2})$ operator in the $\ket{0}, \ket{1}$ basis. The qubits+ancilla state is then $ (\alpha \ket{00} + \beta \ket{11})\ket{0} =  \alpha \ket{000} + \beta \ket{110} $. 
If there is no bit flip, one obtains:
$ CNOT_{23}CNOT_{13} (\alpha \ket{000} + \beta \ket{110}) = (\alpha \ket{00} + \beta \ket{11})\ket{0}$, where the ancilla stays in state $\ket{0}$, with parity +1.
If there is a bit flip, leading to $ \alpha \ket{01} + \beta \ket{10}$, the parity check results in: 
$ CNOT_{23}CNOT_{13} (\alpha \ket{010} + \beta \ket{100}) = (\alpha \ket{01} + \beta \ket{10})\ket{1}$, where the ancilla changes to state $\ket{1}$, with parity -1.  

As another specific example, consider checking for phase-flips with the XX $(\sigma_{x1}\sigma_{x2})$ operator  in the $\ket{+}, \ket{-}$ basis. Phase-flips are sign changes of the code word, $ \alpha \ket{00} - \beta \ket{11}$, and look like bit-flips along the X-axis, corresponding to $ \ket{+} \rightarrow  \ket{-}$ .
The qubits+ancilla state is then $ (\alpha \ket{00} + \beta \ket{11})\ket{+} =  \alpha \ket{00+} + \beta \ket{11+} $. 
If there is no phase flip, one obtains:
$ CNOT_{42}CNOT_{41} (\alpha \ket{00+} + \beta \ket{11+}) = (\alpha \ket{00} + \beta \ket{11})\ket{+}$, where the ancilla stays in state $\ket{+}$, with parity +1.
If there is a phase flip, leading to $ \alpha \ket{00} - \beta \ket{11}$, the parity check results in: 
$ CNOT_{42}CNOT_{41}(\alpha \ket{00+} - \beta \ket{11+}) = (\alpha \ket{00} - \beta \ket{11})\ket{-}$, where the ancilla changes to state $\ket{-}$, with parity -1. 

Since ZZ and XX commute, $[ZZ,XX] = 0$, bit-flip and phase-flip parities can be checked (and corrected) independently and the ZZ and XX parity eigenvalues characterise the state, (ZZ,XX): (+,+), (+,-), (-,+) and (-,-).

\subsection{Three qubit code}

To obtain a correctable code, classically the simplest case is to make use of redundancy via code words with three bits: $000$ and $111$, in which case a bit flip can be corrected by a majority vote: $010 \rightarrow 000$, etc.
Quantum mechanically the corresponding protected logic qubit becomes:
\begin{equation}
\ket{\psi} = \alpha \ket{0} + \beta \ket{1}  \rightarrow   \alpha \ket{000} + \beta \ket{111} 
\end{equation} 
Experimental implementation of this coding has been done in ion traps \cite{Schindler2011,Schindler2013,Chiaverini2004}, NV centra \cite{Waldherr2014} and transmon circuits \cite{Riste2015,Reed2012}

\begin{figure}[h]
\center
\includegraphics[width=10cm]{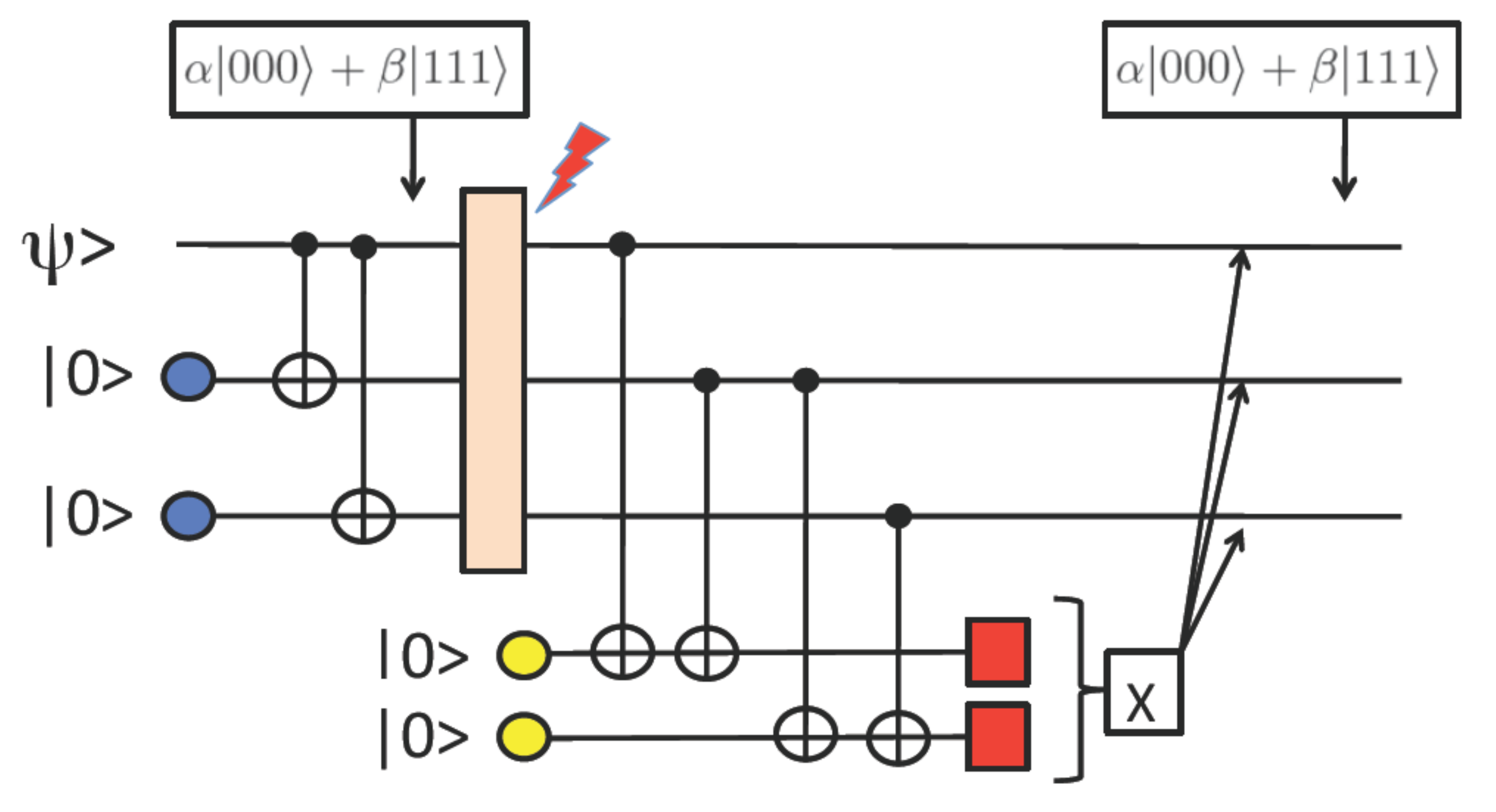}
\caption
{\small Standard bit-flip QEC with two ancillas for detecting and storing the error syndrome, and correcting the error. In this way the code words are not perturbed by measurement and recoding is not needed. The figure illustrates explicit error detection followed by feedforward control using e.g. FPGA electronics. The correction can also be implemented via Toffoli gates.
In the case of phase flips one uses Hadamard gates to transform phase flips into bit flips, checks the parity, and then transforms back.
}
\label{QEC_5q}
\end{figure}

A systematic scheme to correct bit flips in any of the three qubits requires separate ancillas for syndrome measurement, storage and correction, as illustrated in Fig.~\ref{QEC_5q}. This involves precisely the bit-flip ZZ parity-check procedure described in Fig.~\ref{parity2}. 

Defining $\hat U_1 = CNOT_{q1,a1} CNOT_{q2,a1} $ and  $\hat U_2 = CNOT_{q2a2} CNOT_{q3a2} $, we want to calculate $\hat U_2 \hat U_1 (\alpha \ket{000} + \beta \ket{111})\ket{00}$, and characterise the effects of bit flips via the ancilla syndrome. 

Zero bit-flip error:
\begin{eqnarray}
\hat U_2 \hat U_1 (\alpha \ket{000} + \beta \ket{111})\ket{00}  =  (\alpha \ket{000} + \beta \ket{111})\ket{00} 
\end{eqnarray}
On measuring (00), do nothing.

Bit-flip error in qubit 1:
\begin{eqnarray}
\hat U_2 \hat U_1 (\alpha \ket{100} + \beta \ket{011})\ket{00}  =  (\alpha \ket{100} + \beta \ket{011})\ket{10} 
\end{eqnarray}
On measuring (10), X-flip qubit 1.

Bit-flip error in qubit 2:
\begin{eqnarray}
\hat U_2 \hat U_1 (\alpha \ket{010} + \beta \ket{101})\ket{00}  =  (\alpha \ket{010} + \beta \ket{101})\ket{11} 
\end{eqnarray}
On measuring (11), X-flip qubit 2.

Bit-flip error in qubit 3:
\begin{eqnarray}
\hat U_2 \hat U_1 (\alpha \ket{001} + \beta \ket{110})\ket{00}  =  (\alpha \ket{001} + \beta \ket{110})\ket{01} 
\end{eqnarray}
On measuring (01), X-flip qubit 3.

There are typically three ways to apply the correction:\\
1. By feed-forward application of X gates via fast electronics, flipping the faulty qubit; \\
2. By automatic correction via a set of Toffoli (CCNOT, controlled X) gates with suitable truth tables, flipping the faulty qubit. 
This corrects and restores the original coded qubit - there is then no need for re-coding, only the ancillas have to be reset. \\
3. To store the errors in classical memory, and correct at the end.

A single round of QEC for 3 qubits was implemented by Reed et al. \cite{Reed2012}  in a 3 transmon qubit circuit without ancillas, correcting for single bit- or  phase-flips using a Toffoli gate. 
Recently the scheme in Fig.~\ref{QEC_5q} was implemented by Riste et al. \cite{Riste2015}, detecting bit-flip errors in a logical qubit using stabiliser measurements. 
The experiment uses 3 qubits for encoding, 2 ancilla qubits for the syndrome, and fast feed-forward control for correction (Fig.~\ref{QEC_5q}).

Moreover, using their 9-qubit 1D chain, Kelly et al. \cite{Kelly2015a} recently implemented the 3q repetition code with 3 QEC cycles, and extended the work to a 5q code. This work represents first steps toward the 2D-surface code, and will be discussed in some detail below.

Fast electronics makes it possible to perform qubit calibration during repetitive error detection \cite{Kelly2016a}. Moreover, one does not have to apply corrections when errors are detected - it is enough to store the information about errors in classical memory and correct at the end \cite{Fowler2014,Fowler2014b}.

\subsection{Surface codes}

The surface (toric) QEC code was invented by Kitaev \cite{Kitaev1997,Kitaev2006} and is now at the focus of intense development and experimental implementations \cite{IBM_Takita2016,IBM_Takita2017,Versluis2016,Fowler2014,Fowler2014b,Fowler2012,Tomita2014,Terhal2015,Gosh2015}.

\subsubsection{Basic concepts and models}

\begin{figure}[h]
\center
\includegraphics[width=15cm]{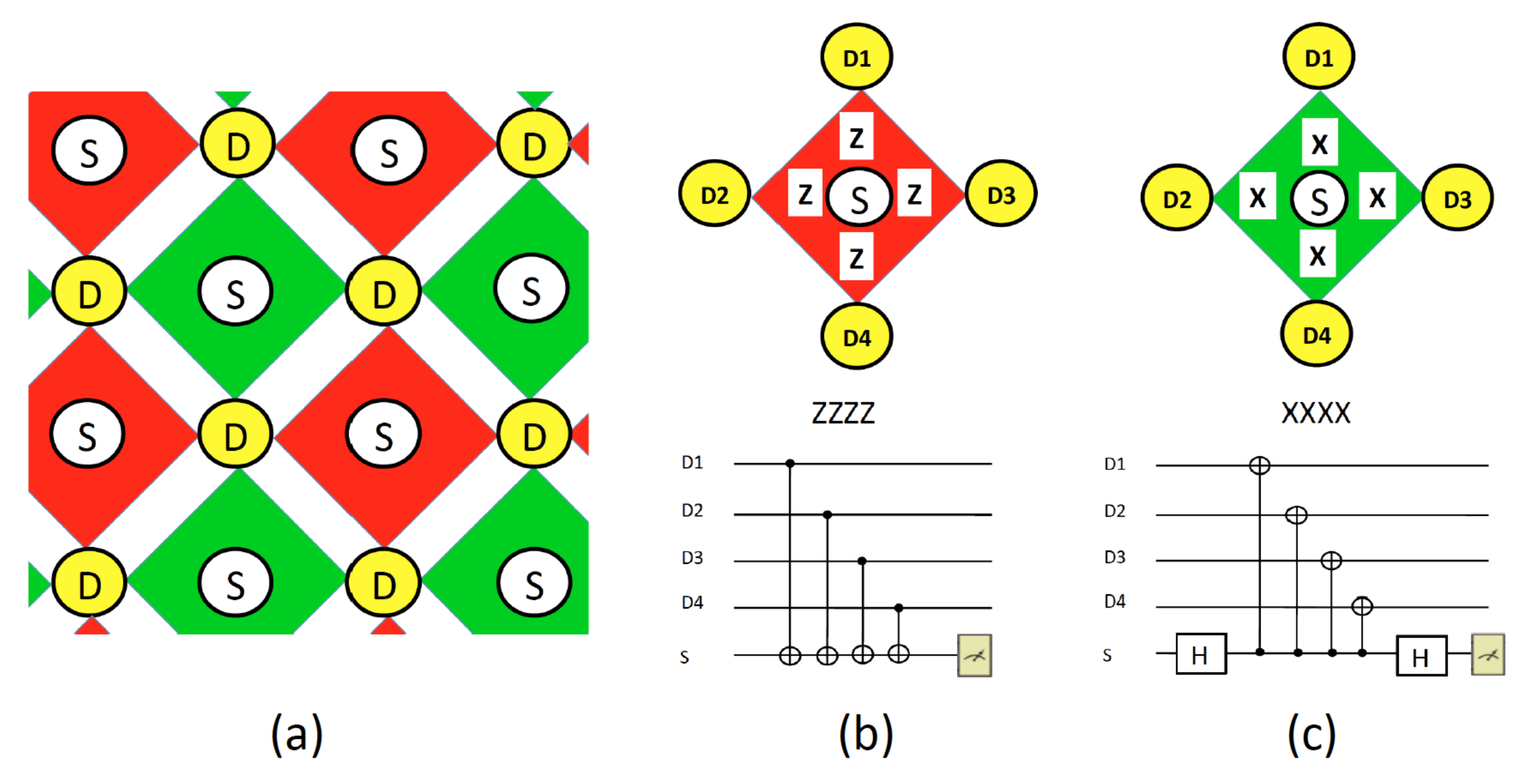}
\caption
{\small (a) Surface code 4 x 4 qubit array layout. (b) Basic 5 qubit plaquette with 4 data qubits (D) and a central ancilla (syndrome, S) qubit for measuring ZZZZ (bit-flip) parities. (c) Basic 5 qubit plaquette with 4 data qubits (D)  and a central ancilla (syndrome, S) qubit for measuring XXXX (phase-flip) parities.
Note the measuring order D1-D2-D3-D4 forming an S-like trace. 
}
\label{Surfcode}
\end{figure}

The surface code is connected with a specific geometrical arrangement (architecture) of qubits (Fig.~\ref{Surfcode}): 4 data qubits (D) at the corners of a square with a syndrome ancilla qubit (S) at the centre.
The central task is to perform 4-qubit parity measurements on the data qubits (D) using ZZZZ and XXXX stabiliser operators and to register the measured (classical) ancilla syndrome (S) eigenvalue, showing whether there has been a bit-flip  $\ket{0} \rightarrow  \ket{1}$ or a phase-flip $ \ket{+} \rightarrow  \ket{-}$ in the 4q data cluster.

\subsubsection{4-qubit parity measurements on a surface code plaquette}

IBM has performed 4-qubit parity measurements that demonstrate the detection of arbitrary single-qubit quantum errors on an effective 4-qubit plaquette (Fig.~\ref{Surfcode}b,c) designed for the surface code using all-microwave control  \cite{IBM_Takita2016,IBM_Takita2017}. In particular, the qubit-qubit couplings are dynamically driven via the CR technique, and the pulse schemes include echo sequences to remove non-ideal-gate errors. Gates were characterised by randomised benchmarking (RB) and correct ZZZZ and XXXX parity assignments were obtained with 0.75-0.80 probability.

\subsubsection{17-qubit design for the surface code}

TU Delft is presently developing and testing a scalable architecture \cite{Versluis2016} for executing the error-correction cycle of a monolithic surface-code fabric (Surface-17 \cite{Tomita2014}), composed of fast-flux-tunable transmon qubits with nearest-neighbor coupling. The scheme combines four key concepts \cite{Versluis2016}: (i) an eight-qubit unit cell as the basis for repetition of quantum hardware and control signals (cutting out a section in Fig.~\ref{Surfcode}a); (ii) pipelining of X- and Z-type stabiliser measurements (Fig.~\ref{Surfcode}b,c); (iii) a fixed set of three frequencies for single-qubit control; and (iv) a fixed set of eight detuning sequences implementing the needed controlled-phase gates. 
The design couples nearest-neighbour data and ancilla transmons in a planar cQED architecture via bus resonators.
The eight-qubit unit cell can be repeated in 2D using vertical interconnects for all input and output signals from the control layers. Each transmon will
have a dedicated flux line allowing control of its transition frequency on nanosecond timescales, a dedicated microwave-drive line, and a dedicated dispersively-coupled resonator for readout.

\subsubsection{Multi 2-qubit parity measurements on a surface code 1D chain}

\begin{figure}[h]
\includegraphics[width=8.3cm]{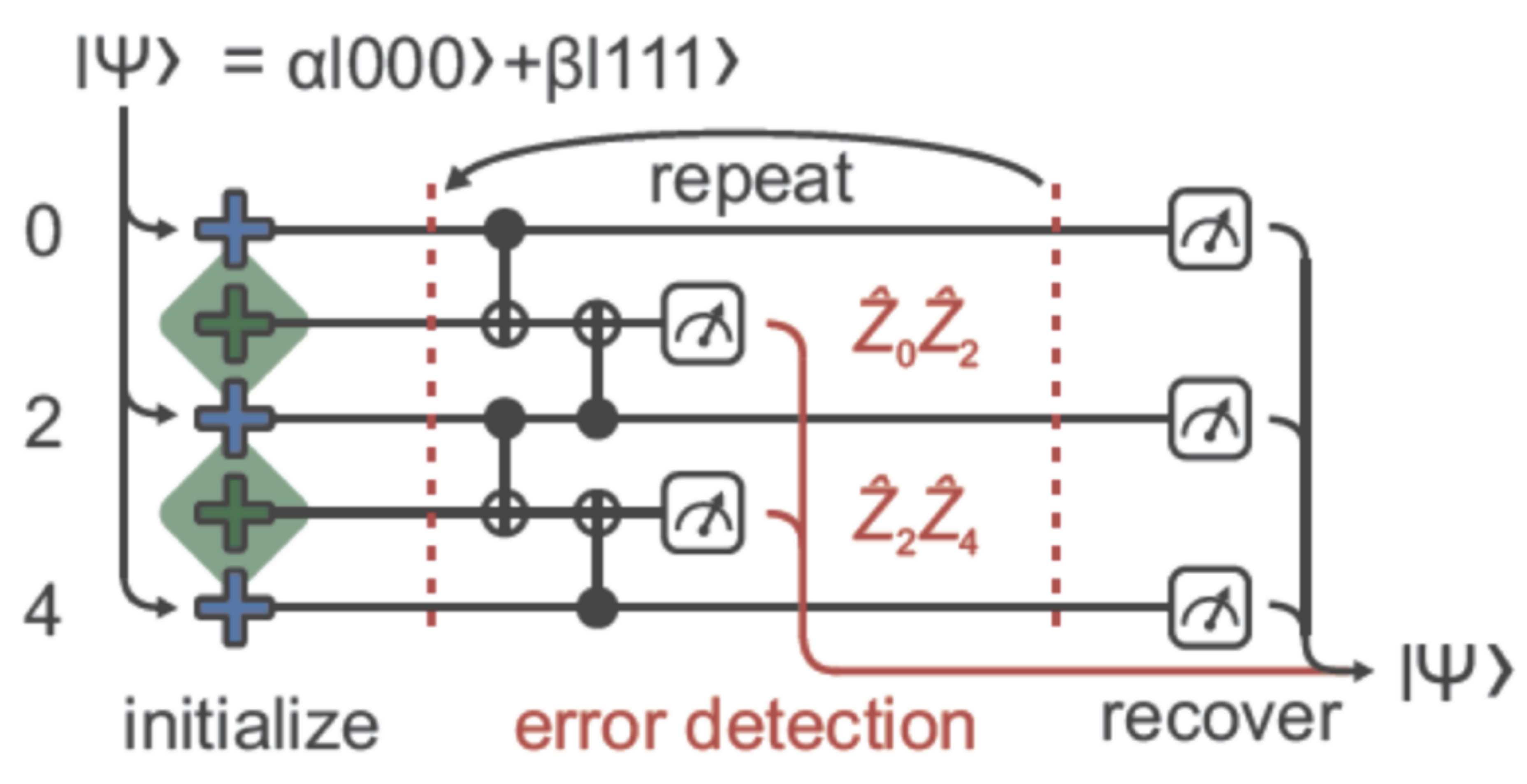} \;\;\;
\includegraphics[width=7.3cm]{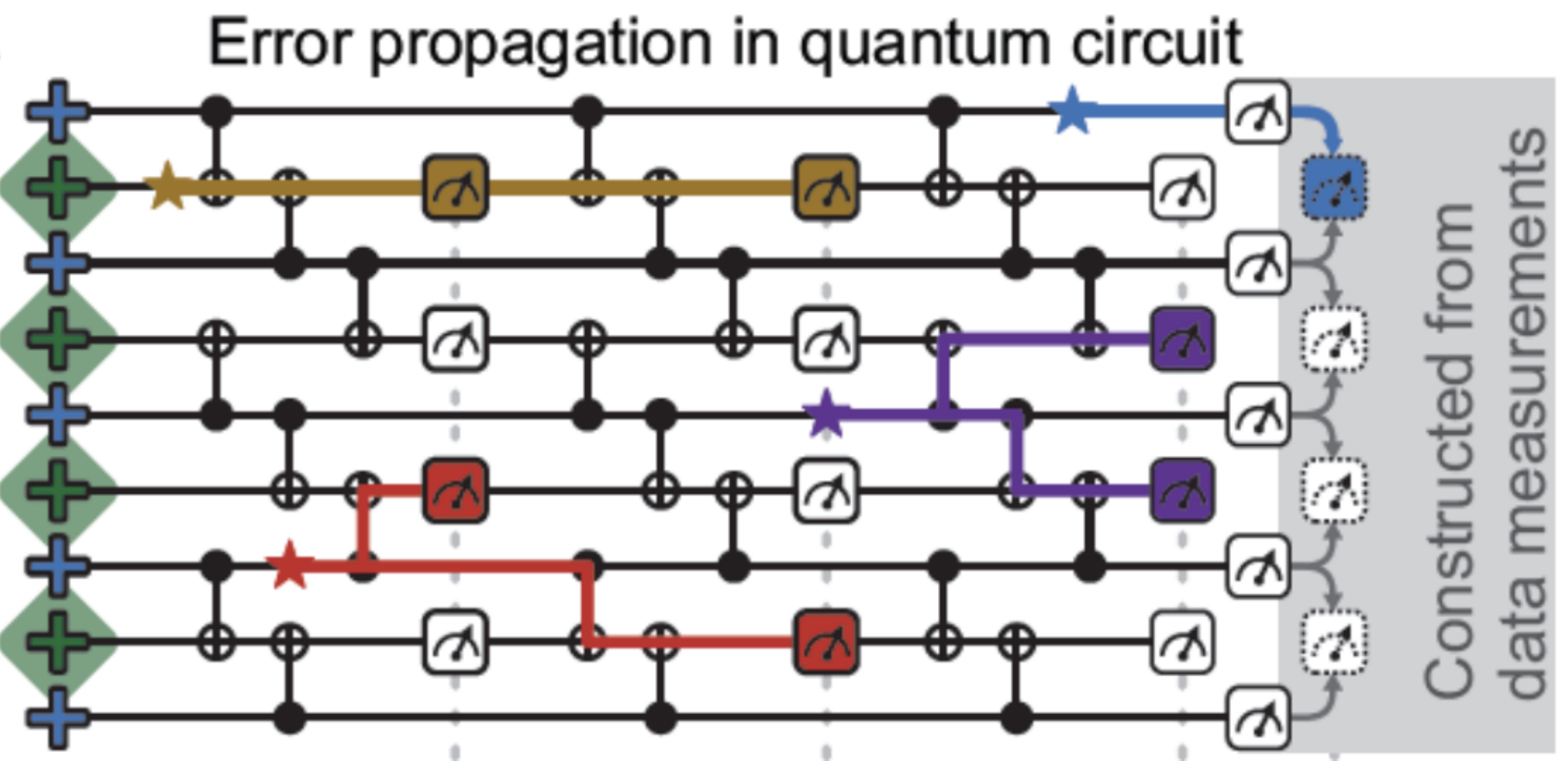}
\caption
{\small (a) Left: 3q repetition code $\alpha \ket{000} +  \beta \ket{111}$: algorithm (same circuit as in Fig.~\ref{QEC_5q}. (b) Right: 5q repetition code $\alpha \ket{00000} +  \beta \ket{11111}$: error propagation and identification. Adapted from \cite{Kelly2015a}.
 }
\label{UCSB_9q_QEC}
\end{figure}

Martinis' group has performed a set of surface-code type of experiments on a linear 1D chain with 9 qubits \cite{Kelly2015a,Kelly2016a} (Fig.~\ref{UCSB_9q_QEC}), tracking errors as they occur by repeatedly performing projective quantum non-demolition parity measurements. This is a first step toward the 2D surface code scheme (Fig.~\ref{Surfcode}), measuring ZZ (bit-flip) parities (Figs.~\ref{parity2}, \ref{QEC_5q}).

The first experiment \cite{Kelly2015a} used 5 qubits (Fig.~\ref{UCSB_9q_QEC}a) with 3 data qubits and 2 measurement ancillas to implement the 3q repetition code (same circuit as in Fig.~\ref{QEC_5q}). The repetition code algorithm uses repeated entangling and measurement operations
to detect bit-flips using the ZZ parity.  The initial state can be reproduced by removing physical errors in software based on qubit measurements during the execution the repetition code.

Figure~\ref{UCSB_9q_QEC}b  shows a 9 qubit quantum circuit with 5 data qubits and 4 measurement ancillas,
for three cycles of the repetition code, with examples of errors \cite{Kelly2015a}. Errors propagate
horizontally in time, and vertically through entangling gates. Different errors
lead to different detection patterns: 
an error on a measurement qubit is detected in two subsequent rounds. 
Data qubit errors are detected on neighbouring measurement qubits in the same or next cycle. 
Data errors after the last round  are detected by constructing the final set
of ZZ eigenvalues from the data qubit measurements. 

To study the ability to preserve quantum states, the data qubits were initialised into a GHZ state, and two rounds of the repetition code were applied. The result shows that the one-dimensional repetition code algorithm allows for preserving the quantum state in the case of no errors, and correcting bit-flip errors otherwise, purely through error detection and classical post-processing. This is similar to the full surface code, avoiding the need for dynamic feedback with quantum gates. 

Relative to a single physical qubit,  the
failure rate in retrieving an input state was reduced by a factor of 2.7 when using
five qubits, and by a factor of 8.5 when using all nine qubits after eight cycles.  Moreover, the preservation of the GHZ state was verified tomographically.

\subsection{Architecture and error correction in 3D cQED}

Recent experimental results for the 3D transmon approach show great promise, and 3D architectures are in the process of being scaled up.

\subsubsection{cQED 3D architecture}

Architectures based on 2D resonator networks coupled by JJs have been proposed and developed by Houck {\it et al.} \cite{Houck2012}. However, the superior coherence times of 3D resonators have recently been exploited by the Yale group to design a 3D architecture \cite{Mirrahimi2014,Brecht2016} which  encodes quantum information in the Hilbert space of a 3D array of 3D superconducting cavities and employs the Josephson junction only to perform operations on states of the cavities. The approach promises to be hardware efficient since there is only one dominant error mechanism - single photon loss of the cavity - while the room in Hilbert space is large enough to accommodate particularly easy-to-correct sub-manifolds. The idea \cite{Brecht2016} is to control the qubit-cavity components well enough that they may serve as logical qubits in an architecture based on multilayer microwave integrated quantum circuit (MMIQC) technology.  (MMIC is already a well-established classical microwave technology).

\subsubsection{Engineering cat-states}

Specially engineered interaction with the environment can become a resource for the generation and
protection of coherent superpositions of multiple, stable, steady states.  Leghtas {\it et al.} \cite{Leghtas2015} have confined the state of a superconducting resonator to the quantum manifold spanned by two coherent states of opposite phases and have observed a Schr\"odinger cat state
spontaneously squeeze out of vacuum before decaying into a classical mixture. This experiment
points toward robustly encoding quantum information in multi-dimensional steady-state manifolds. 

Moreover, Wang {\it et al.} \cite{Wang2016} have realized a two-mode cat state of electromagnetic fields in two
microwave cavities bridged by a superconducting artificial atom, which can also be
viewed as an entangled pair of single-cavity cat states. The work presents full quantum state
tomography of this complex cat state over a Hilbert space exceeding 100 dimensions
via quantum nondemolition measurements of the joint photon number parity. This  paves the way for logical
operations between redundantly encoded qubits for fault-tolerant quantum
computation and communication. 

\subsubsection{Engineering QEC}

The break-even point of QEC is when the lifetime of a qubit exceeds the lifetime of the constituents of the
system.  Yale \cite{Ofek2016} has demonstrated a QEC system that reaches the break-even point
by suppressing the natural errors due to energy loss for a qubit logically encoded in superpositions of Schr\"odinger-cat states of a superconducting resonator. Ofek {\it et al.}  \cite{Ofek2016} implement a full QEC protocol
by using real-time feedback to (i) encode, (ii) monitor naturally occurring errors, (iii) decode and (iv) correct. 
As measured by full quantum process tomography without post-selection, the corrected qubit lifetime is 320 $\mu s$, which is longer than the lifetime of any of the parts of the system. 

\subsubsection{Gates and operations}

Heeres {\it et al.} \cite{Heeres2016} recently demonstrated high-fidelity implementation of a universal set of gates on a qubit encoded into an oscillator using the cat code, targeting the creation and manipulation of a logical qubit encoded in an even-parity four-component cat subspace. Using optimal control techniques, they created a universal set of gates on this  logical qubit, including X and Y rotations by $\pi$ and  $\pi/2$, as well as Hadamard and T gates. This includes the creation of encoding and decoding pulses to transfer bits of quantum information between the transmon subspace and the encoded cat subspace.

A major step nevertheless remains - to perform entangling gate operations on two logical qubits to achieve a universal set of gates for quantum computing.
As discussed in Sect.~\ref{entswap}, entangling two remote quantum systems that never interact directly is an essential building block in quantum information science - it forms a basis for modular architectures of quantum computing.
For protocols relying on using traveling single-photon states as flying qubits, needed for the 3D architecture considered by Yale, efficiently detecting single photons is challenging because of the low energy of microwave quanta. 
Entangling distant qubits by photon detection is a well-established technique in quantum optics.
Narla {\it et al.}  \cite{Narla2016} now report the realisation of a robust form of concurrent remote entanglement based on a novel microwave photon detector implemented in their cQED platform.  This may open the way for a modular 3D-architecture of QIP with superconducting qubits.


\newpage

\section{Quantum simulation of many-body systems}
\label{Qsim}

Quantum systems need quantum systems for efficient simulation \cite{Feynman1982}. Simulation of a large quantum system, say a molecule of medical interest, may be intractable on a classical digital computer due to lack of time and memory, but tractable on quantum computers and simulators in order to achieve necessary accuracy.

Ordinary classical digital computers are basically used for number crunching: encoding and running algorithms that process numbers for various purposes, like numerically solving equations, performing search, classifying data, and optimising approximate solutions. 
Classical digital computers are based on networks of logic gates and memory. The concept of digital quantum computation (QC) and simulation (QS) is similar: it involves circuit models with quantum gates to input, process and output digital quantum information \cite{Brown2010,Georgescu2014,NielsenChuang2010,Lloyd1996,AbramsLloyd1997,AbramsLloyd1999,Aspuru-Guzik2005,Whitfield2011,Bravyi2017,Muller-Blume-Kohout2015}. Both digital QC and QS map mathematical problems onto a quantum representation (Hilbert space), devise sequences of gate operations, and use superposition and entanglement to compute and to achieve speedup.

In contrast, analogue QC and QS are not based on quantum gates, but on direct construction of the physical system Hamiltonian in hardware (HW). There are a number of ways to emulate interesting quantum Hamiltonians in quantum HW. In adiabatic QC (AQC) \cite{Farhi2000,Farhi2001}  one adiabatically follows the development of the ground state when a perturbation is slowly switched on, switching from an initial model Hamiltonian to a final one, implementing the transition to the desired interacting many-body system. 

Finally, quantum annealing (QA) \cite{Boixo2014} is related to AQC in that one emulates a Hamiltonian in hardware. The difference is that one heats up and then cools the system, following its path toward the ground state via classical thermal and quantum tunneling transitions.

\subsection{Basics of quantum simulation}

Basically, a quantum simulator solves the time-dependent Schr\"odinger equation (TDSE) for a system described by a Hamiltonian $\hat H$,
\begin{equation}
i \hbar \frac{\partial}{\partial t}\ket{\psi(t)} = \hat{H}(t) \ket{\psi(t)}
\label{Seq}
\end{equation}
via propagation of an initial state $\ket{\psi(t_0)}$ using the time-evolution operator
\begin{equation}
\ket{\psi(t)} = \hat U(t,t_0) \ket{\psi(t_0)} = \hat T \;e^{-\frac{i}{\hbar} \int_{t_0}^{t} \hat{H}(t') dt'}  \ket{\psi(t_0)}.
\label{psi}
\end{equation}
The initial state $\ket{\psi{(t_0)}}$ represents an essential part of the problem. 
It can be a computational basis state or a superposition of configurations. If the configuration is not an eigenstate, the state will then evolve in time through state space and reflect the dynamics of the system Hamiltonian, and the Fourier spectrum will provide the energies of the eigenstates.
A systematic way to construct the initial state $\ket{\psi(t_0)}$ is to start from a reference state $\ket{\psi_{ref}}$ and add states representing excitations from the reference state:   
\begin{eqnarray}
\label{CI_ansatz}
\ket{\psi(t_0)} =  a_{ref}\ket{\psi_{ref}} + \\ \nonumber
 \;\;\;\;\;\;\;\;\;\; + \sum a_{ni}c_p^+c_q\ket{\psi_{ref}}  + \sum a_{mnji}c_p^+c_q^+c_rc_s\ket{\psi_{ref}} + ....
\label{psi}
\end{eqnarray}
Relation~(\ref{CI_ansatz}) represents a configuration interaction (CI) state including single $a_{ni} \;c_p^+c_q$ and double $a_{mnji} \;c_p^+c_q^+c_rc_s$  excitations, typically adding many-particle correlation effects to a mean-field initial state;  formal inclusion of all possible excitations defines the full CI (FCI) state. Simple approximations for  $\ket{\psi{(t_0)}}$ could be a product state, or a Hartree-Fock determinant.
Advanced approximations can be constructed via coupled-cluster (CC) \cite{CC} or matrix product (MPS) \cite{MPS} reference states. Also note the very recent work by Carleo and Troyer \cite{CarleoTroyer2017}.

 For a nice review and "hands-on" discussion of all the steps needed for simulating the time evolution  of a $H_2$ molecule and extracting the ground state energy, see Whitfield {\it et al.} \cite{Whitfield2011}.

\subsection{Trotterisation}

There are many different approaches that can be used to compute $e^{-i\hat Ht} \ket{\psi(0)}$, and many of them rely on Trotter decompositions involving discretisation of the time evolution \cite{AbramsLloyd1997,Aspuru-Guzik2005,Whitfield2011}. 

Let us for simplicity consider a time-independent Hamiltonian, and set $t_0=0$. (In the following we also set $\hbar = 1$, with energy measured in units of frequency).
\begin{equation}
\ket{\psi(t)} = \hat U(t,0) \ket{\psi(0)} = \;e^{-i \hat{H} t}  \ket{\psi(0)},
\label{psi}
\end{equation}
where
\begin{equation}
\hat H = \sum_{i=1,k} \hat H_i
\label{psi}
\end{equation}
Trotterisation (Lie-Trotter-Suzuki formula)
\begin{eqnarray}
e^{-i \hat{H}t} \approx [e^{-i (\hat{H_1}+\hat{H_2} + .....  + \hat{H_k})t/m}]^m \nonumber \\
 \approx [e^{-i\hat{H_1}t/m}e^{-i\hat{H_2}t/m} ....\; e^{-i\hat{H_k}t/m}]^m 
\label{psi}
\end{eqnarray}
makes it possible to express the time evolution operator as a sequence of operations of the individual terms $\hat H_i$ of the Hamiltonian. These operations can be gate operations in a quantum circuit model, or the application of classical control fields in an analog quantum system, or combinations thereof. In practice one uses higher order Trotter formulas to minimize the errors \cite{Aspuru-Guzik2005,Babbush2015}.

\subsection{Phase estimation}

\begin{figure}[h]
\includegraphics[width=7cm]{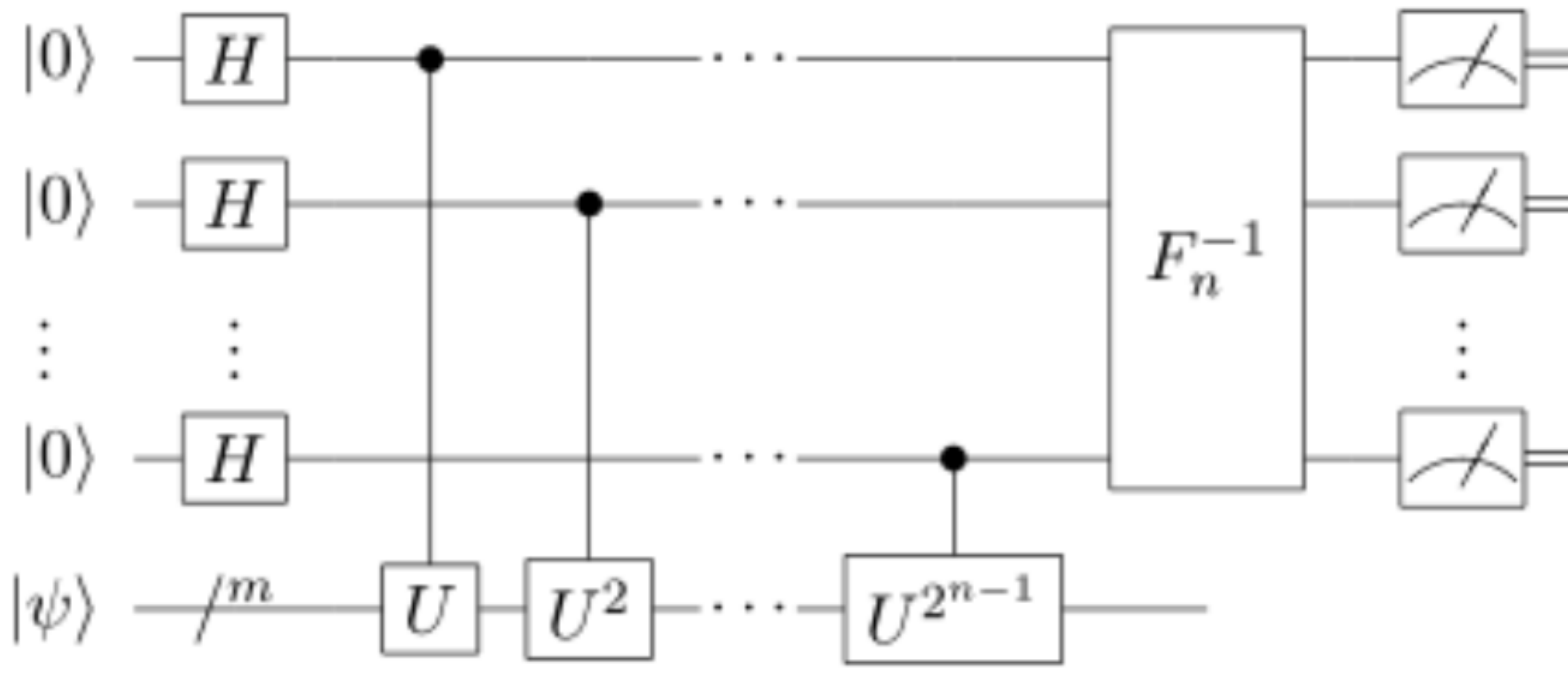} \;\;
\includegraphics[width=8.5cm]{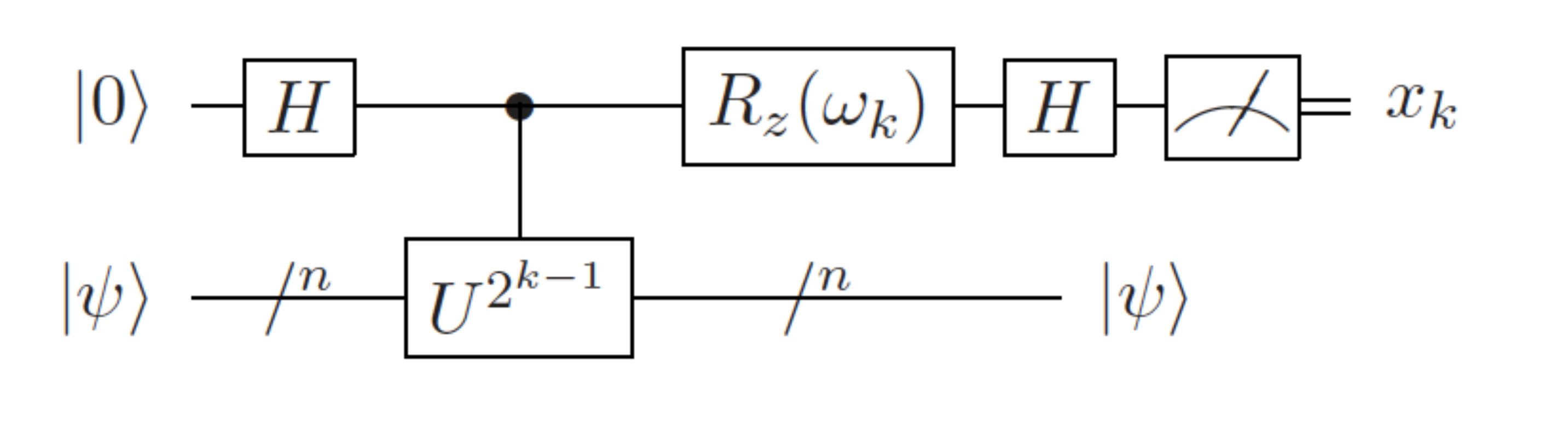}
\caption
{\small (a) The Phase Estimation Algorithm (PEA). (b) The kth iteration of the iterative PEA (IPEA). The feedback angle $k$ depends on the previously measured bits \cite {Dobsicek2007}. }
\label{PEA}
\end{figure}

The time evolution methods solve dynamical simulation problems, and do not directly solve the ground-state energy estimation problem. The phase estimation algorithm (PEA) (see e.g. 
\cite{NielsenChuang2010,AbramsLloyd1999,Aspuru-Guzik2005,Whitfield2011,Dobsicek2007})
provides the connection needed to relate the eigenvalue estimation problem to the dynamical simulation problem.
The left part of Fig.~\ref{PEA}a describes the propagation of the quantum state  $\hat U \ket{\psi} = e^{-i\hat Ht} \ket{\psi}$  in smaller and smaller steps (longer and longer times) to achieve the required accuracy.
The phase information from the time evolution is stored in the ancillas, and the energy spectrum is finally analysed by an inverse quantum Fourier transform (QFT) (right part of Fig.~\ref{PEA}a). 

The multi-qubit state $\ket{\psi}$ itself in Fig.~\ref{PEA}a can be efficiently implemented by a polynomial number of gates (Sect.~\ref{universal}), because the $\hat U$-operators consist of exponentials of polynomial numbers of Pauli operators in the Hamiltonian (cf. Fig.~\ref{2q_gates}e). Calculation of the energy via the PEA is however not necessarily scaling polynomially: it employs Cntrl-$R_z(\theta$ gates (cf. Fig.~\ref{2q_gates}f), and the number depends on the accuracy required. For a large molecule, requiring many qubits and chemical precision, the number of gates will be very large, requiring long coherence times. This will be further discussed in Sect.~\ref{chemistry}.  

The PEA in Fig.~\ref{PEA}a requires as many ancillas as significant bits in the result. To save on the number of ancilla qubits, the PEA can be implemented through an iterative process using only a single ancilla qubit  and classical feedback \cite {Whitfield2011,Dobsicek2007}, as illustrated in (Fig.~\ref{PEA}b).

\subsection{Digital quantum simulation of the quantum Rabi model}

Langford {\it et al.} \cite{Langford2016} have  performed a digital quantum simulation of the quantum Rabi model  (QRM, Eqs.~(\ref{tmon-cQED1}),(\ref{tmon-cQED2})) in the ultra-strong (USC, DSC) regime. 
Building on the proposal in  \cite{Mezzacapo2014b}, they perform a digital simulation of the QRM for arbitrarily
large coupling strength using a physical qubit(transmon)-resonator circuit in the moderate-coupling regime ($g/\omega < 10^{-3}$). The name of the game is to simulate a system where the effective resonator frequency $\omega_{eff} $ can be made  so small that $g/\omega_{eff} \sim 1$, representing the USC/DSC regimes. This was achieved by transforming the Hamiltonian to rotating frames and simulating the effects of the JC and AJC co- and counter-rotating interactions in the qubit-resonator Hamiltonian using up to 90 second-order Trotter steps \cite{Langford2016}. The work demonstrates the expected Schr\"odinger-cat-like entanglement and build-up of large photon numbers and population revivals characteristic of deep USC.
Furthermore, it  opens the door to exploring extreme USC regimes, quantum phase transitions and many-body effects in the Dicke model \cite{Lamata2016}.

\subsection{Digital quantum simulation of spin models}

In digital quantum simulation (DQS) one induces the time evolution of a qubit register in the quantum circuit model by applying a sequence of qubit gates according to a specific protocol. 
DQS has previously been implemented in an ion trap to perform universal digital quantum simulation of spin models  \cite{Lanyon2011}, and now also on a transmon platforms \cite{Barends2015a,Salathe2015,LasHeras2014a} to simulate the dynamics of small spin systems.

\subsubsection{Two-spin Ising and Heisenberg models}

For a two-spin system, the canonical spin models are:
\begin{enumerate}
\item The Ising model:
\begin{equation}
\hat H_I = J \sigma_{1z}  \sigma_{2z} + B \sum \sigma_{iz}   
\label{Ising}
\end{equation}
\item The transverse field Ising model (TIM):
\begin{equation}
\hat H_{TIM} = J \sigma_{1x}  \sigma_{2x} + B \sum \sigma_{iz}  
\label{TIM}
\end{equation}

\item The XY model:
\begin{equation}
\hat H_{XY} = J (\sigma_{1x}  \sigma_{2x} + \sigma_{1y}  \sigma_{2y}) + B \sum \sigma_{iz}   
\label{XY}
\end{equation}
\item The XYZ anisotropic Heisenberg model:
\begin{equation}
\hat H_{XYZ}  = J_x \sigma_{1x}  \sigma_{2x} + J_y \sigma_{1y}  \sigma_{2y} + J_z \sigma_{1z}\sigma_{1z} + B \sum \sigma_{iz}
\label{XYZ}
\end{equation}
\end{enumerate}

The XY interaction $ \frac{1}{2}(\sigma_{1x}  \sigma_{2x} + \sigma_{1y}  \sigma_{2y}) =  \sigma_{1+}  \sigma_{2-} + \sigma_{1-}  \sigma_{2+}$ is naturally implemented via  the $iSWAP$  gate $\hat U_{XY}(t)$  (Eq.~(\ref{H12})), tuning the qubits in and out of resonance,  and can be used to construct a digital decomposition of the model-specific evolution and to extract its full dynamics. 

Salath\'e {\it et al.} \cite{Salathe2015} performed digital quantum simulation of the XY and isotropic Heisenberg XYZ spin models with a four-transmon-qubit circuit quantum electrodynamics setup, using two qubits to represent the two spins. 
The isotropic model Hamiltonian can be written as

\begin{equation}
\hat H_{XYZ}  = \frac{1}{2} ( \hat H_{XY} + \hat H_{XZ}  + \hat H_{YZ} )
\label{}
\end{equation}
and since the three terms commute, the time evolution operator takes the form of a simple product: $\hat U_{XYZ}  = \hat U_{XY} \hat U_{XZ} \hat U_{YZ}$.

\begin{figure}[h]
\center
\includegraphics[width=8cm]{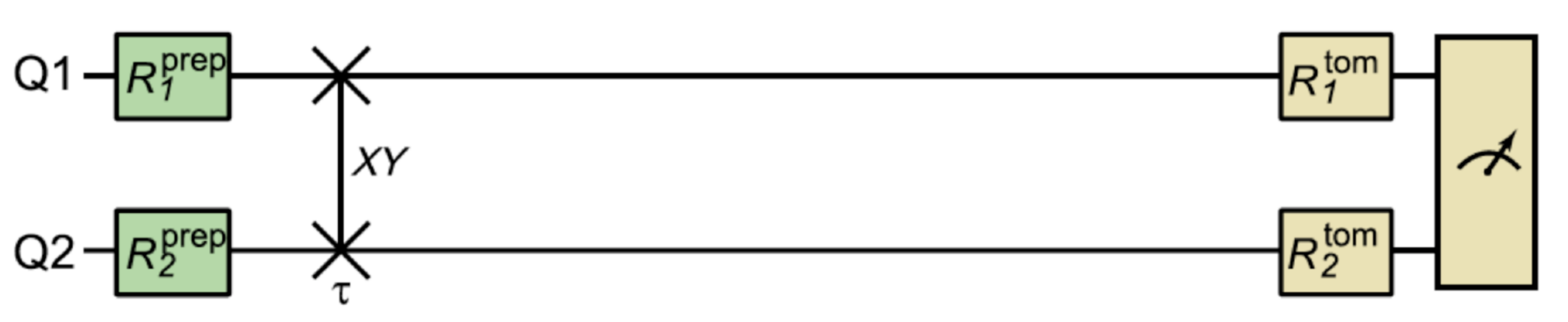}
\includegraphics[width=7.5cm]{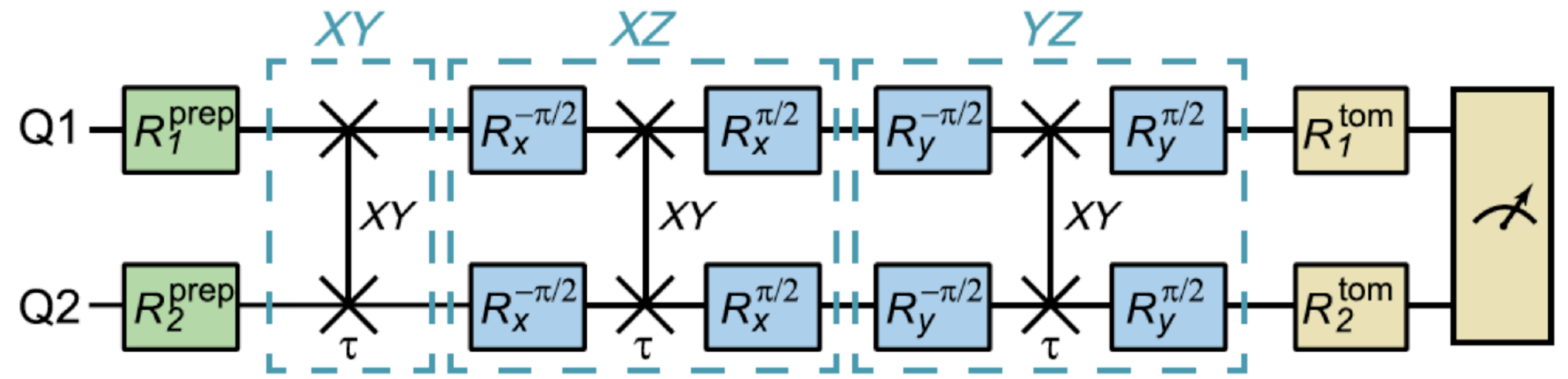}\\
\includegraphics[width=15.5cm]{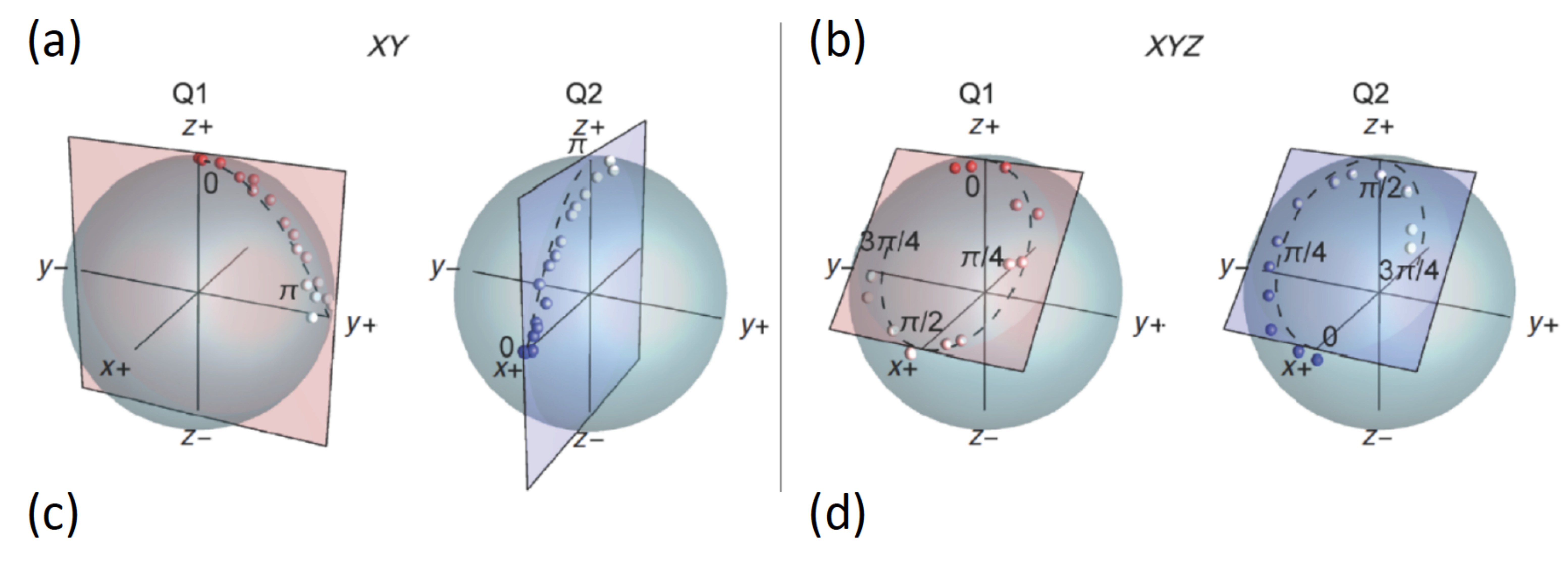}
\caption{ 
(a) Two-spin XY model: Circuit diagram implementing the $\hat U_{XY}(t)$ gate for a certain time $t=\tau$. (b)  Two-spin Heisenberg (XYZ) model:  Circuit diagram implementing the $\hat U_{XYZ}(t)$ gate (Eq.~\ref{Uxyz}) for a certain time $t=\tau$.
(c)  Time evolution $\hat U_{XY}(t)$ of the two-spin XY model: Experimentally determined coordinates of the Bloch vectors. Red
(Q1) and blue (Q2) points are compared to the ideal paths shown as dashed lines in the XY model.  (d) describes the same thing for the Heisenberg (XYZ) model. Adapted from \cite{Salathe2015}.
 }
\label{ETHZ_XYZ}
\end{figure}

Moreover, rotating the computational basis,
\begin{eqnarray}
\label{Uxyz}
\hat U_{XYZ}  = \hat U_{XY} \hat U_{XZ} \hat U_{YZ} \\
= \hat U_{XY}  
R_y(-\pi/2) \hat U_{XY} R_y(\pi/2) 
R_x(-\pi/2)\hat U_{XY} R_x(\pi/2) 
\end{eqnarray}
\begin{equation}
\label{U}
R_x(\pi/2) = \frac{1}{\sqrt{2}}
\left(\begin{array}{cc}
1  &  -i \ \\
 -i   & 1  
\end{array}\right),  \;\;\;
R_y(\pi/2) = \frac{1}{\sqrt{2}}
\left(\begin{array}{cc}
1  &  - 1 \\
 1  & 1
\end{array}\right) 
\end{equation}
makes it possible to apply the XZ and YZ interactions via the XY interaction.

The procedure is shown in Fig.~\ref{ETHZ_XYZ}a,b, displaying the circuit diagrams to digitally simulate (a) the XY interaction acting on the qubits Q1 and Q2 by applying $\hat U_{XY}(t)$ for a  time $\tau$; and  (b) the two-spin Heisenberg (XYZ) interaction by applying $\hat U_{XY}(t)$ for time $\tau$.
Fig.~\ref{ETHZ_XYZ}c  presents the time-development of the spin dynamics under the XY interaction for a characteristic initial two-qubit state $\ket{\psi(0)} = \ket{0}(\ket{0}+\ket{1})/2$ in which the 
spins point in the  +{\bf z} and +{\bf x} directions. 
Since this is not an eigenstate of the Hamiltonian, the spins start to rotate due to the XY-interaction. 
Fig.~\ref{ETHZ_XYZ}d  presents the result of simulating the full Heisenberg model.

\begin{figure}[h]
\center
\includegraphics[width=15cm]{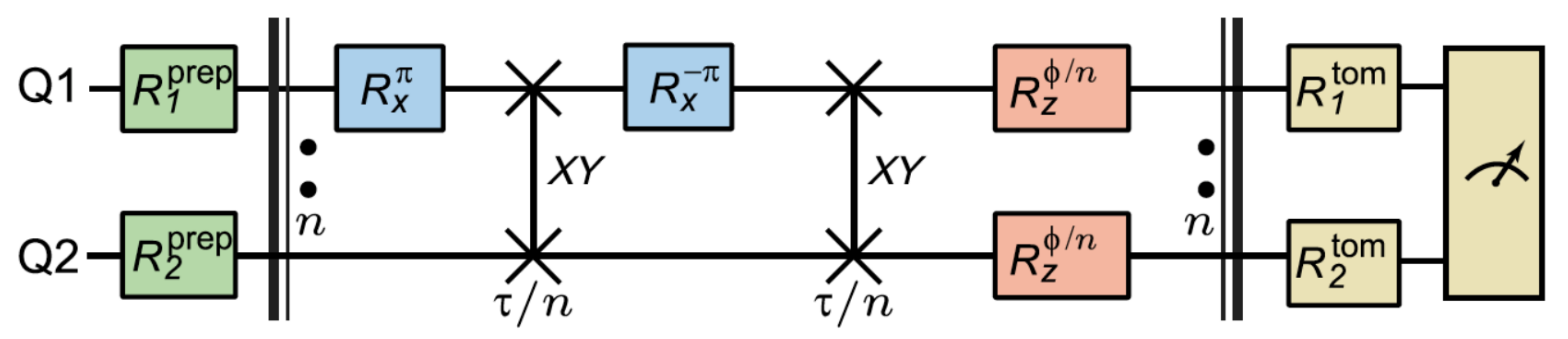}\\
\caption
{\small Protocol to decompose and simulate Ising XY spin dynamics in a homogeneous transverse magnetic field.  Adapted from \cite{Salathe2015}.
 }
\label{ETHZ_TIM}
\end{figure}

Salath\'e {\it et al.} \cite{Salathe2015} also performed digital quantum simulation of the transverse Ising model (TIM), Eq.(\ref{TIM}).
Here the XY and Z parts of the Hamiltonian do not commute, which means that one must implement a split-operator procedure (trotterisation), as shown in Fig.~\ref{ETHZ_TIM}, splitting the evolution over time $\tau$ into $n$ slices. In each Trotter time slice of length $\tau/n$, the $R_x(\pm\pi)$ rotation operators change the sign of the second term in the XY interaction, adding up to the XX interaction, and $R_z(\phi/n)$ implements the Z-part of the Hamiltonian. 
The simulation uses n=3 Trotter slices and demonstrates \cite{Salathe2015} that after three iterations, the z-components of the spins oscillate as expected. 

The approach uses only Clifford gates and is universal and efficient, employing only resources that are polynomial in the number of spins (Sect.~\ref{universal}). An idea of the future challenges can be obtained from a recent investigation of how to simulate the transverse Ising model (TIM) on a quantum computer including error correction with the surface code \cite{You2013}.

\subsubsection{Digitized adiabatic four-spin transverse Ising model}

An experiment with a 9-qubit superconducting circuit was recently carried out by Barends  {\it et al.}  \cite{Barends2016}.
They probed the adiabatic evolutions, and quantified the success of the algorithm for random spin problems, 
approximating the solutions to both frustrated Ising problems and problems with more complex interactions. 
The approach is compatible with small-scale systems as well as future error-corrected quantum computers.

\begin{figure}[h]
\includegraphics[width=16cm]{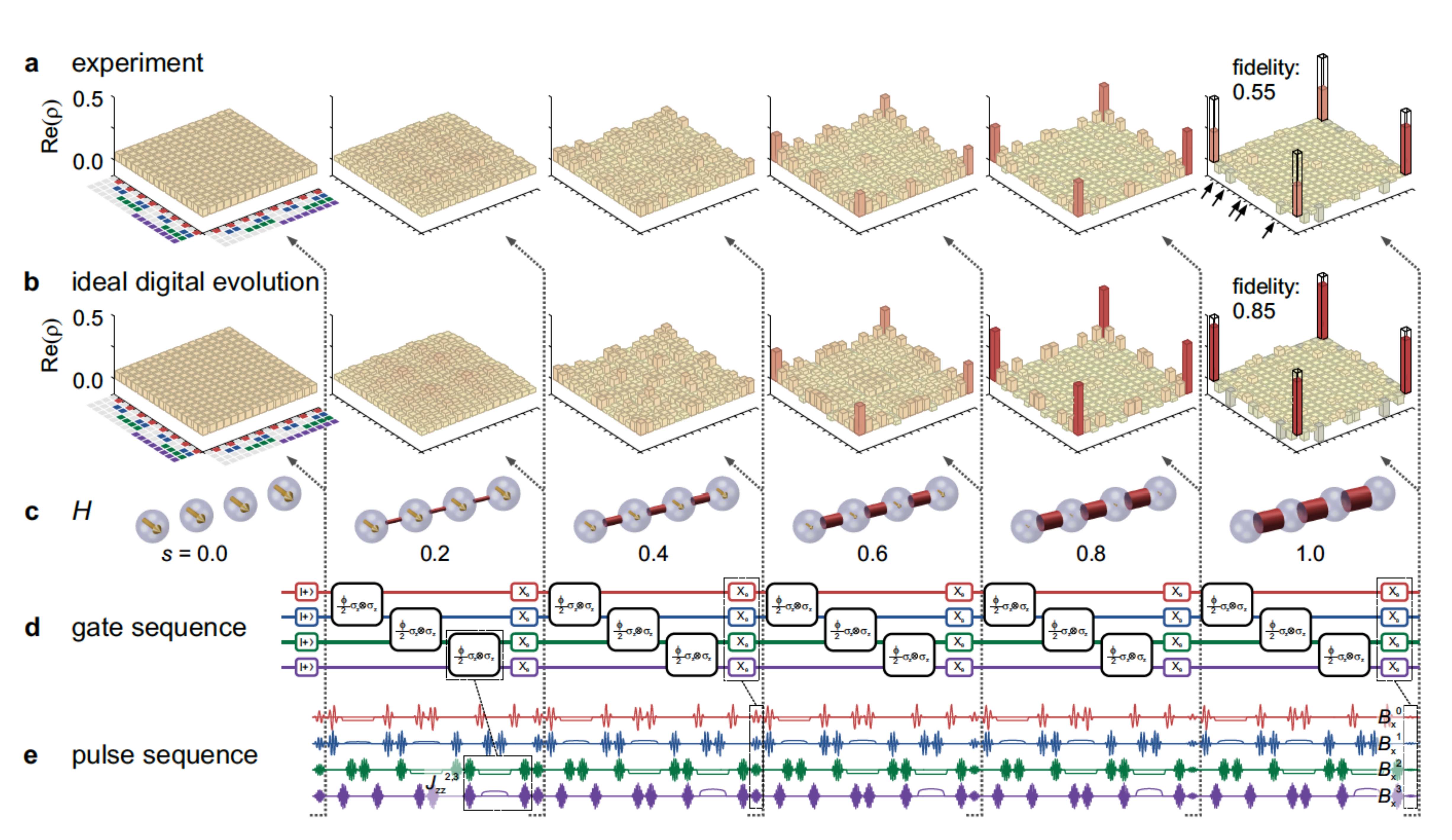}
\caption
{\small Quantum state tomography of the digital evolution into a GHZ state. A four qubit system is adiabatically
evolved from an initial Hamiltonian where all spins are aligned along the X axis to a problem Hamiltonian with equal ferromagnetic couplings between adjacent qubits.  The displayed five step algorithm is 2.1 $\mu s $ long. Implementations of zz coupling and local X-fields are highlighted. Adapted from  \cite{Barends2016}.} 
\label{Google4qferro}
\end{figure}

The digital quantum simulation (DQS) involved the following two Hamiltonians:
\begin{equation}
\hat H_I = - B_{x,I} \sum_i \sigma_{ix}   
\label{I}
\end{equation}
describing noninteracting spins in an external field in the x-direction, and 
\begin{equation}
\hat H_P =  -  \sum_i (B_{iz} \sigma_{iz} + B_{ix} \sigma_{ix} )   - \sum_i  (J_{zz}^{i,i+1} \sigma_{iz}  \sigma_{i+1z} + J_{xx}^{i,i+1} \sigma_{ix}  \sigma_{i+1x})
\label{P}
\end{equation}
describing a range of Ising-type spin Hamiltonians.  For the analogue quantum simulation (AQS) part, these were combined:
\begin{equation}
\hat H(s) =  (1-s)\hat H_I +  s \hat H_P
\end{equation}
to allow one to perform DQS for a series of Hamiltonians from non-interacting spins (s=0) to a range of interacting spin models (s=1), and to follow the evolution of the density matrix.
Figure \ref{Google4qferro} shows an application where a four qubit system is stepwise
evolved from an initial Hamiltonian $H_I$, where all spins are aligned along the x-axis, to a problem Hamiltonian $H_P$ with equal ferromagnetic couplings between adjacent qubits, described by a 4-qubit GHZ state. 

Barends  {\it et al.} \cite{Barends2016} also investigated digital evolutions of random stoquastic and non-stoquastic problems.
Quantum stochastic calculus  concerns generalisation of the Langevin equation to quantum systems.
Because of the relation to stochastic processes one has adopted the term "stoquastic" to refer to quantum Hamiltonians where
all off-diagonal matrix elements in the standard basis are real and non-positive \cite{Bravyi2008}.
Stoquastic Hamiltonians are very common in physics. Among spin-1/2 models, the well-studied ferromagnetic
Heisenberg models and the quantum transverse Ising model  \cite{Farhi2001}  are stoquastic. Another example is a Heisenberg antiferromagnet
on a cubic lattice.

Barends  {\it et al.} chose to investigate a stoquastic frustrated Ising Hamiltonian having random local X and Z fields, and random zz couplings.  Non-stoquastic problems have additional random xx couplings. The results show that the system can find the ground states of both stoquastic and non-stoquastic Hamiltonians with comparable performance.

\subsection{Digital quantum simulation of fermionic models}

Computational physics, chemistry and materials science deal with the structure and dynamics of electronic systems: atoms, molecules, solids, liquids, soft matter, etc. To describe theses systems one needs the full machinery of quantum many-body theory involving fermionic and bosonic particles and excitations. So far a we have been working with 2-level (spin) systems coupled to bosonic modes. However, to describe electronic systems, the fermionic anti-commutation rules have to be built in. One way to do this was invented a long time ago  in the form of the Jordan-Wigner (JW) transformation \cite{JordanWigner1928}. One then works in the occupation-number representation and keeps track of parity under permutations via the the anti-commutation rules of a set of auxiliary Pauli $\sigma$ operators embedded in the fermionic creation and annihilation operators. In this way the number of $\sigma$ operators scales as $O(n)$, i.e. as the number of qubits.

Bravyi and Kitaev \cite{BravyiKitaev2000} derived an alternative (BK) transformation, using the qubits for storing parities rather than occupation numbers. This scheme also maps the fermionic operators on products of Pauli $\sigma$ operators. One advantage, however, is that  the number of $\sigma$ operators scales as $O($log$~n)$, which will be important for simulation of large systems that require large numbers of qubits.

These methods have been developed theoretically and simulated classically over the last 15 years  \cite{Whitfield2011,Ortiz2001,Somma2002,Lamata2011,Seeley2012,GarciaAlvarez2015,LasHeras2015}, but never explored experimentally, until now. The first experimental applications ever, with superconducting circuits, have recently been published, implementing digital simulation of the Fermi-Hubbard model \cite{Barends2015a} and the ground state binding curve of the hydrogen molecule, $H_2$ \cite{OMalley2016} (see further Sect.~\ref{chemistry_PEA}). 

For illustration of the approach to an elementary fermonic many-body system, consider a closed-shell atom or molecule. The general second-quantised Hamiltonian is given by:
\begin{equation}
\hat H =   \sum_{pq} h_{pq}c^+_p c_q + \frac{1}{2}\sum_{pqrs} h_{pqrs}c^+_p c^+_q c_r c_s 
\label{Atom}
\end{equation}
where the first term describes the single-particle kinetic and potential energies, and the second term the 2-body Coulomb interaction. The indices refer to the set of basis orbitals (fermionic modes) used to expand the Hamiltonian.

The simplest possible case is the ground state of a 2-electron system with a minimal basis of 2 states: a $He$ atom with $1s\uparrow$  $1s\downarrow$, or a $H_2$ molecule with $1\sigma\uparrow$ $1\sigma\downarrow$. The Hartree Hamiltonian is then given by:
\begin{equation}
\hat H =   h_{1}c^+_1 c_1 + h_{2}c^+_2 c_2 + V_{12}c^+_1 c_1c^+_2c_2  
\label{Hatom2}
\end{equation}
where the Hartree term can be written as $V_{12} \; n_{1\uparrow} n_{2\downarrow}$, on the form of a Hubbard onsite interaction (here is only one site).

The JW transformation becomes
\begin{eqnarray}
c^+_1 =  I \otimes  \sigma^+  \\ 
c^+_2 =   \sigma^+ \otimes  \sigma_{z}    \\ 
c_1 =        I \otimes  \sigma^-     \\ 
c_2 =        \sigma^- \otimes  \sigma_{z} 
\label{Hatom}
\end{eqnarray}
Worked out in detail, one obtains \cite{Whitfield2011}
\begin{eqnarray}
c^+_1 c_1 =  \frac{1}{2} (I - \sigma_{z1})  \\ 
c^+_2c_2 =   \frac{1}{2} (I -  \sigma_{z2}\sigma_{z1})   \\
c^+_1c_1c^+_2c_2 =   \frac{1}{4} (I -  \sigma_{z1} - \sigma_{z2}\sigma_{z1} + \sigma_{z2}) 
\label{Hatom}
\end{eqnarray}
The Hamiltonian then finally becomes \cite{Whitfield2011}:
\begin{eqnarray}
\hat H = h - h/2 \;(\sigma_{z1} + \sigma_{z2}  \sigma_{z1}) 
+ V/4 \; (1 - \sigma_{z1}  +  \sigma_{z2} -  \sigma_{z2} \sigma_{z1} )
\label{XY}
\end{eqnarray}

The evolution operator corresponding to the parts of the Hamiltonian with $\sigma_{z} \otimes \sigma_{z}$ products, $\hat U = exp[{-i\frac{\theta}{2}\sigma_{z} \otimes \sigma_{z}}]$ can be implemented by a quantum circuit of the form shown in Fig.~\ref{2q_gates}e.

At the next level, the simplest possible case is still the ground state of a 2-electron $He$ atom, but now with a slightly extended basis of 4 states, 1: $1s\uparrow$, 2: $1s\downarrow$, 3: $2s\uparrow$ and 4: $2s\downarrow$. At this level one can begin to investigate effects of correlation on the ground state energy. Taking into account that in the ground state only states 1 and 2 are occupied, the Hamiltonian can be written as:
\begin{eqnarray}
\hat H =   \hat H_1 + \hat H_2 \\
\hat H_1 =   h_{11}c^+_1 c_1 + h_{22}c^+_2 c_2  \\
\hat H_2 = h_{1221}c^+_1c^+_2c_2  c_1 +     h_{1243}(c^+_1c^+_2c_4  c_3 + c^+_3c^+_4c_2  c_1) 
\label{Hatom4}
\end{eqnarray}
The interaction term can be rewritten as
\begin{eqnarray}
\hat H_2 = V_{12}c^+_1 c_1c^+_2c_2  +  V_{1324}(c^+_1c_3 c^+_2c_4   + c^+_3c_1 c^+_4c_2)
\label{H2atom4}
\end{eqnarray}
emphasizing the physical meaning of the terms: the first term describes the direct Coulomb interaction $V_{12}$ between the  $1s\uparrow$ and  $1s\downarrow$ electrons, while the second term describes the (radial) correlation energy generated by interacting virtual  $(1s\uparrow)^{-1} (2s\uparrow)$ and $(1s\downarrow)^{-1} (2s\downarrow)$ electron-hole pair excitations \cite{Comment_He1s2p}.

The correlation term involves all four states, and therefore all four qubits. The result of applying the JW transformation then leads to the appearance of an interaction Hamiltonian of the form $\sigma_{\alpha} \otimes \sigma_{\beta} \otimes \sigma_{\gamma} \otimes \sigma_{\delta}$. A detailed analysis \cite{Whitfield2011} shows that only the terms $\sigma_{x} \sigma_{x} \sigma_{y} \sigma_{y}$, $ \sigma_{y} \sigma_{y} \sigma_{x} \sigma_{x}$, $\sigma_{y} \sigma_{x} \sigma_{x} \sigma_{y} $, $\sigma_{x}  \sigma_{y} \sigma_{y} \sigma_{x}$ need to be considered.
These can be generated by an evolution operator of the form $\hat U = exp[{-i\frac{\theta}{2}\sigma_{z} \otimes \sigma_{z}\otimes \sigma_{z}\otimes \sigma_{z}}]$ together with suitable qubit rotations changing the computational basis (see Table A3 in \cite{Whitfield2011}). 

This method - the full approach with trotterisation and phase estimation - was recently applied by O'Malley {\it et al.}  \cite{OMalley2016} to the calculation the $H_2$ binding energy curve, as discussed in Sect.~\ref{chemistry}

\subsection{Analogue/adiabatic quantum simulation}

In analogue quantum simulation (AQS) one induces the time evolution of a qubit register by application of a sequence of time-dependent external fields and internal interactions \cite{Brown2010,Kendon2010}. AQS has so far been implemented experimentally in 
ion traps \cite{Gerritsma2010,Britton2012,Islam2013}, 
ultra-cold atoms in optical lattices \cite{Bloch2012,Schneider2012,Fukuhara2013,Aidelsburger2013,Miyake2013} 
and photonics circuits \cite{Walther2014,Shadbolt2014,Lund2014,Lund2017,Spagnolo2014,Bentivegna2015,Bentivegna2016,Caruso2016,Crespi2016,Qiang2016,Inagaki2016}, 
to perform simulation of various physical models involving spins, bosons and fermions.

With the advent of useful and powerful JJ-based qubit circuits, there is a recent surge of proposals involving superconducting qubits \cite{Macha2014,Mezzacapo2014b,Houck2012,Schmidt2013,Barrett2013,Wiese2014,Dalmonte2015,Zhang2013,Ashhab2014,Mei2013,Stojanovic2014,Zippilli2015,Mostame2012,Egger2013,Geller2015a} 
as well  experimental results \cite{Chen2014b,Roushan2014,Braumuller2017,Eichler2015,Li2013,Neill2016,Roushan2016,Hacohen-Gourgy2015,Kirchmair2013,Potocnik2017} for superconducting circuits.

\subsection{Digital-analogue quantum simulation}

By digital-analogue quantum simulation we denote methods that control the time evolution (operator) by both (i) applying circuit-based digital gates (DQS) and (ii) evolving the Hamiltonian parameters in time (AQS). If the time scales are widely separated, e.g. the analogue evolution being adiabatic, the calculation becomes a number of complete DQS calculations for a series of adiabatic Hamiltonian time steps.

Recently a fermionic 4-mode problem was performed experimentally with a superconducting system  by Barends {\it et al.} \cite{Barends2015a}, implementing a scheme of Las Heras {\it et al.} \cite{LasHeras2015}.
The experiment involved time evolutions with constant interactions, with up to four fermionic modes encoded in four qubits, using the Jordan-Wigner transformation.
The time evolution involved over 300 single-qubit and two-qubit gates, reaching global fidelities limited by gate errors in an intuitive error model. 
Barends {\it et al.} \cite{Barends2015a} also introduced time-dependence in the model Hamiltonian, by slowly ramping the hopping interaction from zero to the strength of the  on-site fermion-fermion interaction, switching the system from localised to itinerant fermions, observing elements of a dynamic phase transition.
The experiment, as well as that in \cite{Barends2016}, therefore may present a step on the path to creating an analogue-digital quantum simulator using discrete fermionic modes combined with discrete or continuous bosonic modes \cite{Garcia-Alvarez2016}.

In the general case with no separation of time scales, the combined evolution needs to be implemented via the full evolution operator with a time-dependent Hamiltonian. 
In the end combinations of analogue and digital simulation schemes may be the most powerful ones, driving or inducing selected terms in the Hamiltonian that allow sets of gates not possible in the undriven physical system. \\


\newpage

\section{Toward quantum chemistry simulation}
\label{chemistry}

Quantum chemistry is traditionally a testing ground for classical high-performance computing (HPC)  \cite{Pople1999} and is currently at the focus of investigations of various advanced approximation methods in many-body physics \cite{Verstraete2010,Khoromskaia2015,Szalay2015}. The simulation of quantum chemistry is one of the most anticipated applications of quantum computing, but the scaling of known upper bounds on the complexity of these algorithms is very demanding. 
Hamiltonian problems with 2-body interactions have been shown to be QMA-hard \cite{Kempe2004}, which means that it is fundamentally impossible to calculate the exact ground state electron structure of large molecules \cite{Osborne2012,Rassolov2008,Schuch2009,Aaronson2009,Whitfield2013,Whitfield2014}. This implies that the quantum chemical problem  of ground state energy search is non-polynomial (actually exponential) in time with respect to the system size.
Consequently, like in the classical case, also quantum simulation of molecular electronic structure must build on advanced approximation schemes in order for full configuration interaction (FCI) to be tractable in high-accuracy calculations already for fairly small molecules.

Quantum simulation methods are now the  targets of an emerging field of quantum HPC both theoretically \cite{OMalley2016,Wecker2014,Haner2016a,Haner2016b,Bauer2016,Reiher2016,Aspuru-Guzik2005,Whitfield2011,Babbush2015,Seeley2012,Poulin2015,Peruzzo2014,Tranter2015,Mueck2015,Love2014,TolouiLove2013,Babbush2014,Yung2014,Hastings2015,McClean2014,Zalka1998,Kassal2008,Lu2011,Pritchett2010,Sornborger2012,Lanyon2010,Sugisaki2016} and experimentally \cite{OMalley2016,Peruzzo2014,Lu2011,Lanyon2010,Du2010,Wang2015}. 
Even if the number of gates "only" scales polynomially, the required number may still be prohibitive to reach chemical accuracy in near-future applications. 
However, applying the intuition of classical quantum chemistry to QS may reduce the computational complexity. A natural way forward is to make use of  state-of-the-art classical approximate treatments of FCI and apply these to quantum algorithms  \cite{OMalley2016,Babbush2015}. Also, it is very important to have realistic ideas of the computational effort needed for specific molecules, to be able to address cases that are tractable with present resources.
Fortunately, the recent development has been quite dramatic \cite{Mueck2015}. Combined with the rapid development of superconducting multi-qubit systems, it now seems possible to go beyond toy models and implement larger-scale calculations on real hardware systems \cite{OMalley2016,Babbush2015,Peruzzo2014}.

\subsection{Hamiltonian ground-state energy estimation}

The quantum phase estimation algorithm PEA efficiently finds the eigenvalue of a given eigenvector but requires fully coherent evolution. For large systems requiring many qubits and gate operations, the coherence time may eventually become too short. To alleviate this problem,  Peruzzo et al. \cite{Peruzzo2014} introduced an alternative to the PEA that significantly reduces the requirements for coherent evolution. 
They have developed a reconfigurable quantum processing unit, which efficiently calculates the expectation value of a Hamiltonian, providing an exponential speedup over exact diagonalisation, the only known exact method of solution to the problem on a traditional computer. 
The calculation is mainly classical but uses a quantum subroutine for exponential speedup of the critical step of quantum energy estimation.
The power of the approach derives from the fact that quantum hardware can store a global quantum state with exponentially fewer resources than required by classical hardware, and as a result the QMA-hard N-representability problem (constraining the two-electron reduced density matrix to represent an N-electron density matrix) \cite{Liu2007} does not arise.

\subsubsection{Quantum energy estimation}

The quantum energy estimation (QEE) algorithm computes the expectation value  $<\hat H> = \bra{\psi}\hat H \ket{\psi} $  of a given Hamiltonian  $ \hat H  $ with respect to a given state $\ket{\psi}$  \cite{Peruzzo2014}. 

As discussed in Sect.~\ref{Qsim}, after JW or BK transformations \cite{Tranter2015} the second quantized  Hamiltonian for electronic physical systems can be written in terms of Pauli operators as 
\begin{equation}
\hat H =   \sum_{i\alpha} h_{i\alpha} \sigma_{i\alpha}   + \sum_{i\alpha,j\beta}  h_{i\alpha,j\beta} \sigma_{i\alpha}  \sigma_{j\beta} + ....
\label{P}
\end{equation}
with expectation value:
\begin{equation}
<\hat H>  \; =   \sum_{i\alpha} h_{i\alpha} <\sigma_{i\alpha}>   + \sum_{i\alpha,j\beta}  h_{i\alpha,j\beta} <\sigma_{i\alpha}  \sigma_{j\beta}> + ....
\label{P}
\end{equation}
The coefficients are determined using a classical quantum chemistry package.

The expectation value of a tensor product $<\sigma_{i\alpha}  \sigma_{j\beta} \sigma_{k\gamma} .....>$ of an arbitrary
number of Pauli operators can be estimated by local measurement of each qubit \cite{NielsenChuang2010}, independent measurements that can be performed in parallel. The advantage of this approach \cite{Peruzzo2014} is then that the coherence time to make a single measurement after preparing the state is $O(1)$. The disadvantage relative  to the PEA is that the scaling in the total number of operations as a function of the desired precision is quadratically worse \cite{Peruzzo2014}. The scaling will also reflect the number of state preparation repetitions required, whereas in PEA the number of state preparation steps is constant. 

In the end, however, the QEE dramatically reduces the coherence time requirement, 
while maintaining an exponential advantage over the classical case by adding only a polynomial number of repetitions
with respect to QPE \cite{Peruzzo2014}.

\subsubsection {Quantum variational eigensolver}

The quantum variational eigensolver (QVE)  \cite{Peruzzo2014} is based on the Ritz variational principle, finding the minimum of the expectation value of the Hamiltonian under variation of the trial state function: (i) prepare the trial state $\ket{\psi}$; (ii) compute the Rayleigh-Ritz quotient $ <H_i>=\bra{\psi}\hat H_i \ket{\psi} / \bra{\psi}\ket{\psi} $ of all the terms in the Hamiltonian using the QEE as a subroutine; (iii) calculate $ \sum_i{<H_i>}$; (iv) compare the resulting energy with the previous runs and feed back new parameters for the trial state. Note that the only step that is quantum is step (iii) - the other steps are prepared using a classical computer.

The issue now concerns state preparation. One example of a quantum state parameterised by a
polynomial number of parameters for which there is no known efficient classical implementation is the unitary coupled cluster ansatz (UCC)  \cite{CC,Peruzzo2014}

\begin{equation}
\ket{\psi} = e^{T-T^\dagger} \ket{\psi_{ref}} 
\label{CC}
\end{equation}
where $ \ket{\psi_{ref}}$ is some reference state, usually the Hartree Fock
ground state, and T is the cluster operator for an N electron system, defined by operators
\begin{equation}
T = T_1 + T_2 + T_3 + .... + T_N 
\label{CCT}
\end{equation}
producing $1,2,3, ...., N$ electron-hole pairs from the N-electron reference state. Explicity for $T_1$ and $T_2$: 
\begin{eqnarray}
T_1 =   \sum_{pq} t_{pq}c^+_p c_q \label{T1} \\
T_2 = \sum_{pqrs} t_{pqrs}c^+_p c^+_q c_r c_s \label{T2}
\label{CCT1T2}
\end{eqnarray}
The series in Eqs.~(\ref{T1}),(\ref{T2}) generate in principle all possible configurations for FCI, producing all possible ground and excited state correlations. 

In real molecules, often a limited number of these correlations produce the bulk of the interaction energy due to the Coulomb repulsion. The problem is that to achieve the accuracy needed for describing realistic molecular chemical energy surfaces and accurately predicting chemical reaction paths, a large number of small correlations are needed to build up to the final accurate result. This is QMA-hard, i.e. intractable for both classical and quantum computers. It therefore becomes a question of useful approximations. Again, note here the recent work by Carleo and Troyer \cite{CarleoTroyer2017}.

In the case of the two-electron $H_2$ and $He$-$H^+$ molecules, $N=2$. The cluster operators are then limited to $T_1$ and $T_2$  in Eq.~(\ref{CC}) and it is possible to apply the full machinery with suitable approximations and to obtain chemical accuracy.  

\subsubsection{H-H ground-state energy curve}
 
We will now describe an experimental application of the QVE to the problem of the ground-state energy curve of the hydrogen molecule \cite{OMalley2016}. 
 
For a 2-electron system, the Hamiltonian reduces to
\begin{equation}
\hat H =   \sum_{i\alpha} h_{i\alpha}(R)\; \sigma_{i\alpha}   + \sum_{i\alpha,j\beta}  h_{i\alpha,j\beta}(R) \;\sigma_{i\alpha}  \sigma_{j\beta}
\label{P}
\end{equation}
or equivalently
\begin{equation}
\hat H =  g_0 {\bf1} + g_1 Z_0 + g_2 Z_1 + g_3 Z_0Z_1 + g_4 X_0X_1 + g_5 Y_0Y_1
\label{H2BK}
\end{equation}
where the set of parameters $g_i = g_i(R)$ depends on the $H$-$H$ distance and is obtained from the expectation values of the Hamiltonian terms evaluated on a classical computer using the basis (reference) states. 

We discussed quantum state preparation in general in Sect.~\ref{Qsim}, and the coupled-cluster approach above.
In the QVE, the state $\ket{\psi(\theta)}$ is parameterised by the action of a quantum circuit $ \hat U(\theta)$ on an initial state 
$\ket{\psi_{ref}}$, i.e. $\ket{\psi(\theta)} =  \hat U(\theta) \ket{\psi_{ref}}$. Even if $\ket{\psi(\theta)}$ is a simple product state
and $ \hat U(\theta)$ is a very shallow circuit, $\ket{\psi_{ref}}$ can contain complex many-body correlations and span an exponential number of standard basis states.

\begin{figure}[h]
\center
\includegraphics[width=16cm]{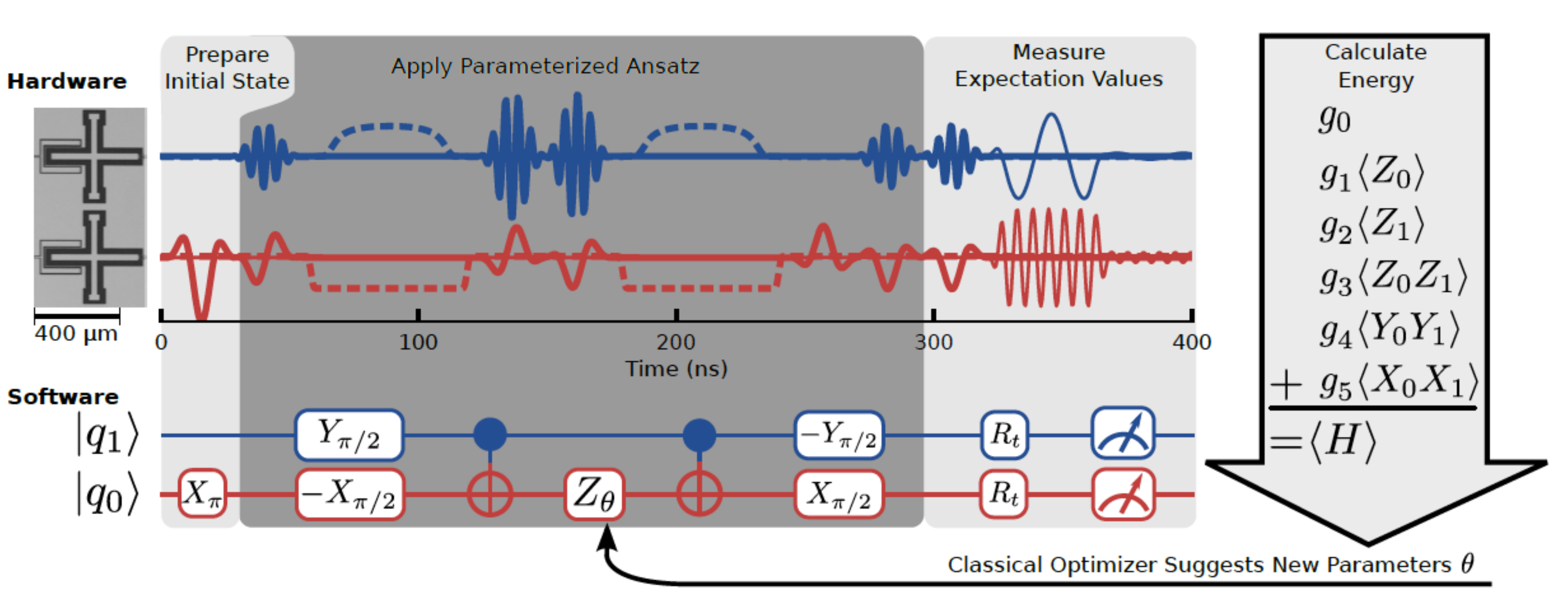}
\caption
{\small Schematic of the application of the quantum variational  eigensolver (QVE) to the $H2$ ground state energy. The "hardware part" (top panel) shows two Xmon transmon qubits and microwave pulse sequences to perform single-qubit rotations (thick lines), dc pulses for two-qubit entangling gates (dashed lines), and microwave spectroscopy tones for qubit measurements (thin lines). The "software" quantum circuit diagram (bottom panel) shows preparation of the Hartree-Fock state, followed by application of the unitary coupled cluster ansatz (UCC) in Eqs.~(\ref{CC}),(\ref{CCT1T2})  and efficient partial tomography (R$_t$) to measure the expectation values in Eq.~(\ref{P}). Finally, the total energy is computed via the QEE protocol according to Eq.~(\ref{P}) and provided to a classical optimiser that suggests new parameters $\theta$ for the time evolution operator $ \hat U(\theta)$ (right panel). Adapted from \cite{OMalley2016}.
}
\label{OMalley1}
\end{figure}

The unitary coupled cluster approach states that the ground state of Eq.~(\ref{CC}) can be expressed as
\begin{equation}
\ket{\psi(\theta)} = \hat U(\theta) \ket{\psi_{HF}} = e^{-i\theta X_0Y_1} \ket{01}  
\label{P}
\end{equation}
where  $\ket{01} $ is the Hartree-Fock (mean-field) state of molecular hydrogen in the representation of Eq.~(\ref{CC}). The
gate model circuit that performs this unitary mapping is shown in the software section of Fig.~(\ref{OMalley1}). 

\begin{figure}[h]
\center
\includegraphics[width=9cm]{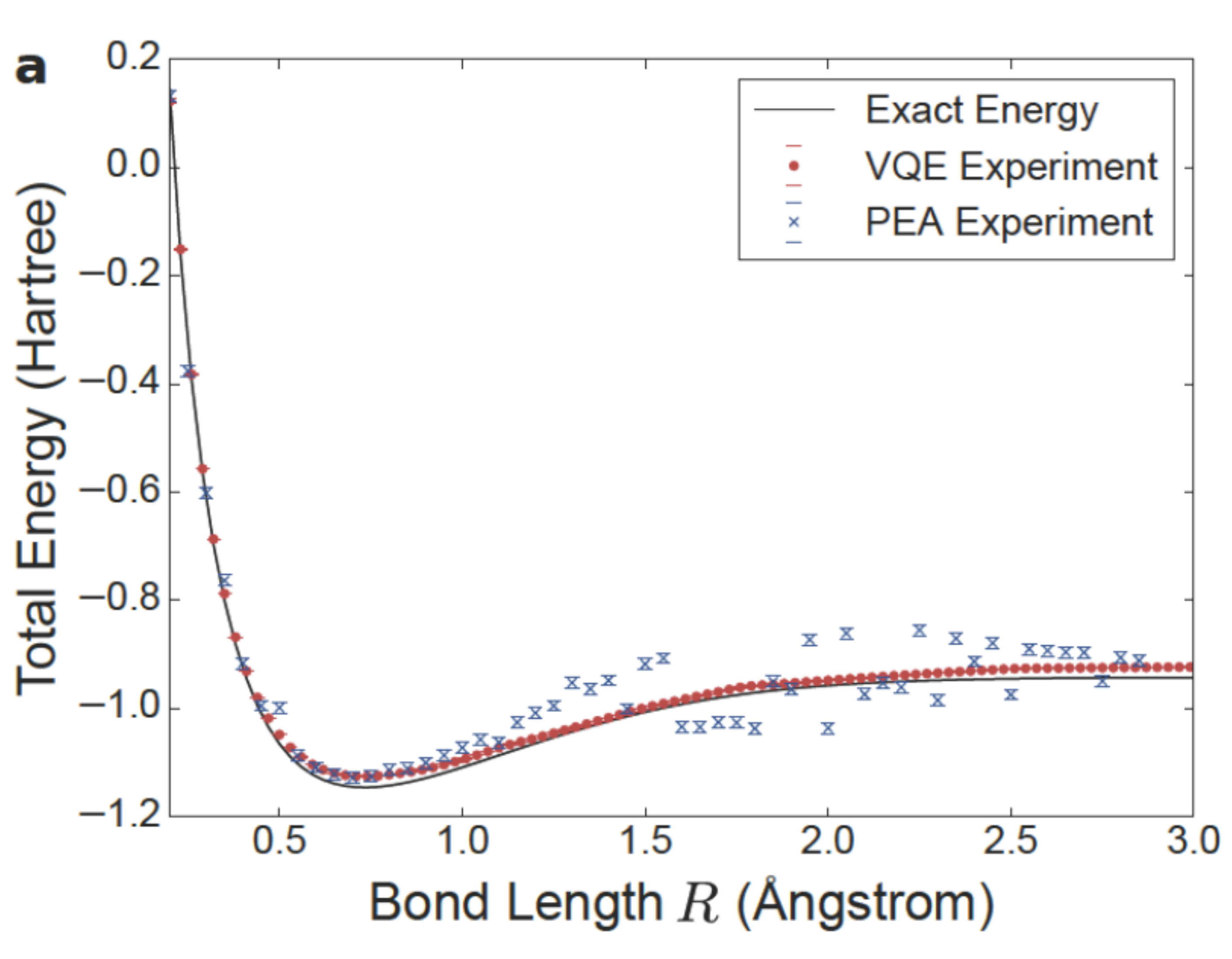}
\caption
{\small Total energy curve  of molecular hydrogen as determined by both QVE and PEA. The QVE approach shows dissociation energy error of $(8\pm5) \times 10^{-4}$ hartree (error bars on QVE data are smaller than markers). The PEA approach shows dissociation energy error of $(1\pm1) \times 10^{-2}$ hartree. Adapted from \cite{OMalley2016}.}
\label{OMalley2}
\end{figure}

The total bonding energy curve  in Fig.~(\ref{OMalley2}) demonstrates chemical accuracy (better than $10^{-3}$ hartree), which is a very important result. In contrast, the calculation using  the full canonical protocol of trotterisation plus quantum phase estimation (PEA) turns out much less accurate, amply demonstrating that the fully quantum approach is very demanding on coherence time.

\subsubsection{He-H$^+$ ground-state energy curve}
\label{chemistry_PEA}
The QVE was originally applied to the helium-hydride cation $He-H^+$ problem on a 2-qubit photonic processor  by Peruzzo et al. \cite{Peruzzo2014} .
Recently, Wang et al. \cite{Wang2015} applied the IPEA to the $He-H^+$ problem using a solid-state quantum
register realised in a nitrogen-vacancy centre (NVC) in diamond, reporting an energy uncertainty (relative to the model basis) of the order of
$10^{-14}$ hartree, 10 orders of magnitude below the desired chemical precision. 
However, this remarkable precision only refers to the 13 iterations of the IPEA itself. The fundamental propagation of the quantum state using the time evolution operator 
was done on a classical computer. In a fully quantum approach trotterisation will be necessary to create the state, 
which will put high demands on precision and coherence time. With present hardware systems, this will drastically reduce the accuracy, as is clear from the experimental results for $H_2$ in Fig.~\ref{OMalley2}  \cite{OMalley2016}.

\subsubsection{Ground-state energy curves for H$_2$, LiH and BeH$_2$} 
 
The IBM group  has recently demonstrated \cite{Kandala2017} the experimental optimisation of up to six-qubit Hamiltonian problems with over a hundred Pauli terms, determining the ground state energy curves of H$_2$, LiH and BeH$_2$. Instead of the UCC, they use  trial states parameterised by quantum gates that are tailored to the available physical devices. 
This hardware-efficient approach \cite{Kandala2017} does not rely on the accurate implementation of specific two-qubit gates - it can be used with any time-evolution operator that generates sufficient entanglement. In the present experiments, two-qubit cross-resonance (CR) gates were used as components of the entanglers.
This is in contrast to UCC trial states that require high-fidelity quantum gates approximating a unitary operator designed to describe a theoretical ansatz.

\subsection{Toward large-scale simulations}

\subsubsection{From high-level language to hardware instructions}

In order to assess the computational capability of even a small quantum systems for producing total-energy surfaces with chemical accuracy, it is necessary to develop simulation SW all the way from high-level language programs down to HW-specific instructions. 
H\"aner et al. \cite{Haner2016a} have recently developed a SW methodology for compiling quantum programs  that goes beyond the simulators that have been developed so far.
The approach involves an embedded domain-specific language by representing quantum types and operations
through types and functions existing in a classical host language,  underlining the roles of quantum computers as special purpose accelerators for existing classical codes.  In the near term, a quantum SW architecture will allow the control of small-scale quantum devices and enable the testing, design, and development of components on both the HW and SW sides \cite{Haner2016a}.

\subsubsection{Quantum computer emulation}

H\"aner et al. \cite{Haner2016b} have introduced the concept of a quantum computer emulator as a component of a SW framework for quantum computing.
A QC emulator is an interface (HW or SW) that makes the user believe that he/she is operating a quantum computer even if the calculations are performed classically. This can enable a significant performance advantage by avoiding simulating essentially classical boolean logic by quantum gate operations. 
H\"aner et al. \cite{Haner2016b} describe various optimisation approaches and present benchmarking results, establishing the superiority of quantum computer emulators in terms of performance. 
The results show \cite{Haner2016b} that emulating quantum programs allows one to test and debug large quantum circuits at a cost that is substantially reduced when compared to the simulation approaches which have been taken so far. The advantage
is already substantial for operations such as the quantum Fourier transforms, and grows to many orders of magnitude
for arithmetic operations, since emulation avoids simulating ancilla qubits (needed for reversible arithmetic) at an exponential cost. Emulation will thus be a crucial tool for testing, debugging and evaluating the performance of quantum algorithms involving arithmetic operations.

\subsubsection{Electronic structure calculations - molecules}

To describe the function of an enzyme from first principles is a computationally hard problem.
While at present a quantitative understanding of chemical processes involving complex open-shell species remains beyond
the capability of classical computer simulations, the work of Reiher et al. \cite{Reiher2016} shows that quantum computers used as accelerators to classical computers could be used to elucidate this mechanism using a manageable amount of memory and time.
In this context a quantum computer would be used to obtain, validate, or correct the energies of intermediates
and transition states and thus give accurate activation energies for various transitions. 
In particular, Reiher et al. \cite{Reiher2016}  show how a quantum computer can be employed to elucidate reaction mechanisms in complex chemical systems, using the open problem of biological nitrogen fixation in nitrogenase as
an example. Detailed resource estimates show that, even when taking into account the substantial overhead of
quantum error correction, and the need to compile into discrete gate sets, the necessary computations
can be performed in reasonable time on small quantum computers. This demonstrates that
quantum computers will realistically be able to tackle important problems in chemistry that are
both scientifically and economically significant. 

However, the required quantum computing resources are comparable to that needed for Shor's factoring
algorithm for interesting numbers, both in terms of number of gates and physical qubits  \cite{Reiher2016}.
The complexity of these simulations is thus typical of that required for other targets for quantum computing, requiring robust qubits with long coherence time.

\subsubsection{Electronic structure of strongly correlated materials}
Using a hybrid quantum-classical algorithm that incorporates the power of a small quantum computer into a framework of classical embedding
algorithms \cite{Bauer2016,Dallaire-Demers2016}, the electronic structure of complex correlated materials can be efficiently tackled using a quantum computer. The quantum computer solves a small effective quantum impurity problem that is self-consistently determined via a feedback loop between the quantum and classical computation. Use of a quantum computer enables much larger and more accurate simulations than with any known classical algorithm. This will allow many open questions in quantum materials to be resolved once small quantum computers with around one hundred long-lived (logical) qubits become available. \\


\newpage

 \section{Adiabatic quantum optimisation}
 
 It is not known to what extent coherence and entanglement are essential for AQC. (Except, since AQC has been shown to be equivalent to DQC \cite{Aharonov2004}, one would perhaps expect that coherence and entanglement are needed also for AQC to provide optimum processing speed).

\subsection{Adiabatic quantum algorithms}

\begin{figure}[h]
\center
\includegraphics[width=5.9cm]{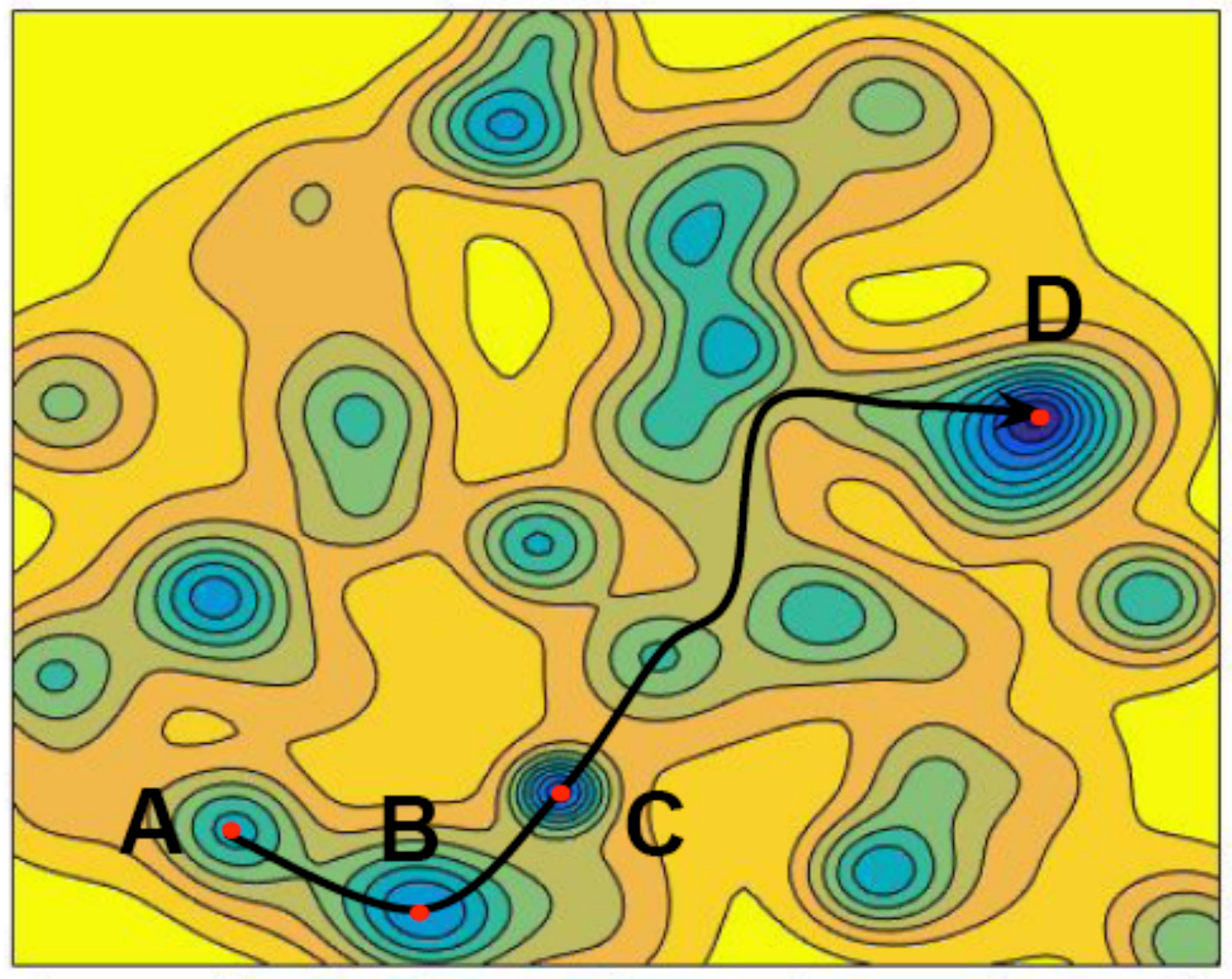}
\includegraphics[width=5.5cm]{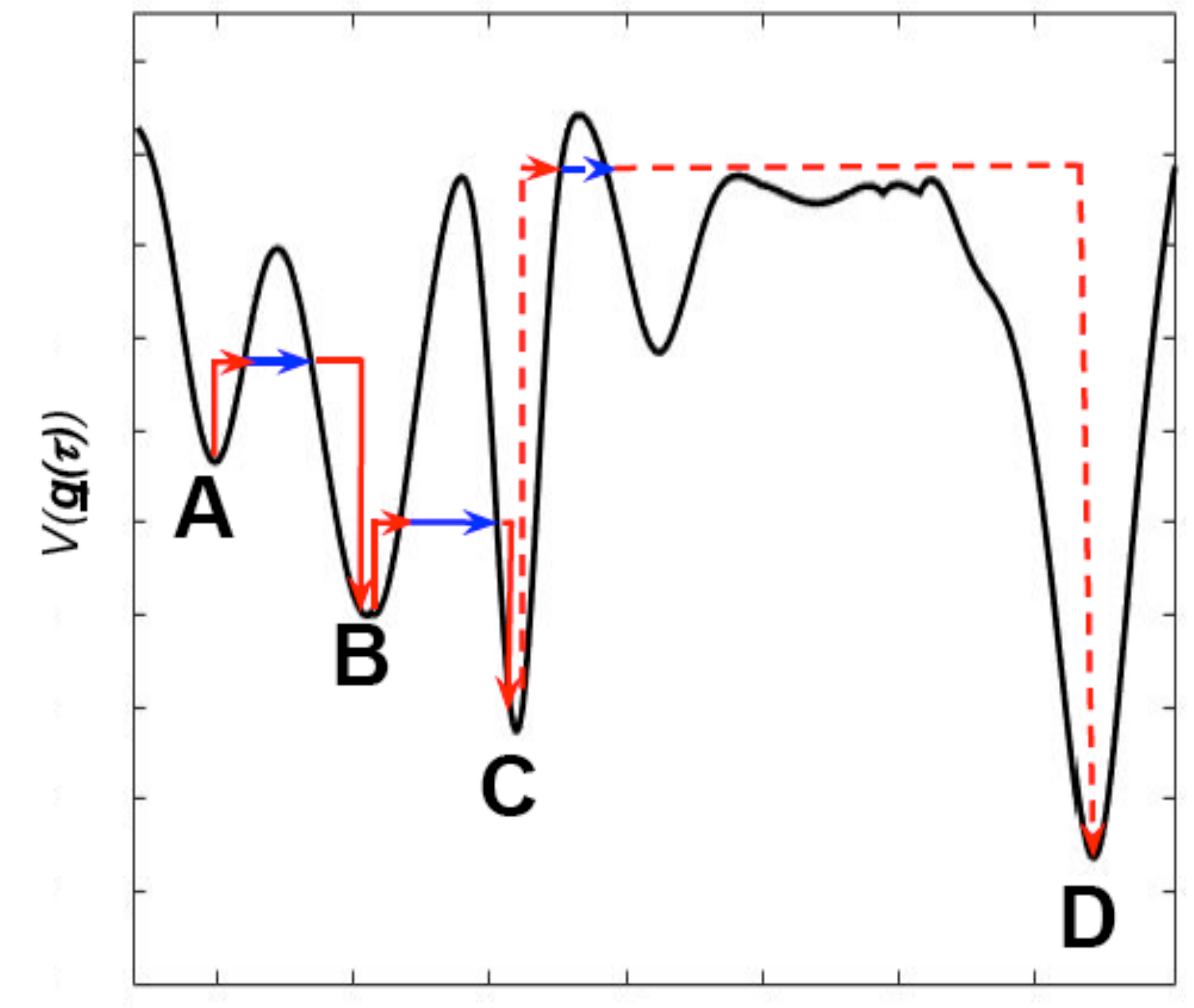}
\caption
{\small Quantum adiabatic optimisation (AQO) and quantum annealing (QA). $\hat H(t) =  A(t) \hat H_0 +  B(t) \hat H_T$. Spin-glass (Ising) type of cost function defining the energy landscape $H_T = \sum_{ij} a_{ij} x_i x_j + \sum_i b_i x_i  $. Adapted from \cite{Denchev2015} }
\label{QAfig}
\end{figure}

Adiabatic quantum optimisation (AQO)  is an adiabatic form of analogue quantum computing/simulation \cite{Farhi2000,Farhi2001,Aharonov2004,Kempe2004,Farhi2014,Das2008,Bapst2013,Biamonte2011}. AQO is in principle universal \cite{Aharonov2004} and equivalent to the digital circuit model.
AQO refers to zero temperature.
In AQO one considers the time evolution $\ket{\psi(t)} = \hat U(t,0) \ket{\psi(0)}$
with a time-dependent Hamiltonian of the form:
\begin{equation}
\hat H(t) =  [1-s(t)]\hat H_0 +  s(t) \hat H_T
\end{equation}
$s(t)$ is a scalar function of time  running from 0 to 1, controlling the adiabatic switching.
The starting Hamiltonian $H_0$ is given on a simple form, and the target Hamiltonian $H_T$ is designed to encode the problem under consideration, often defined by an Ising type of Hamiltonian. 
One then searches for an optimal path on the ground state energy surface toward a global energy minimum representing a final solution $\ket{\psi(t_f)}$ described by the target Hamiltonian $H_T$ (Fig.~\ref{QAfig}).

The philosophy behind AQO is that a wide range of problems in science can be formulated as Ising models (see e.g. Refs. \cite{Amin2016,Lucas2014,Dattani2014,Hen2014a,Hen2014b,Perdomo2012,Bian2013,Gaitan2014}), and therefore AQO could be a practical way to proceed toward addressing hard spin-glass type of problems with quantum computing. The vision is that gate-driven DQS will need extensive error correction and is, at best, a distant option, while AQO may be able to shorten that long path. However, there is seldom any free lunch - in the end calibration and embedding issues may limit the real power of AQO.

\subsection{Quantum annealing}
\label{QA}

In Quantum Annealing (QA) the temperature of a system is lowered during the time-evolution of the Hamiltonian until the system gets trapped in an energy minimum, preferably a global minimum, but more typically a  local one \cite{Ronnow2014,Buynk2014,Johnson2011,Boixo2013,Boixo2014}.

Quantum annealing can be looked upon in typically two ways:\\ 
(i) As an extension of classical Simulated Annealing (SA) by including quantum tunneling in addition to thermal hopping over the barriers (Fig.~\ref{QA}) \cite{Das2008,Nishimori2015} .\\
(ii) As a version of Adiabatic QC/QS applied to real systems influenced by noise and imperfections \cite{Bapst2013} .

Traditionally, SA and QA have been performed with software running on classical machines, solving both classical and quantum problems, e.g. using quantum Monte Carlo (QMC), including descriptions of quantum tunneling \cite{Nishimori2015}. Recently, however, the D-Wave machines (DW1-108 qubits; DW2-504 qubits; D-Wave 2X-1152 qubits) have been used to perform annealing in hardware. The mission is to perform QA (actually AQO), and the goal has been to gain decisive speedup over classical machines because the hardware is intended to function as quantum circuits and to be able to profit from coherence and entanglement.

The D-Wave Systems machines are built top-down - scaling up is based on flux qubits and circuits with short coherence time \cite{Harris2010,Buynk2014,Johnson2011}.   The technology is based on classical Nb RSFQ circuits combined with Nb rf-SQUID qubits, and forms the basis of the current D-Wave processors. The architecture is based on a cross-bar network of communication buses allowing (limited) coupling of distant qubits. The qubits are operated by varying the dc-bias, changing the qubit energies and qubit-qubit couplings.

As a result, the coherence and entanglement properties have to be investigated by performing various types of experiments on the machines and their components: Physics experiments on the hardware  \cite{Harris2010,Lanting2014}, and "benchmarking" of the performance by running a range of QA schemes  \cite{Ronnow2014,Denchev2015,Boixo2014,Boixo2013,Crowley2014,Dickson2013,Boixo2016a,Albash2015a,Albash2015b,Pudenz2014,Pudenz2015}.

During the last three years, the topic has rapidly evolved, and by now a certain common understanding and consensus has been reached.
Based on the discussion in some recent papers
\cite{Denchev2015,Zintchenko2015,King2015,Isakov2015,Martin-Mayor2015,Selby2014,Selby2015,Aaronson2015b,Katzgraber2015,Hen2015,King2017,Mandra2017},
the situation can be summed up in the following way:

\begin{itemize}
\item  The behaviour of the D-Wave machines is consistent with quantum annealing.
\item  No scaling advantage (Òquantum speedupÓ) has so far been seen \cite{Katzgraber2015,King2017,Mandra2017}.
\item  QA is efficient in quickly finding good solutions as long as barriers are narrow, but ultimately gets stuck once broad barriers are encountered
\item   The Google D-Wave 2X results showing million-times speedup \cite{Denchev2015} are for native instances that perfectly fit the hardware graph of the device \cite{Zintchenko2015}.
\item   For generic problems that do not map well onto the hardware of a QA, performance will suffer significantly  \cite{Zintchenko2015,Katzgraber2015,King2017,Mandra2017}.
\item   Even more efficient classical optimisation algorithms exist for these problems (mentioned in \cite{Denchev2015}), which outperform the current D-Wave 2X device for most problem instances \cite{Isakov2015,Selby2014,Selby2015}. However, the race is on \cite{King2017,Mandra2017}.
\item  With improved engineering, especially faster annealing and readout, the time to perform a quantum annealing run can be reduced by a factor 100x over the current generation QA devices \cite{Zintchenko2015}. 
\item  However, misspecification of the cost function due to calibration inaccuracies is a challenge that may hamper the performance of analogue QA devices\cite{Zintchenko2015}. 
\item  Another challenge is the embedding of problems into the native hardware architecture with limited connectivity. 
\item  There is the open question of quantum speedup in analogue QA \cite{Katzgraber2015,Hen2015}.
\item  QA error correction has been demonstrated and may pave a path toward large scale noise-protected AQO devices \cite{Pudenz2014,Pudenz2015}.
\item  Typically, classically computationally hard problems  also seem to be  hard problems for QA devices  \cite{Katzgraber2015}.
\item  Improved machine calibration, noise reduction, optimisation of the QA schedule, larger system sizes and tailored spin-glass problems may be needed for demonstrating quantum speedup. However what is hard may not be easy to judge  \cite{King2017,Mandra2017}.
\item  It remains to see what the newest D-Wave 2000Q system can do with 2000 qubits.
\end{itemize}


\newpage

\section{Perspectives}

\subsection{Looking back}

Once upon a time, in 2010, there was a major European flagship project proposing to marry QIPC to classical high-performance computing (HPC). The effort failed (like earlier the Swedish flagship Wasa, 1628), for a number of good reasons:
(i) The HPC people had essentially no knowledge of what QIPC might be good for, and the QIPC side had no really convincing arguments; (ii) There was no clear and convincing focus on scalable quantum hardware and software; (iii) Solid-state circuits were still at an embryonic level, demonstrating some limited basic QIP functionality only at the 2-3 qubit level.

Since then, there has been dramatic progress in the way of superconducting devices and systems.
It is now possible to build a variety of superconducting Josephson junction multi-qubit  platforms able to seriously address proof-of-principles quantum simulations of significant interest for future Materials Science and Chemistry, as well as smaller-scale Physics problems (e.g. quantum magnetism) where classical computers already now cannot provide solutions. 
Moreover, D-Wave Systems now operates 2000 qubit systems for quantum annealing. One can expect these systems to develop toward better coherence. This means that there will be a range of systems and problems that can be investigated from both bottom-up and top-down points of view. 

Fortunately this message has now reached the European Union - a Quantum Technologies flagship will be launched in January 2019.

\subsection{Looking around}

The focus of the present review has so far been on superconducting devices and hybrid systems. It is now time to broaden the scope a bit and discuss the wider perspective including a number of emerging solid-state quantum technologies.

\subsubsection*{Spins implanted in semiconductors} 
Originally, the solid-state approach to quantum computers was a silicon-based nuclear spin quantum computer \cite{Kane1998}.
This line of research has been very active ever since, with extensive efforts on technology for implanting spin impurities in semiconductors, in particular silicon. 
Presently there is great progress at the 1- and 2-qubit level \cite{Muhonen2014,Veldhorst2014,Dehollain2014,Kalra2014,Morello2015,Christle2015,Widmann2015}
with reported qubit lifetimes up to 30 seconds \cite{Muhonen2014} and robust 2-qubit gates \cite{Kalra2014}. 
Nevertheless, experience suggests that it may take quite some time to build multi-qubit systems. Possible routes may be on-chip coupling of implanted impurity spin arrays \cite{Singh2016}, or photonic coupling of individual spin qubits.

From a QIP point of view, however, the most advanced spin systems involve multi-qubit NV-centres in diamond \cite{Childress_Hanson2013,Doherty2013}, with demonstrations of quantum error correction (QEC) \cite{Waldherr2014} and digital quantum simulation (DQS) \cite{Wang2015}. There are also advanced plans for large-scale QIP in diamond \cite{Nemoto2016b}.

\subsubsection*{Interfaces and networks}

In the future Quantum Internet \cite{Herbst2015,Kimble2008} interfaces between stationary qubits and photons will be critically important. Such interfaces involve the entanglement of qubits with single-photon emitters \cite{Matsuda2016} and are typically based on semiconductor quantum dots \cite{Gao2015,Zadeh2016} or NV-centres \cite{Gao2015,Nemoto2016a}. Important experimental steps toward large-scale quantum networks have recently been taken through demonstrations of loophole-free Bell tests \cite{Hensen2015,Hensen2016}, quantum network memory \cite{Reiserer2016}, perfect state transfer of an entangled photonic qubit \cite{Chapman2016}, and digital photonic QIP \cite{Barz2014a,Barz2014b,Greganti2016}.

\subsubsection*{Sensors} 

The future ultimate sensor may be a quantum computer at the tip of a scanning probe - "SPQ" - where e.g. the measured dephasing of the quantum device provides the information. Presently the quantum device is typically a diamond nanocrystal with an NV centre working as an advanced NMR probe with a built-in quantum pre-processor (see e.g. \cite{Chipaux2015,Kaufmann2013,McGuinness2014,Staudacher2015,Lovchinsky2016} and refs. therein).
 
\subsubsection*{Microfabricated ion traps}
There is intense work on scaling up microfabricated ion traps \cite{Cho2015,Brandl2016}, with applications to benchmarking \cite{Riofrio2016}, computing \cite{Monz2016,Debnath2016,Martinez2016a} and simulation \cite{Martinez2016b,Bohnet2016}. The NIST simulation \cite{Bohnet2016} involves up to 219 ions (qubits) with global control, simulating spin-dynamics in a classically tractable 2D Ising model, laying the groundwork for the classically hard case of the transverse Ising model with variable range interactions. 

\subsubsection*{Non-equilibrium quantum dynamics}

Strongly driven non-equilibrium Floquet systems can exhibit persistent time correlations at an emergent
subharmonic frequency - referred to as "discrete time crystal" (DTC). Very recently discrete
time translational symmetry breaking into a DTC has been observed experimentally in strongly driven spin systems, in an ion trap with $10$ ions \cite{ZhangMonroe2017}, and in an NV crystal with $10^6$ disordered nuclear spins \cite{ChoiLukin2017}. The work creates opportunities for exploring dynamical phases of matter and to control interacting, disordered many-body systems. This includes topological phases that might be used for quantum information tasks (see  \cite{ChoiLukin2017,ZhangMonroe2017} and refs. therein).

\subsubsection*{Quantum "cloud computing"}

In May 2016, IBM launched a cloud-enabled quantum computing platform, called IBM Quantum Experience, to allow users to run algorithms and experiments on a 5-qubit superconducting transmon-cQED quantum processor. One year later, in May 2017, the web-connected platform was expanded to 16 qubits \cite{IBM-Q2017,Castelvecchi2017,IBM_github2017}, and a 17-qubit platform was announceed to go commercial in the near future.
Although even a 16-qubit quantum simulator remains a "toy", to make it available online for experimentation represents a kind of "intellectual crowd funding" - it provides an interesting new game, and paves the way for new applications. Moreover, it may put focus on the issue of which platform (and/or which nation in the world) is going to win the race for the quantum computer. In the author's opinion this is a hyped non-issue - the dominance of nations, and the fate of the world, will not depend on quantum computers in a long time to come. Nevertheless, serious comparisons of scalable alternatives \cite{Linke2017} are very important and useful. In the end, hybrid platforms has always been the name of the game - the problem is to develop efficient communication interfaces.

\subsection{Looking ahead}

The field of experimental and applied quantum information processing with superconducting Josephson junction-based circuits and systems is now preparing for scaling up to levels of Quantum Supremacy.
Within a few years there will be well-controlled coherent platforms with 20-50 qubits addressing a range of algorithms and benchmarking protocols, comparing  favourably against the best classical systems and algorithms.
And in view of the commitments of various groups around the world, in 5-7 years there will most likely be coherent functional 100 qubit superconducting systems claiming Quantum Supremacy. 

Despite the experimental evidence of  Majorana bound states, there is probably a long way to go to demonstrate functional Majorana qubits, and to manipulate them in efficient ways. The situation reminds a bit of the situation with superconducting qubits 20 years ago: highly developed theory but only emerging experimental evidence for potential superconducting qubits. Regarding Majoranas, only time will tell.

Quantum computing, emulation and simulation have finally become serious endeavours at a practical level, promising approximate solutions to hard problems with quantum speedup. To describe the structure and function of complex biological molecules, computational chemistry has developed multi-scale modelling (see e.g. Warshel  \cite{Warshel2014}). This means embedding quantum problems in a classical molecular and dielectric environment to make the full problem (more) tractable, only involving the quantum solver when the physics is manifestly quantum. The catalytic centre of an enzyme, or the protein-producing core a ribosome, do need a full quantum-chemical treatment. In contrast, the surrounding molecular cage and solvent can be described by classical molecular dynamics and dielectric screening. In a similar way, a quantum computer must necessarily be embedded in a classical HPC environment, forming a hybrid classical system with a quantum accelerator. 
Since this is the way Biology and Life works, it looks like a good idea to pursue computational science in a similar way.


\section *{Acknowledgement} This work has been supported by the European Commission under contract ICT- FP7 600927 ScaleQIT, and by Chalmers University of Technology. The author is grateful for illuminating and helpful discussions with Michel Devoret, John Martinis, Will Oliver, Vitaly Shumeiko and Frank Wilhelm.


\newpage

\section *{References}


\begin{thebibliography}{99}
\bibitem{Comment_He1s2p}
Actually, to get the best possible correlation energy with this toy model, one should use a $1s2p$ basis: the angular (dipole) correlation energy generated by interacting virtual  $(1s\uparrow)^{-1} (2p\uparrow)$ and $(1s\downarrow)^{-1} (2p\downarrow)$ electron-hole pair excitations is larger than the radial one.
\end{thebibliography}
\end{document}